\pdfoutput=1
\documentclass[%
aip,
 amsmath,amssymb,floatfix,
 reprint,%
]{revtex4-1}

\usepackage{url}
\usepackage{breakurl}
\usepackage[breaklinks]{hyperref}

\usepackage[caption=false]{subfig}

\usepackage{graphicx}
\usepackage{dcolumn}
\usepackage{bm}

\usepackage[utf8]{inputenc}
\usepackage[T1]{fontenc}
\usepackage{mathptmx}
\usepackage{etoolbox}

\usepackage{enumerate} 
\usepackage{physics}
\usepackage{mathrsfs}

\usepackage{algorithm}
\usepackage{algpseudocode}

\usepackage{booktabs}
\usepackage{makecell}

\newcommand{\heh}{\ensuremath{\text{HeH}^+} }
\newcommand{\lih}{\ensuremath{\text{LiH}} }
\newcommand{\ethylene}{\ensuremath{\text{C}_2\text{H}_4} }
\newcommand{\paranitro}{\ensuremath{\text{C}_6\text{H}_6\text{N}_2\text{O}_2} }
\newcommand{\paranitrocomma}{\ensuremath{\text{C}_6\text{H}_6\text{N}_2\text{O}_2,} }

\newcommand{\bc}{\ensuremath{\mathbf{c}}}
\newcommand{\bC}{\ensuremath{\mathbf{C}}}
\newcommand{\h}{\ensuremath{\mathbf{h}}}
 \newcommand{\p}{\ensuremath{\mathbf{p}}}
\newcommand{\bA}{\ensuremath{\mathbf{A}}}
\newcommand{\bHess}{\ensuremath{\mathbf{Hess}}}
\newcommand{\bbeta}{\ensuremath{\boldsymbol{\beta}}}
\newcommand{\bgamma}{\ensuremath{\boldsymbol{\gamma}}}

\newcommand{\btheta}{\ensuremath{\boldsymbol{\theta}}}

\makeatletter
\def\@email#1#2{%
 \endgroup
 \patchcmd{\titleblock@produce}
  {\frontmatter@RRAPformat}
  {\frontmatter@RRAPformat{\produce@RRAP{*#1\href{mailto:#2}{#2}}}\frontmatter@RRAPformat}
  {}{}
}%
\makeatother

\begin{document}
\title{Scalable learning of potentials to predict time-dependent Hartree-Fock dynamics}
  \author{Harish~S. Bhat}
  \email{hbhat@ucmerced.edu}
  \affiliation{Applied Mathematics, University of California, Merced, 5200 N. Lake Rd, Merced, CA 95343 USA}
  \author{Prachi Gupta}\affiliation{NERSC, Lawrence Berkeley National Laboratory}
  \author{Christine~M. Isborn}
  \affiliation{Chemistry \& Biochemistry, University of California, Merced}


\begin{abstract}
We propose a framework to learn the time-dependent Hartree-Fock (TDHF) inter-electronic potential of a molecule from its electron density dynamics.   Though the entire TDHF Hamiltonian, including the inter-electronic potential, can be computed from first principles, we use this problem as a testbed to develop  strategies that can be applied to learn \emph{a priori} unknown terms that arise in other methods/approaches to quantum dynamics, e.g., emerging problems such as learning exchange-correlation potentials for time-dependent density functional theory.  We develop, train, and test three models of the TDHF inter-electronic potential, each parameterized by a four-index tensor of size up to $60 \times 60 \times 60 \times 60$.  Two of the models preserve Hermitian symmetry, while one model preserves an eight-fold permutation symmetry that implies Hermitian symmetry.  Across seven different molecular systems, we find that accounting for the deeper eight-fold symmetry  leads to the best-performing model across three metrics: training efficiency, test set predictive power, and direct comparison of true and learned inter-electronic potentials.  All three models, when trained on ensembles of field-free trajectories, generate accurate electron dynamics predictions even in a field-on regime that lies outside the training set.  To enable our models to scale to large molecular systems, we derive expressions for Jacobian-vector products that enable iterative, matrix-free training.
\end{abstract}

\maketitle


\section{Introduction}
\label{sect:intro}
To understand phenomena such as charge transfer, spectroscopy, and the response of a molecule to an applied laser field, it is essential to be able to compute the dynamics of electrons.  For such systems, the governing equation is the time-dependent electronic Schr\"{o}dinger equation (TDSE):
\begin{equation}
\label{eqn:TDSE}
\imath \partial_t \Psi(\mathbf{r},t) = \widehat H (\mathbf{r},t) \Psi(\mathbf{r},t).
\end{equation}
Here $\imath = \sqrt{-1}$ denotes the imaginary unit, and $\widehat H(\mathbf{r},t)$ is the electronic Hamiltonian operator that operates on the time-dependent many-body electronic wave function $\Psi(\mathbf{r},t)$, where $\mathbf{r}$ represents the spatial and spin coordinates of all electrons.  The word \emph{electronic} signifies that other particles (i.e., protons and more generally, nuclei) are held fixed as part of the Born-Oppenheimer approximation, and we solve for the electronic wave function in the field of fixed nuclei.  For real, multi-electron molecular systems of interest, direct, grid-based numerical solution of (\ref{eqn:TDSE}) is not possible.  Additionally, for many properties of interest, the full wave function $\Psi$ is unnecessary; the $1$-electron reduced density $\rho$, a much lower-dimensional quantity (itself computable from $\Psi$), would suffice.

In principle, time-dependent density functional theory (TDDFT) enables one to compute the reduced electron density $\rho$, without first solving the TDSE for $\Psi$  \cite{Ullrich2012}.  However, in TDDFT, the density-dependent exchange-correlation potential (itself a part of the overall inter-electronic potential) is not known and must be approximated \cite{Gross2012,Maitra16_220901}. TDDFT is widely used to simulate electron dynamics and excited state properties, but in some cases the approximations employed for the exchange-correlation potential lead to inaccurate electron dynamics and charge transfer \cite{elliott2012universal,Rappoport2012,Habenicht2014,Provorse2015,lacombe2020developing,Ranka2023}.  Learning a more accurate Hamiltonian, including the unknown density-dependent electron potential of TDDFT, would lead to advances in the simulation of the molecular properties and excited state dynamics.

At the time of writing, there are only three works we know of that seek to learn the TDDFT exchange-correlation 
potential from data consisting of temporal snapshots of the electron density \cite{Suzuki2020,bhat2021dynamic,Yang_2023}.  These works focus on one-dimensional model problems; extending the methods to three-dimensional systems of chemical/physical interest will be a large-scale undertaking from both a theoretical and computational point of view.  We lack an understanding of questions such as: (i) how should we form our training sets? (ii) how should we construct our model potentials? and (iii) how should we quantify errors?  The literature on learning potentials for quantum molecular systems is tilted heavily toward static, time-independent problems \cite{Snyder2012,Bartok2017,Chandrasekaran2019,Schleder2019,Jorgensen2019,Smith2019,Ceriotti2019} and offers little guidance on the answers to these questions, especially for molecular systems with, e.g., dozens of electrons.

The broad goal of this paper is to extend our understanding of how to learn potentials from time series observations of quantum systems.  We develop a learning framework for dynamics governed by the time-dependent Hartree-Fock (TDHF) equation.  We focus on TDHF for four reasons.  First, for TDHF, we know \emph{a priori} how to compute the entire Hamiltonian, including the potential term we seek to learn.  Hence we can compare our learned potentials against the ground truth.  Second, TDHF is numerically tractable even for large molecular systems, so there is no \emph{a priori} obstacle that stops us from generating training data.  Third, Hartree-Fock theory incorporates an exact treatment of electron exchange.  Consequently, the inter-electronic potential is purely a function of a $2$-electron integral tensor $(i j | k l)$---see (\ref{eqn:twoelec})---that enjoys an $8$-fold permutation symmetry.  Though existing density functionals that approximate electron exchange based on the local density do not have this symmetry, we expect an exact exchange-correlation functional to have symmetries similar to what we see in Hartree-Fock theory.  This motivates studying how to incorporate symmetries into a machine learning model of a potential.  Fourth, Hartree-Fock is itself the first model (approximating $\Psi$ using a single Slater determinant) in a sequence of increasingly accurate models that culminates in full configuration interaction (in which $\Psi$ is expanded using all possible Slater determinants excited from the Hartree-Fock reference determinant).  Time-dependent full configuration interaction yields numerical solutions of (\ref{eqn:TDSE}) that are exact within a finite-dimensional basis set \cite{mcweeny1989methods}.

Therefore, by studying and solving a machine learning problem for TDHF, we aim to produce insights that can be abstracted and generalized to learning problems that involve more accurate models for quantum dynamics.  The present study aims to guide future efforts to learn \emph{a priori unknown terms} that play important roles in time-dependent equations of motion.  This includes, but is not limited to, learning the exchange-correlation potential for TDDFT\cite{Suzuki2020,bhat2021dynamic,Yang_2023}, closure terms for time-dependent reduced density matrix functional theory and similar theories\cite{blum2012,AkbariA.2012Citt,FerreNicolas2016RDMF,bhat2024incorporating},  and surrogate models for the Kadanoff-Baym equations\cite{Bassi2023}.

The TDHF equation
\begin{equation}
\label{eqn:TDHF}
\imath\frac{d P(t)}{d t} =  \big [   H(P(t),t), P(t) \big],
\end{equation}
can be derived from the TDSE by approximating $\Psi$ as a single Slater determinant computed from a finite set of spin orbitals \cite{szabo2012modern}.  Here $P$ is the $1$-electron reduced density matrix corresponding to $\Psi$, while the Hamiltonian matrix $H$ is the operator $\widehat H$ from (\ref{eqn:TDSE}) expressed in a finite-dimensional basis  \cite{mclachlan1964time,mcweeny1989methods}.  Note that (\ref{eqn:TDHF}) only holds if one expresses $P$ and $H$ in orthonormal bases.  The full, time-dependent Hamiltonian in (\ref{eqn:TDHF}) is
\begin{equation}
\label{eqn:timedepham}
H(P(t),t) = H(P(t)) + V_{\text{ext}}(t).
\end{equation}
On the right-hand side, the field-free Hamiltonian $H(P(t))$ includes the kinetic energy of the electrons, the potential due to interaction with the fixed nuclei, and the density dependent inter-electronic repulsion term, also known as the Fock matrix in Hartree-Fock theory.  In (\ref{eqn:timedepham}), $V_{\text{ext}}(t)$ is an optional, external, time-dependent potential, e.g., an applied electric field.
In our prior work \cite{bhat2020machine,Gupta2022}, we initiated the project of learning a model $\widetilde{H}(P)$ of the true field-free Hamiltonian $H(P)$ from time series observations $\{P(t_j)\}$ of electron density matrices.

Throughout the remainder of this paper, if we write $H(\cdot, \cdot)$ with two arguments, then we are referring to the full, time-dependent Hamiltonian, the left-hand side of (\ref{eqn:timedepham}).  If we write $H(\cdot)$ with only one argument, as in $H(P(t))$ or $H(P)$, then we are referring to the field-free Hamiltonian.  

Here we retain one feature of our prior work: we \emph{train} our model $\widetilde{H}(P)$ on field-free trajectories, but we then \emph{test} our models by quantifying how well they predict field-on dynamics.    This is one way of measuring how well our models generalize to settings outside the training data.  Further details are given in Sections \ref{sect:traintest} and \ref{sect:properr}.

In the present work, we move beyond prior work in three ways: (i) scaling our methods to much larger molecular systems while maintaining accuracy, (ii) training using high-quality ensembles (instead of single trajectories), and (iii) incorporating an $8$-fold symmetry that was ignored in prior work.  We now explain these distinctions in turn.

First, we pay close attention to scalability to systems such as paranitroaniline  ($\paranitro$) in the STO-3G basis set, where the density matrices $P(t)$ are of size $60 \times 60$.  For $\paranitro$, the methods developed in this paper enable learning an inter-electronic potential parameterized by a tensor with more than $1.6$ million entries, with which we can propagate electron dynamics with an overall error on the order of $10^{-7}$---see Figure \ref{fig:LinftyPropErrors}.  This is six orders of magnitude better than what we can achieve using methods from our prior work.   We are unaware of other methods that can scale to this system size.  In the present work, we propose three models and derive expressions for Jacobian-vector products that enable matrix-free, iterative training.  We formulate our models in such a way that they preserve symmetries using linear algebra.  The combination of these techniques leads to scalability.

Second, in the present work, we train models using ensembles of field-free trajectories.  In our past work, we only explored training on single field-free trajectories.  We also pay close attention to the accuracy of these trajectories, employing a higher-order numerical method that, for the largest molecular systems we study, avoids spurious oscillations.  We hypothesize that the ensemble more thoroughly explores the inter-electronic potential landscape.  As we show, models trained on this ensemble have superior predictive power as measured by the ability to generate field-on trajectories $\widetilde{P}(t)$ that match true field-on trajectories $P(t)$.

Third, developing symmetry-constrained models lies at the core of the present work.  For a variety of machine learning problems in the sciences, models that incorporate symmetries have achieved superior results\cite{zhang2018end,Kondor2019,eismann2021hierarchical,satorras2021n,Klus_2021,Villar2021,li2024unifying}; much of this literature focuses on continuous symmetry groups in real space.  We present approaches tailored to the Hartree-Fock setting, but our methods can be generalized to other matrix/tensor-valued regression problems with permutation symmetries.  Two of the models we propose preserve Hermitian symmetry.  Another preserves a deeper $8$-fold permutation symmetry that has not been exploited in prior work.  We express our model in a basis for the space of all $4$-index tensors that satisfy this symmetry, ensuring symmetry preservation.

For the largest molecular system we study, we find that the $8$-fold symmetry-preserving ensemble-trained model dramatically outperforms all other models in terms of propagation error.  For all molecular systems studied here, the $8$-fold symmetry-preserving model leads to the smallest difference between learned and true Hamiltonians.  While incorporating known symmetries into the model clearly makes physical sense, our results show that it benefits the machine learning enterprise as well.  The $8$-fold symmetry-preserving model uses approximately one-eighth as many parameters as other models we consider.  For all systems, the $8$-fold symmetry-preserving model requires orders of magnitude fewer iterations to train.  These findings have important consequences for future machine learning efforts that involve time-dependent quantum systems trained on time-dependent data.

\section{Methods}
\label{sect:meth}
Our learning approach is guided by the following question: what training data and what type of model do we need in order to learn a TDHF inter-electronic potential that yields the smallest test set error?  

\subsection{Choice of initial condition and time-stepping method}
\label{sect:gendata}
In order to train and test our models, we must have on hand a reliable set of electron dynamics trajectories.  For a particular molecule and basis set, we apply standard electronic structure methods to compute the ground truth field-free Hamiltonian $H(P)$ and an initial condition $P(0)$ with an applied delta-kick perturbation.  Before doing anything further, we evolve the density matrix forward in time to $P(2 \delta t)$ with time step $\delta t = 8.268 \times 10^{-2}$ a.u. (atomic units).  We then relabel this density matrix $P(2 \delta t)$ as $P(0)$ and use this as our initial condition for all subsequent modeling/testing.  We use a modified version of the Gaussian electronic structure code \cite{GaussianDV} to compute the Hamiltonian $H(P)$ and the matrices $P(0)$ through $P(2 \delta t)$.  

To understand why we relabel $P(2 \delta t)$ as $P(0)$, note that in our past \cite{bhat2020machine,Gupta2022} and present work, our learning procedures depend on first estimating $dP/dt$ from $P(t)$ and then fitting $\imath dP/dt$ to $[H,P]$.  That is, we minimize the residual between the left- and right-hand sides of (\ref{eqn:TDHF}).  As expected given the nature of the Dirac delta, our numerical estimates of $dP/dt$ are much more accurate if we move past the time of the initial delta-kick perturbation.  This improves all learning procedures we have tested.

Equipped with $H(P)$ and $P(0)$, our task is to numerically integrate  (\ref{eqn:TDHF}) to produce a trajectory $\{P(t_j)\}$ of electron density matrices on a temporal grid $t_j = j \Delta t$ for $j = 0, \ldots, J$.  In our prior work \cite{bhat2020machine,Gupta2022}, we employed the second-order modified midpoint unitary transformation (MMUT) method \cite{Li2005}.  In the present work, for reasons we now explain, we employ a fourth-order scheme\cite{Casas2006} that we abbreviate as CI4.  We detail both schemes in Supplementary Material (SM) Section \ref{sect:propschemes}.  Note that the CI4 method is based on a fourth-order Magnus expansion.  In a comparative study of time-stepping methods for the time-dependent Kohn-Sham equations, a fourth-order Magnus integrator emerged as the clear winner\cite{Castro2018}.

\begin{figure*}[p]
\begin{center}
\includegraphics[width=6.125in]{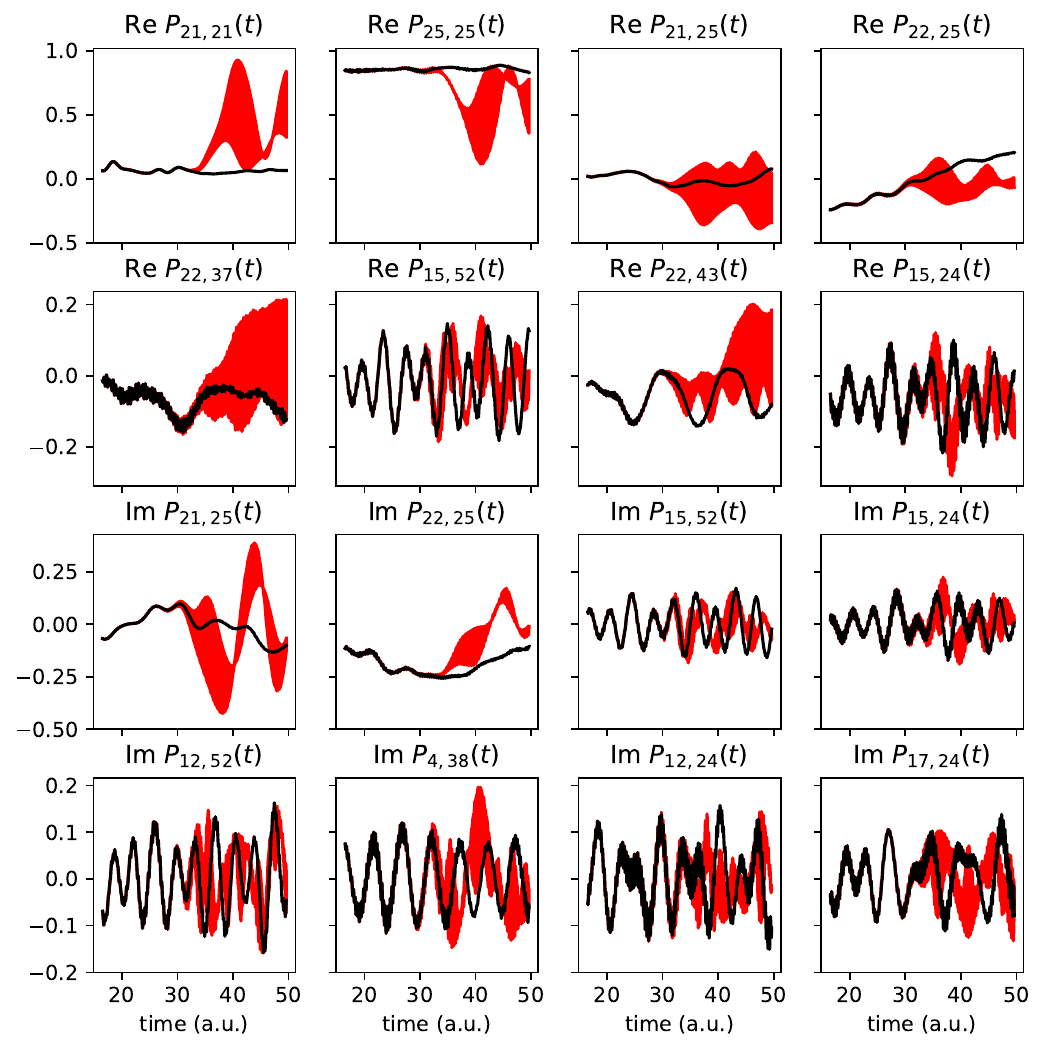}\\
\includegraphics[width=3in]{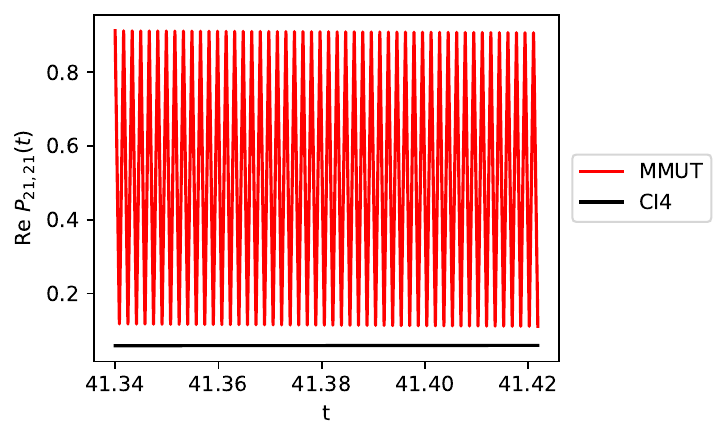}
\end{center}
\vspace{-0.6cm}
\caption{Preliminary long-term propagation test using the exact Hamiltonian and two different propagation methods, the second-order MMUT method (red) and the fourth-order CI4 method (black).  The bottom-most plot shows a zoomed in view of the real part of one density matrix element as a function of time, clearly showing that MMUT leads to large-amplitude spurious oscillations that are not present in the CI4 solution.}
\label{fig:prelimtest}
\end{figure*}

We present in Figure \ref{fig:prelimtest} the results of solving (\ref{eqn:TDHF}) for \paranitro in the STO-3G basis using both schemes, MMUT (red curves) and CI4 (black curves). We set $\Delta t = 8.268 \times 10^{-4}$ a.u. and propagate (\ref{eqn:TDHF}) for $200000$ steps; we use only the field-free Hamiltonian $H(P)$ and delta-kick initial condition $P(0)$ computed as described above, with no external field.  The $P(t)$ matrix for this system is of size $60 \times 60$.  We find that both schemes yield comparable results for the first $20000$ time steps, and hence omit these from our plots.  

From time step $20000$ to time step $60000$ (corresponding to physical times between $16.536$ a.u. and $49.608$ a.u.), we see from the top half of Figure \ref{fig:prelimtest} that the MMUT (red) trajectories diverge substantially from the CI4 (black) trajectories.  In the $4 \times 4$ array of plots, we show the $8$ components of both $\Re P(t)$ and $\Im P(t)$ that yield the greatest disagreement between the two schemes.  Importantly, the CI4 trajectories retain their overall structure until the final time step of $200000$ (not shown).  In contrast, the MMUT trajectories develop spurious oscillations at a frequency that greatly exceeds any frequencies exhibited during the first $20000$ time steps.

To see this more clearly, in the bottom-most plot of Figure \ref{fig:prelimtest}, we have zoomed in on the real part of the $P_{21,21}(t)$ trajectory between $50000$ and $50100$ time steps.  The large-scale, high-frequency oscillations in the MMUT curve should be viewed in sharp contrast to the totally flat behavior of the CI4 solution.  We find similar behavior for other systems of similar size, e.g., \ethylene in 6-31+$\text{G}^*$ (with $P(t)$ of size $46 \times 46$) and \lih in 6-311++$\text{G}^{**}$ (with $P(t)$ of size $29 \times 29$).  Experiments with MMUT showed that decreasing $\Delta t$ by a factor of $10$ delays the onset of this divergence between trajectories, but at prohibitive computational cost to reach the same final integration time. 

Therefore, for all subsequent calculations, we use CI4\cite{Casas2006} to solve (\ref{eqn:TDHF}). 
This includes generation of training and test data using either ground truth Hamiltonians $H(P)$ or learned Hamiltonians $\widetilde{H}(P)$.  We continue this thread in Section \ref{sect:traintest} below.

\subsection{Linear models}
\label{sect:model}
We begin with a set of atomic orbitals $\{ \phi_j \}$. Each $\phi_j$ is a function of a spatial coordinate $\mathbf{r} \in \mathbb{R}^3$; each $\phi_j$ is an approximate solution of Schr\"odinger's equation.  The atomic orbitals (AO) form a basis used to express molecular orbitals, and so the collection $\{ \phi_j \}$ is often referred to as a basis set.  Hartree-Fock electronic structure theory yields an expression for the ground truth Hamiltonian $H^\text{AO}$ (in the AO basis), a matrix-valued function of the electron density matrix $P^\text{AO}$ (in the AO basis).  For any complex quantity $Z$, let $\overline{Z}$ denote its complex conjugate. 
 Then the $(i,j)$-th component of $H^\text{AO}$ is
\begin{equation}
\label{eqn:gthamAO}
H^{\text{AO}}(P^\text{AO})_{i j} = H^{\text{core.AO}}_{i j} + \sum_{k l} \left[ (i j| l k) - \frac{1}{2} (i k | l j) \right] P^\text{AO}_{k l},
\end{equation}
where $H^\text{core.AO}$ comprises the kinetic and electron-nuclear part of the Hamiltonian matrix (again, in the AO basis), and $(i j | k l)$ is the $2$-electron integral defined by
\begin{equation}
\label{eqn:twoelec}
(i j | k l)  = \iint \overline{\phi_{i}(\mathbf{r}_1)} \phi_j(\mathbf{r}_1) H_{12} \overline{\phi_k(\mathbf{r}_2)}  \phi_{l}(\mathbf{r}_2)   d \mathbf{r}_2  d \mathbf{r}_1.
\end{equation}
Given the AO basis, we form the canonically orthogonalized (CO) basis \citep[\S 3.4.5]{szabo2012modern}.  Let $S$ be the overlap matrix with $S_{ij} = \int \overline{\phi_{i}(\mathbf{r})} \phi_{j}(\mathbf{r}) \, d\mathbf{r}$.  As $S$ is Hermitian, it can be diagonalized: $S = U s U^\dagger$ where $U$ is orthogonal and $s$ is real and diagonal.  Let $\mathcal{X} = U s^{-1/2}$, define $\phi_j' = \sum_{i} \phi_i \mathcal{X}_{ij}$ and check that $\int \overline{\phi_j'(\mathbf{r})} \phi_{\ell}'(\mathbf{r}) \, d\mathbf{r} = \delta_{j\ell}$, showing that the CO basis $\{ \phi_j' \}$ is orthonormal.  One can show that the electron density matrices and Hamiltonians in the CO and AO bases are related via 
\begin{subequations}
\label{eqn:AOCO}
\begin{align}
\label{eqn:PAOCO}
P &= \mathcal{X}^{-1} P^{\text{AO}} \mathcal{X}^{-\dagger} \\
\label{eqn:HAOCO}
H &= \mathcal{X}^\dagger H^{\text{AO}} \mathcal{X}.
\end{align}
\end{subequations}
Let us use the shorthand $\mathcal{E}$ for the following transformation of a critical tensor---the tensor in square brackets in the Hamiltonian (\ref{eqn:gthamAO})---into the CO basis:
\begin{equation}
\label{eqn:mathcale}
\mathcal{E}_{abcd} = \sum_{ijkl} \overline{\mathcal{X}_{ia}} \mathcal{X}_{jb} \left[ (i j| l k) - \frac{1}{2} (i k | l j) \right] \mathcal{X}_{kc} \overline{\mathcal{X}_{l d}}.
\end{equation}
Then (\ref{eqn:AOCO}) and (\ref{eqn:gthamAO}) together give us the ground truth Hamiltonian in the CO basis:
\begin{equation}
\label{eqn:gthamCO}
H(P)_{ab} = H^{\text{core}}_{ab} + \sum_{cd} \mathcal{E}_{abcd} P_{cd}.
\end{equation}
In this paper, unless specified otherwise via $^\text{AO}$ superscripts, all density and Hamiltonian matrices ($P$ and $H$) are in the CO basis.  We use the CO basis because it is orthonormal, and the TDHF equation (\ref{eqn:TDHF}) holds only in an orthonormal basis.  The AO basis is not guaranteed to be orthonormal---in practice, it is not.  Abstractly, we think of (\ref{eqn:gthamCO}) as 
\begin{equation}
\label{eqn:gtham}
H(P) = H^{\text{core}} + H_1(P),
\end{equation}
where $H_1(P)$ is the Hermitian, density-dependent, inter-electronic potential.  In this work, we are interested in methods that scale well to $N=60$ and beyond.  At $N=60$, the Hermitian matrix $H^{\text{core}}$ is determined by $N^2 = 3600$ real numbers.  The inter-electronic potential $H_1(P)$ is determined by the $4$-index tensor $\mathcal{E}$; accounting for symmetries that we describe further below, $\mathcal{E}$ is determined by just over $1.6 \times 10^6$ real numbers.  As high-dimensionality manifests most prominently in the $H_1(P)$ term, we focus on learning this inter-electronic potential in the present work.    In prior work, we have found that we can learn the constant term $H^{\text{core}}$ very accurately  \citep{bhat2020machine, Gupta2022}.

Purely for instructive/motivational purposes, we describe a basic linear model for $H(P)$.  We use boldfaced lowercase $\p \in \mathbb{C}^{N^2}$ and $\h \in \mathbb{C}^{N^2}$ to denote vectorized or flattened versions of $P$ and $H$, respectively.  Then a basic linear model is
\begin{equation}
\label{eqn:simplemodel}
\widetilde{\h} = \mathbf{b}^\text{core} + \p \mathbf{b}.
\end{equation}
Here we have an intercept $\mathbf{b}^\text{core}$, a complex vector of shape $N^2 \times 1$, and a complex coefficient matrix $\mathbf{b}$ of shape $N^2 \times N^2$.   A key problem with (\ref{eqn:simplemodel}) is that after we reshape $\widetilde{\h}$ into an $N \times N$ matrix $\widetilde{H}$, we have no guarantee that $\widetilde{H}$ is Hermitian, as is required to obtain physically meaningful results.

To build a physically accurate model, we must ensure that $\widetilde{H}$ is Hermitian.  One way to do this is to model only \emph{diagonal and upper-triangular elements} of $\widetilde{H}$, and then use symmetry to fill in the lower-triangular elements.  We followed this approach in our earlier work \citep{bhat2020machine, Gupta2022}.  However, our prior approach of manually tracking diagonal and upper-triangular entries of $\widetilde{H}$ causes our prior methods to scale poorly for large $N$.

We use $\widetilde{H}_1(P)$ to denote a model of the density-dependent, inter-electronic potential in (\ref{eqn:gtham}).  We present three scalable $\widetilde{H}_1(P)$ model, in increasing order of mathematical complexity.  All three are linear in the entries of $P$.  All three preserve Hermitian symmetry.  The third model preserves an additional $8$-fold permutation symmetry that we detail below.

All three $\widetilde{H}_1(P)$ models we present can be computed using standard numerical linear algebra operations, without manually looping over particular sets of indices as in our previous work.  As numerical linear algebra is highly optimized on both CPU and GPU architectures, this improves scalability of the model for larger $N$.

\subsubsection{Tied Regression}
Observe that if one wants to convert a matrix $M$ into a Hermitian matrix, one can simply use $(M + M^\dagger)/2$, i.e., the Hermitian part of $M$.  This motivates our first model.  With $P^R$ and $P^I$ denoting the real and imaginary parts of an electron density matrix $P$, we introduce the tied regression model:
\begin{subequations}
\label{eqn:tiedmodel}
\begin{align}
\label{eqn:realtarget}
R_{kl}(P) &= H^\text{core}_{kl} + \sum_{i,j} P^R_{ij} \beta_{ijkl}   \\
\label{eqn:imagtarget}
Q_{kl}(P) &=  \sum_{i,j} P^I_{ij} \beta_{ijkl} \\
\label{eqn:hammodel}
\widetilde{H}(P) &= \frac{1}{2}[R(P^R) + R(P^R)^T] +  \frac{\imath}{2}[Q(P^I) - Q(P^I)^T].
\end{align}
\end{subequations}
Here $R$, $Q$, $H^\text{core}$, and $\bbeta$ are all real; $^T$ denotes ordinary matrix transpose.  The idea is that $R$ and $Q$ respectively model the real and imaginary parts of the Hamiltonian.  As in the ground truth Hamiltonian, we allow $R$ and $Q$ to depend, respectively, on the real and imaginary parts of the density matrix $P$.  We do not need an intercept term in $Q$ because, as noted above, the constant Hamiltonian term $H^{\text{core}}$ is real.  Let $M = R + \imath Q$ be the presymmetrized model; the model (\ref{eqn:hammodel}) then consists of the symmetrization of $M$, i.e., $\widetilde{H} = (M + M^\dagger)/2$.

We refer to this as the \emph{tied regression} model because the coefficient tensor $\bbeta$ that multiplies $P^R$ in (\ref{eqn:realtarget}) is \emph{also used} to multiply $P^I$ in (\ref{eqn:imagtarget}).  In preliminary work, we began with a model that had
\begin{equation}
\label{eqn:oldimagtarget}
Q_{kl}(P) =  \sum_{i,j} P^I_{ij} \gamma^{1}_{ijkl}
\end{equation}
instead of (\ref{eqn:imagtarget}); for this model, the task is to learn both $\bbeta$ and $\bgamma$.  Model (\ref{eqn:tiedmodel}) effectively \emph{ties the parameters $\bbeta$ and $\bgamma$ to one another} (via $\bbeta = \bgamma$), reducing the number of real parameters from $2N^4$ to $N^4$, and thus reducing training times.  In SM Section \ref{sect:tiedgt}, we show that there exists $\bbeta$ such that the tied model (\ref{eqn:tiedmodel}) matches the ground truth Hamiltonian (in the CO basis).  Thus there is nothing gained by considering a model in which (\ref{eqn:oldimagtarget}) replaces (\ref{eqn:imagtarget}).

\subsubsection{Regression with Hermitian Representation}
\label{sect:reghermrep}
Though the symmetrization approach described above is simple, it is also potentially wasteful.  In (\ref{eqn:realtarget}-\ref{eqn:imagtarget}), we build full $N \times N$ models of $R$ and $Q$, but in (\ref{eqn:hammodel}), we only use the symmetric part of $R$ and the antisymmetric part of $Q$.  This begs the question of whether we can model only those parts of $R$ and $Q$ to begin with.  We especially seek to do this with linear/multilinear algebraic methods that lead to fast numerical implementations.

We begin by introducing a basis for the space of real symmetric matrices.  Given such a matrix of size $N \times N$, including the diagonal, there are a total of $N(N+1)/2$ upper-triangular entries.  To each such entry with index $(i,j)$, let us associate a symmetric matrix that is identically zero except for entries of $1$ at locations $(i,j)$ and $(j,i)$.  In case $i=j$, the matrix will have only one $1$ at position $(i,i)$.  In this way, to each integer $k$ from $1$ to $N(N+1)/2$, we associate an $N \times N$ matrix $B^R_k$.  Stacking these matrices, we form a three-index tensor $B^R$ of size $N \times N \times N(N+1)/2$.  By construction, these matrices form a basis for the space of real symmetric matrices; any such matrix $S$ can be written as
\begin{equation}
\label{eqn:realrepexample}
S_{i j} = \sum_{k=1}^{N(N+1)/2} B^R_{i j k} v_k
\end{equation}
for some $\mathbf{v} \in \mathbb{R}^{N(N+1)/2}$.  Continuing, suppose we wish to form a real, $4$-index tensor $\bbeta$ such that $\beta_{a,b,c,d} = \beta_{a,b,d,c}$ for all $1 \leq a,b,c,d \leq N$.  For each fixed choice of $(a,b)$, $\beta_{a,b,:,:}$ is a real symmetric matrix that can be represented using $B^R$ as in (\ref{eqn:realrepexample}).  Thus $\bbeta$ is determined by a $3$-index tensor $\mathbf{v}$ such that
\begin{equation}
\label{eqn:realrepexample2}
\beta_{abcd} = \sum_{k=1}^{N(N+1)/2} B^R_{c d k} v_{a b k}.
\end{equation}
Following a similar construction, we can create a basis for the space of real antisymmetric matrices.  Now we exclude the diagonal, so there are a total of $N(N-1)/2$ upper-triangular entries.  To each such entry with index $(i,j)$, we associate an antisymmetric matrix that is identically zero except for entries of $1$ and $-1$ at locations $(i,j)$ and $(j,i)$, respectively.  In this way, to each integer $k$ from $1$ to $N(N-1)/2$, we associate an $N \times N$ matrix $B^I_k$.  Stacking these matrices, we form a three-index tensor $B^I$ of size $N \times N \times N(N-1)/2$.

Now suppose we wish to form a real, $4$-index tensor $\bgamma$ such that $\gamma_{a,b,c,d} = -\gamma_{a,b,d,c}$ for all $1 \leq a,b,c,d \leq N$.  Following the same logic as above, $\bgamma$ is determined by a $3$-index tensor $\mathbf{w}$ such that
\begin{equation}
\label{eqn:realrepexample3}
\gamma_{abcd} = \sum_{k=1}^{N(N-1)/2} B^I_{c d k} w_{a b k}.
\end{equation}
We now use (\ref{eqn:realrepexample2}) and (\ref{eqn:realrepexample3}) to build our Hermitian representation model
\begin{equation}
\label{eqn:hermrepmodel}
\widetilde{H}(P)_{kl} = H^\text{core}_{kl} + \sum_{i,j} P^R_{ij} \beta_{ijkl} + \imath \sum_{i,j} P^I_{ij} \gamma_{ijkl}.
\end{equation}
The tensors $B^R$ and $B^I$ are known and constant, depending only on the dimension $N$.  Treating $H^\text{core}$ as known, the coefficients of this model amount to the entries of $\mathbf{v}$ and $\mathbf{w}$, of size $N \times N \times N(N+1)/2$ and $N \times N \times N(N-1)/2$, respectively.  There are a total of $N^4$ entries, meaning this model has precisely the same size as our tied regression model (\ref{eqn:tiedmodel}).

\begin{algorithm}[H]
\caption{Populating the sparse tensor $\mathscr{S}$ that enables representation of $4$-index tensors with $8$-fold symmetry}\label{algo:pop}
\begin{algorithmic}[1]
\Require $m=0$ and empty $N \times N \times N \times N$ tensor $\mathscr{B}$
\For{$i,j,k,l = 1, \dots, N$}
\If{$\mathscr{B}(i,j,k,l) = 0$}
\State $m \gets m+1$
\For{$c = 1, \ldots, 8$}
\State $\mathscr{B}(\sigma^c(i,j,k,l)) \gets 1$
\State append the $5$-tuple $(\sigma^c(i,j,k,l),m)$ to the index list of sparse tensor $\mathscr{S}$
\EndFor
\EndIf
\EndFor
\end{algorithmic}
\end{algorithm}

This model \emph{does not} tie $\bbeta$ and $\bgamma$ together. Let $:$ denote matrix-tensor contraction, so that
\[
\sum_{i,j} P^R_{ij} \beta_{ijkl} = [P^R : \bbeta]_{kl}.
\]
Then we can rewrite (\ref{eqn:hermrepmodel}) as $\widetilde{H}(P) = R + \imath Q$ with
$R = H^\text{core} + P^R : \bbeta$ and $Q = P^I : \bgamma$.  By virtue of constructing $\bbeta$ and $\bgamma$ using (\ref{eqn:realrepexample2}-\ref{eqn:realrepexample3}), we guarantee that $R$ and $Q$ are symmetric and antisymmetric, respectively.  Thus $\widetilde{H}(P)$ is Hermitian.  We do not need a final symmetrization step, as in (\ref{eqn:hammodel}).

In SM Section \ref{sect:hermgt}, we show that there exists $\bbeta$, $\bgamma$ such that the Hermitian representation model (\ref{eqn:hermrepmodel}) matches the ground truth Hamiltonian (in the CO basis).

\subsubsection{Eight-Fold Symmetry-Preserving Regression}
\label{sect:eightfoldsymmdef}
The Hermitian symmetry preserved by the first two models (\ref{eqn:tiedmodel}) and (\ref{eqn:hermrepmodel}) is a consequence of a deeper set of symmetries.   Assuming real atomic orbitals $\{ \phi_j \}$, the $4$-index tensor $(i j | k l )$ defined in (\ref{eqn:twoelec}) satisfies\cite{BrailsfordHall1971} the $8$-fold symmetry
\begin{multline}
\label{eqn:eightfold}
(ij|kl)=(ji|lk)=(kl|ij)=(lk|ji)\\
=(ji|kl)=(lk|ij)=(ij|lk)=(kl|ji).
\end{multline}
Based on these observations, let us attempt to model the $4$-index tensor $(i j | k l)$ itself.  The first part of our procedure is analogous to how we formed bases for the spaces of real symmetric and antisymmetric matrices in Section \ref{sect:reghermrep}.

Here we form a basis $\mathscr{S}$ for the space of $N \times N \times N \times N$ tensors that satisfy the $8$-fold symmetry (\ref{eqn:eightfold})---this space has dimension\cite{BrailsfordHall1971}
\begin{equation}
\label{eqn:BHNT}
n_T = \frac{1}{8} N(N+1)(N^2 + N + 2).
\end{equation}
To store this basis efficiently, we let $\mathscr{S}$ be a sparse tensor of dimension $N \times N \times N \times N \times n_T$, determined entirely by a list of indices where its value is $1$.

To populate $\mathscr{S}$, we follow Algorithm \ref{algo:pop}. For $m = 1, 2, \ldots, 8$, let $\sigma^m(i,j,k,l)$ be the $m$-th permutation seen in (\ref{eqn:eightfold}).  For instance, $\sigma^4(i,j,k,l) = (l,k,j,i)$. Let $\mathcal{O}(i,j,k,l)$ denote $\{ \sigma^m(i,j,k,l) \mid m = 1, \ldots, 8\}$, i.e., the orbit of the $4$-tuple $(i,j,k,l)$ under the group $\{ \sigma^m \}_{m=1}^{8}$.  In Algorithm \ref{algo:pop}, each value of $m$ corresponds to one unique orbit $\mathcal{O}(i,j,k,l)$.  After Algorithm \ref{algo:pop} terminates, $\mathscr{S}_{:,:,:,:,m}$ (letting the first four indices of $\mathscr{S}$ be free but fixing the last index at $m$) will be the $m$-th element of the basis, an $N \times N \times N \times N$ tensor that satisfies the symmetry (\ref{eqn:eightfold}).  In the algorithm, the purpose of the binary tensor $\mathscr{B}$ is to signal whether we have already captured the orbit $\mathcal{O}(i,j,k,l)$.  Upon the algorithm's termination, $\mathscr{B}$ will be identically equal to $1$ and $m$ will equal $n_T$.   With this framework, we build the $8$-fold symmetry-preserving model
\begin{subequations}
\label{eqn:eightfoldmodel}
\begin{align}
\label{eqn:eemodel}
\tau_{ijkl} &= \sum_{m=1}^{n_T} \mathscr{S}_{ijklm} \beta_m \\
\label{eqn:efmodel}
\widetilde{H}(P)_{i j} &= H^\text{core}_{i j} + \sum_{k l} \left[\tau_{i j l k} - \frac{1}{2} \tau_{i k l j} \right] P_{k l}.
\end{align}
\end{subequations}
As $\tau$ is a linear combination of basis elements from $\mathscr{S}$, it is clear that $\tau$ will satisfy the $8$-fold symmetry (\ref{eqn:eightfold}).
In this model, as long as $P$ is Hermitian, so is $H$:
\begin{multline*}
\overline{\widetilde{H}(P)_{ji}} = \overline{H^\text{core}_{ji}} + \sum_{k l} \left[\tau_{j i l k} - \frac{1}{2} \tau_{j k l i} \right] \overline{P_{k l}} \\
= H^\text{core}_{ij} + \sum_{l k} \left[\tau_{i j k l} - \frac{1}{2} \tau_{i l k j} \right] P_{l k} = \widetilde{H}(P)_{i j}.
\end{multline*}
For the model (\ref{eqn:eightfoldmodel}), the parameters consist of $\bbeta \in \mathbb{R}^{n_T}$.  This is roughly $1/8$-th as many parameters as needed in the models (\ref{eqn:tiedmodel}) and (\ref{eqn:hermrepmodel}).  In SM Section \ref{sect:eightfoldgt}, we derive $\bbeta$ such that the model (\ref{eqn:eightfoldmodel}) matches the ground truth Hamiltonian.

\subsection{Generation of data sets for training and testing}
\label{sect:traintest}
To train and test our models, we generate data using the following procedures, taking care to keep these procedures uniform across all molecular systems that we study.  At a high level, we emphasize that we train models using \textbf{field-free} trajectories, either a single such trajectory or an ensemble.  We then test our models using \textbf{field-on} trajectories.

In what follows, $H(P)$ can either be the ground truth Hamiltonian, or it can be one of the $\widetilde{H}(P)$ models described above.  When we numerically solve (\ref{eqn:TDHF}) in any setting, we monitor $P(t)$ at every time step to ensure that it remains Hermitian and idempotent, a requirement for Hartree-Fock electron density matrices.

\textbf{Single field-free trajectory}: As described in the first three paragraphs of Section \ref{sect:gendata} above, we take as our initial condition an electron density matrix $P(0)$ that has been advanced slightly past the time of an initial delta-kick perturbation.  We choose the delta-kick perturbation because it is maximally delocalized in Fourier space (equivalently, maximally localized in time), leading to trajectories that explore more modes of oscillation, which in turn helps our learning procedures to reconstruct the global inter-electronic potential.  Using this $P(0)$ and a field-free Hamiltonian $H(P)$, we apply the CI4 method to numerically solve (\ref{eqn:TDHF}) and compute a \emph{single field-free trajectory} $\{P(t_j)\}_{j=0}^J$.  Here and in what follows, we use the temporal grid $t_j = j \Delta t$ and $J = 200000$ integration steps.  Unless specified otherwise, $\Delta t = 8.268 \times 10^{-4}$ a.u.

When we use this trajectory for training, we also compute time-derivatives $\{ \dot{P}(t_j)\}_{j=2}^{J-2}$ using the fourth-order centered-difference formula
\begin{equation}
\label{eqn:loss4point}
\dot{P}(t_j) = \frac{-P(t_{j+2})+8 P(t_{j+1})-8 P(t_{j-1})+P(t_{j-2})}{12\Delta t} + O(\Delta t^4).
\end{equation}
We then restrict the trajectory to match the points in time at which we have time-derivatives, i.e., we pair $\{ \dot{P}(t_j)\}_{j=2}^{J-2}$ with $\{ P(t_j)\}_{j=2}^{J-2}$ and use the pair for training.

\textbf{Ensemble of field-free trajectories}: We take the $P(0)$ described in the previous paragraph and perturb it using a random Hermitian matrix $\mathscr{H}$, creating a new initial condition $\mathscr{P}(0) = P(0) + \varepsilon \mathscr{H}$.  To generate $\mathscr{H}$, we start with $2 N^2$ independent samples from the standard normal distribution, from which we form two real matrices $\mathscr{D}_R$ and $\mathscr{D}_I$ both of size $N \times N$.  Let $^\dagger$ denote conjugate transpose.  We take $\mathscr{D} = \mathscr{D}_R + \imath \mathscr{D}_I$ and then set $\mathscr{H} = (\mathscr{D} + \mathscr{D}^\dagger)/2$.  In this paper, we set $\varepsilon = (10/N^2) \sum_{i,j} |P_{ij}(0)|$, i.e., $10$ times the mean absolute value of $P(0)$.  Though $\mathscr{P}(0)$ is Hermitian, we must modify it slightly to be idempotent.  We compute the eigendecomposition $\mathscr{P}(0) = V D V^\dagger$.  Any entries of $D$ greater than $1/2$ are set to $1$, while entries at most equal to $1/2$ are set to $0$.  With this modified $D$, the resulting $\mathscr{P}(0) = V D V^\dagger$ is guaranteed to be idempotent.

Using the above procedure, we generate a cloud of Hermitian, idempotent initial conditions $\{ \mathscr{P}_k(0) \}_{k=1}^{100}$.  Given a field-free Hamiltonian $H(P)$ and this cloud of initial conditions, we use CI4 (with $\Delta t = 8.268 \times 10^{-4}$ a.u.) to numerically integrate forward in time for $J=20000$ steps.  With $t_j = j \Delta t$, this yields a cloud of trajectories $\{ \mathscr{P}_k(t_j) \}_{k=1}^{100}$ for $j = 0, \ldots, J$.  We apply (\ref{eqn:loss4point}) to compute time-derivatives $\{ \dot{\mathscr{P}}_k (t_j)\}_{k=1}^{100}$ on the equispaced grid $t_j = j \Delta t$ for $j = 2, \ldots, J-2$.  We restrict $\{ \mathscr{P}_k (t_j)\}_{k=1}^{100}$ to the same temporal grid, so that trajectories and their time-derivatives are matched.

\textbf{Single field-on trajectory}: Given a field-free Hamiltonian $H(P)$, we form the time-dependent Hamiltonian $H(P, t) = H(P) + V_{\text{ext}}(t)$ with external forcing term $V_{\text{ext}}(t) =  E_z \sin(\omega t)\mu_z$.  Here $E_z$ is the applied electric field strength in the $z$ direction, $\omega$ is the electric field frequency, and $\mu_z$ is the dipole moment matrix in the $z$ direction. For this study, the electric field is applied only for one cycle ($3.55 \text{fs}$) starting at $t=0$, with $\omega = 0.0428$ a.u.$^{-1}$ and $E_z = 0.05$ a.u.  For field-on trajectories, we take $P(0)$ to be the ground state density matrix with no perturbations (i.e., no delta-kick perturbation).  With $H(P,t)$ and $P(0)$ thus defined, we use CI4 with fixed time step $\Delta t$ to numerically solve (\ref{eqn:TDHF}) and compute the \emph{single field-on} trajectory $\{P(t_j)\}_{j=0}^J$.  Unless we specify otherwise, $\Delta t = 8.268 \times 10^{-4}$ a.u.  Because we do not use field-on trajectories for model training, we do not compute $\dot{P}$.

\subsection{Training: Mathematical Formulation}
\label{sect:training}
For each $\widetilde{H}$ model we have proposed, let $\btheta$ denote the vector of model parameters.  For the tied regression model (\ref{eqn:tiedmodel}) and the $8$-fold symmetry-preserving model (\ref{eqn:eightfoldmodel}), $\btheta = \bbeta$.  For the Hermitian representation model (\ref{eqn:hermrepmodel}), $\btheta = (\mathbf{v},\mathbf{w})$.  In this section, we will emphasize the parametric dependence of each $\widetilde{H}$ model on its parameters $\btheta$ by writing $\widetilde{H}(P; \btheta)$.

To train a given $\widetilde{H}(P; \btheta)$ model for a given molecular system (choice of molecule plus basis set), we use one of two training sets: the single field-free trajectory or the ensemble of field-free trajectories.  As detailed above in Section \ref{sect:traintest}, we generated both training sets by solving (\ref{eqn:TDHF}) with the true Hamiltonian (\ref{eqn:gthamCO}).  Both training sets are of the form $\{ ( P^{(j)}, \dot{P}^{(j)} ) \}_{j=1}^{n_\text{train}}$; each pair in this set consists of an electron density matrix and its time-derivative, taken from the same trajectory and evaluated at the same time.

Given one of the three models (\ref{eqn:tiedmodel}), (\ref{eqn:hermrepmodel}), or (\ref{eqn:eightfoldmodel}), and either training set, the residual associated with the $j$-th data point is the difference between the left- and right-hand sides of the TDHF equation (\ref{eqn:TDHF}) with the time-derivative $dP/dt$ approximated via $\dot{P}$ defined in (\ref{eqn:loss4point}):
\begin{equation}
\label{eqn:res}
S^{(j)}(\btheta) = \imath\Dot{P}^{(j)} - \big [   \widetilde{H}(P^{(j)}; \btheta), P^{(j)} \big]
\end{equation}
We then take as our loss function the sum of squared magnitudes of all residuals:
\begin{equation}
\label{eqn:loss}
\mathcal{L}(\btheta) = \sum_{j=1}^{n_{\text{train}}} \sum_{a,b} | S^{(j)}_{ab}(\btheta) |^2.
\end{equation}
If we were to use the ground truth Hamiltonian $H(P)$ instead of a model $\widetilde{H}(P)$, the loss would be practically zero. Therefore, to train the model, we seek $\btheta^\star = \min_{\btheta}  \mathcal{L}(\btheta)$.  As each of our models $\widetilde{H}(P)$ is linear in the parameters $\btheta$, so is the residual (\ref{eqn:res}). Hence the loss (\ref{eqn:loss}) is quadratic in $\btheta$; quadratic minimization problems have been well-studied.  To leverage known techniques, we put our problem into standard form.  As the residual is linear in the parameters $\btheta$,  we can write it as
\begin{equation}
\label{eqn:Sexpan}
S^{(j)}(\btheta) = \underbrace{ \begin{bmatrix} \nabla_{\btheta} S^{(j)}\end{bmatrix} }_{A^j} \btheta + \underbrace{ S^{(j)}(\mathbf{0}) }_{-\bc^j}
\end{equation}
Let $|\btheta|$ denote the length of the vector $\btheta$.  In (\ref{eqn:Sexpan}), we interpret $S^{(j)}$ as a flattened vector in $\mathbb{C}^{N^2}$, so that the Jacobian $A^j = \nabla_{\btheta} S^{(j)}$ is a matrix of shape $N^2 \times |\btheta|$.  We treat $\bc^j$ as a column vector of shape $N^2 \times 1$.  Stacking  objects vertically, we can rewrite the loss (\ref{eqn:loss}) as
\begin{equation}
\label{eqn:loss3}
\mathcal{L}(\btheta) = \left\| \begin{bmatrix} A^1 \\ \vdots \\ A^{n_{\text{train}}} \end{bmatrix} \btheta - \begin{bmatrix} \bc^1 \\ \vdots \\ \bc^{n_{\text{train}}} \end{bmatrix} \right\|^2 = \| \bA \btheta - \bC \|^2.
\end{equation}
The upshot is that training our models is equivalent to solving a standard linear least-squares minimization problem: given the $(n_{\text{train}} N^2) \times |\btheta|$ matrix $\mathbf{A}$ and the $(n_{\text{train}} N^2) \times 1$ vector $\mathbf{C}$, find $\btheta$ that minimizes (\ref{eqn:loss3}).  The norm $\| \cdot \|$ in (\ref{eqn:loss3}) is the standard Euclidean norm in $\mathbb{C}^{n_{\text{train}} N^2}$.

\subsubsection{Gradient- and Hessian-based minimization of the loss}
If we take the gradient of the loss (\ref{eqn:loss3}) with respect to $\btheta$ and set it equal to zero, the resulting solution can be expressed in terms of (i) the Hessian of the loss and (ii) the gradient of the loss at $\btheta = \mathbf{0}$:
\begin{subequations}
\label{eqn:overallhessapproach}
\begin{align}
\label{eqn:thetanormal2}
\btheta^\star &= -(\bHess \ \mathcal{L})^{-1} \left[ \nabla_{\btheta} \mathcal{L} (\mathbf{0}) \right] \\
\label{eqn:hessloss}
\bHess \ \mathcal{L} &= \Re ( \bA^\dagger \bA + \bA^t \overline{\bA} ) \\
\label{eqn:gradlossatzero}
\nabla_{\btheta} \mathcal{L} (\mathbf{0}) &= -\Re (\bA^\dagger \bC + \bA^t \overline{\bC}).
\end{align}
\end{subequations}
As $\bHess$ is of size $| \btheta | \times | \btheta |$, we can only use (\ref{eqn:thetanormal2}) to train our models when the number of parameters $| \btheta |$ is sufficiently small.  When $\bHess$ has eigenvalues near zero, we replace the matrix inverse with the Moore-Penrose pseudoinverse\cite{Hansen1998} with threshold $10^{-12}$.  Here we treat problems for which $N$ can be as large as $60$; in this case, for the models (\ref{eqn:tiedmodel}) and (\ref{eqn:hermrepmodel}), $| \btheta | = N^4 = 12.96 \times 10^6$.  Assuming $8$-byte double precision floats, a Hessian of size $N^4 \times N^4$ would consume over $1$ petabyte, motivating the use of matrix-free methods.

\subsubsection{Matrix-free minimization of the loss}
\label{sect:lsmr}
Returning to the loss (\ref{eqn:loss3}), we see that the critical objects are the Jacobians $A^j = \nabla_{\btheta} S^{(j)}$, which can be written as
$A^j_{k,l,m} = {\partial S^{(j)}_{kl}} / {\partial \theta_m}$.
Thus $\bA$ is a $4$-index tensor of dimension $n_{\text{train}} \times N \times N \times | \btheta |$.  Since we will not (in general) be able to store $\bA$ in memory, we will present efficient numerical algorithms to compute the left and right products of $\bA$ against tensors of appropriate sizes.  We will use these algorithms as inputs to the matrix-free method LSMR \cite{Fong2011}, and thereby compute minimizers of the loss (\ref{eqn:loss3}).

Note that in situations where $\bA$ does not have full column rank, LSMR returns the solution $\btheta^\star$ with minimal $2$-norm; this is particularly useful for problems such as ours, where $\bHess$ may have nearly zero eigenvalues and high condition number.  Our prior work combatted these problems by combining Hessian-based training with a Tikhonov or ridge regularization approach \cite{Gupta2022}.  For the present work, when we tried adding such regularization terms to the loss, we found that results did not improve, likely due to the implicit regularization already present in LSMR.

Please see SM Section \ref{sect:modelderivatives} for detailed expressions of Jacobian-vector products (both left and right) for all three models (\ref{eqn:tiedmodel}), (\ref{eqn:hermrepmodel}), and (\ref{eqn:eightfoldmodel}). In SM, we also explain how we implement these products efficiently, so that we can train our models even when they contain tens of millions of parameters.  For model (\ref{eqn:eightfoldmodel}), we also provide (in SM Section \ref{sect:eightfoldderivs}) expressions for the gradient and Hessian of the loss $\mathcal{L}$, which are used to train model (\ref{eqn:eightfoldmodel}) for small molecular systems.

\begin{figure*}[t]
\centering
\subfloat[Field-free propagation errors]
{\includegraphics[width=0.4\textwidth]{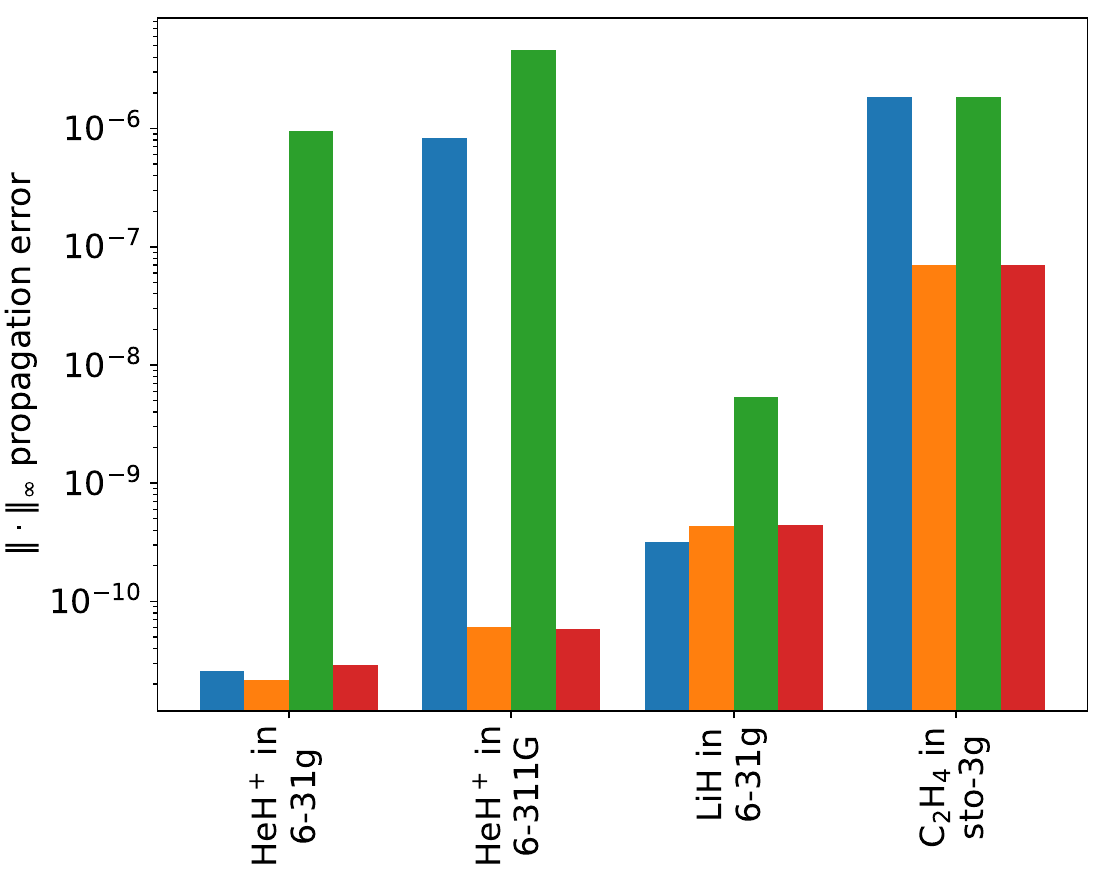}}
\subfloat[Field-on propagation errors]
{\includegraphics[width=0.4\textwidth]{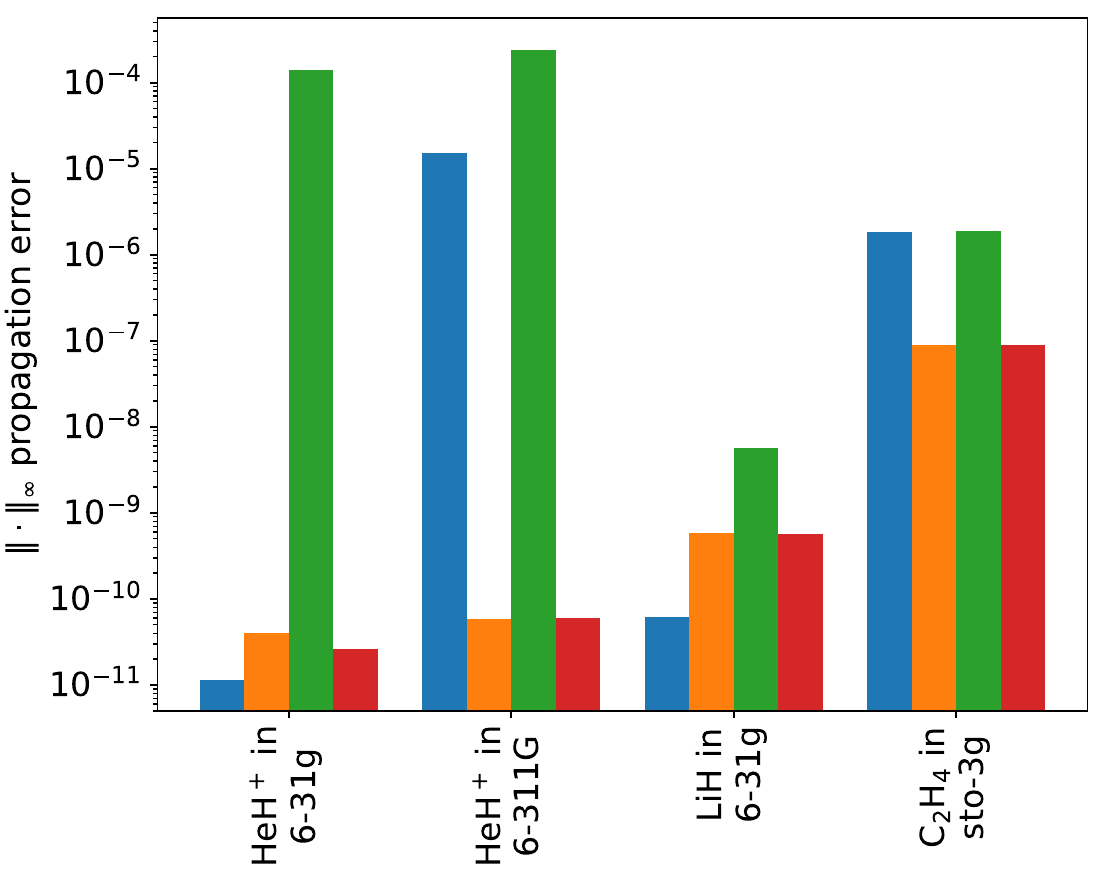}}
\subfloat{\raisebox{2.25cm}{\includegraphics[width=0.15\textwidth]{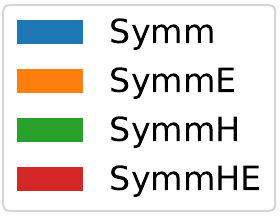}}}
\caption{For the four smaller molecular systems, we plot $\infty$-norm propagation errors (\ref{eqn:inftyerr}) for the $8$-fold symmetry-preserving model (\ref{eqn:eightfoldmodel}) trained using LSMR (Symm, SymmE) and also using the exact Hessian-based approach (SymmH, SymmHE) in (\ref{eqn:thetanormal2}).  The suffix ``E'' indicates that the model has been trained on an ensemble of field-free trajectories; otherwise, the model has been trained on a single field-free trajectory.  The results show that the iterative LSMR method leads to models that compare very favorably against those trained via the exact Hessian-based approach.  This is true for both field-free (training, left) and field-on (test, right) propagation errors.}
\label{fig:HessWash}
\end{figure*}

\subsection{Training: Computational Workflow}
For models (\ref{eqn:tiedmodel}) and (\ref{eqn:hermrepmodel}), the $N^4$ scaling of the number of parameters leads us to use the iterative, matrix-free LSMR method to train.  For the $8$-fold symmetry-preserving model (\ref{eqn:eightfoldmodel}), we train using LSMR for all molecular systems---though this is only necessary when $N \geq 29$, i.e., for molecular systems (v), (vi), and (vii).  Molecular systems (i), (ii), (iii), and (iv) are sufficiently small such that we can train model (\ref{eqn:eightfoldmodel}) using the exact Hessian-based formula (\ref{eqn:thetanormal2}).  We carry this out for the purposes of comparison.

In turn, we train all models on the single field-free trajectory and on the ensemble of field-free trajectories described in Section \ref{sect:traintest}.  There are small differences in how we treat the smaller molecular systems (i-iv) versus the larger systems (v-vii).  Let us first describe these differences for training on single trajectory data.  For the smaller systems, when we use LSMR, we use the entire matched data set $\{P(t_j), \dot{P}(t_j)\}_{j=2}^{j=J-2}$ with $J = 200000$.  For the larger systems, we take every $10$th snapshot in time, so that our training set is of size $20000$.

The ensemble training set initially consists of a cloud of $100$ matched trajectories $\{ \mathscr{P}_k(t_j), \dot{\mathscr{P}}_k(t_j) \}_{k=1}^{100}$ for $j=2, \ldots, J-2$ with $J=20000$.  For the smaller molecular systems, we take every $50$th snapshot from this cloud, for a total of $40000$ temporal snapshots ($400$ snapshots from each of the $100$ trajectories).  We \emph{augment} this using $40000$ snapshots from the single field-free trajectory and its time-derivative, i.e., we augment using every $5$th snapshot from $\{P(t_j), \dot{P}(t_j)\}_{j=2}^{j=199998}$.  In total, for smaller molecular systems, our ensemble training set consists of $80000$ snapshots.  For the larger molecular systems, we take every $100$th snapshot from the cloud and augment with every $10$th snapshot from the single field-free trajectory, leading to a total ensemble training set size of $40000$  snapshots. 

We use \textbf{Tied}, \textbf{Herm}, and \textbf{Symm} to denote the respective $\widetilde{H}(P)$ models (\ref{eqn:tiedmodel}), (\ref{eqn:hermrepmodel}), and (\ref{eqn:eightfoldmodel}), each \emph{trained on the single field-free trajectory} using the matrix-free LSMR algorithm.  We use \textbf{TiedE}, \textbf{HermE}, and \textbf{SymmE} to denote the respective $\widetilde{H}(P)$ models (\ref{eqn:tiedmodel}), (\ref{eqn:hermrepmodel}), and (\ref{eqn:eightfoldmodel}), each \emph{trained on the ensemble of field-free trajectories} using the matrix-free LSMR algorithm.  Finally, we use \textbf{SymmH} and \textbf{SymmHE} to denote the $8$-fold symmetry-preserving $\widetilde{H}(P)$ model (\ref{eqn:eightfoldmodel}), \emph{trained via the exact Hessian-based formula} (\ref{eqn:thetanormal2}), on the single trajectory or ensemble training sets, respectively. 

In summary, we obtain $8$ trained models for each of the $4$ smaller molecular systems (i), (ii), (iii), and (v).  We obtain $6$ trained models for each of the $3$ larger molecular systems (iv), (vi), and (vii).

\section{Results}
\label{sect:results}
We test our models on seven molecular systems, each consisting of a molecule together with a particular atom-centered Gaussian basis set: (i) \heh with 6-31G ($N=4$), (ii) \heh with 6-311G ($N=14$), (iii) \lih with 6-31G ($N=11$), (iv) \ethylene with STO-3G ($N = 14$), (v) \lih with 6-311++$\text{G}^{**}$ ($N = 29$), (vi) \ethylene with 6-31+$\text{G}^*$ ($N = 46$) and (vii) \paranitro with STO-3G ($N = 60$).  In SM Section \ref{sect:coordinates}, we provide the Cartesian coordinates for each molecule.

\begin{figure*}[t]
\subfloat[Field-free propagation errors]{%
\includegraphics[width=0.4\textwidth]{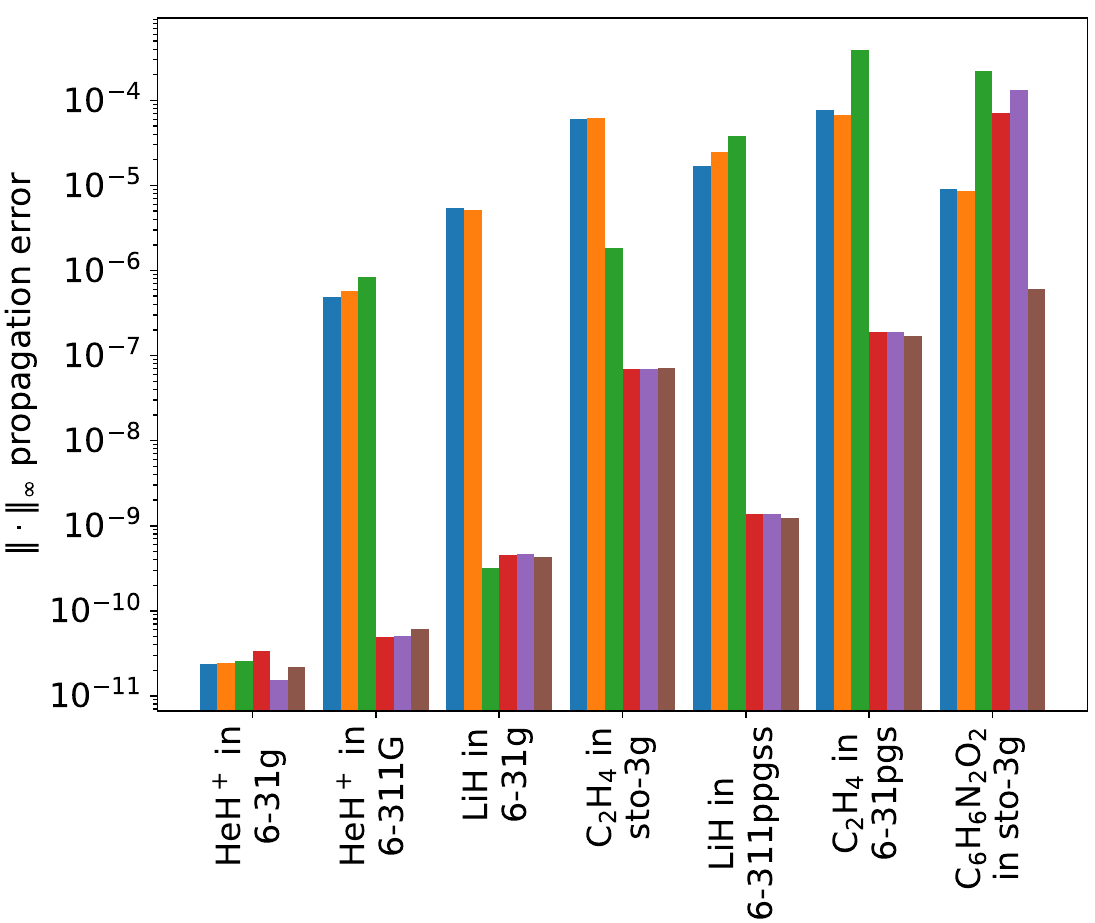}\label{fig:LinftyPropErrorsa}%
}
\hfill
\subfloat[Field-on propagation errors]{%
\includegraphics[width=0.4\textwidth]{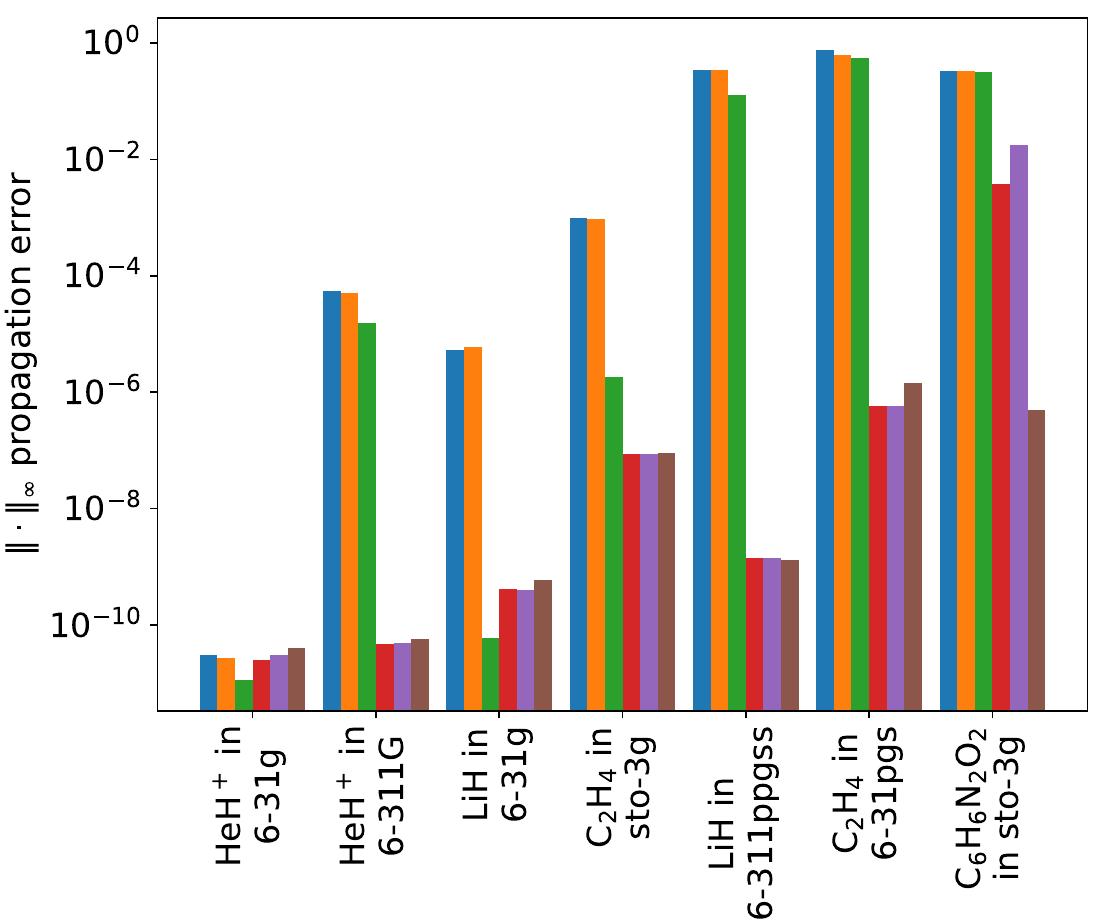}\label{fig:LinftyPropErrorsb}%
}
\hfill
\subfloat{\raisebox{2cm}{\includegraphics[width=0.1\textwidth]{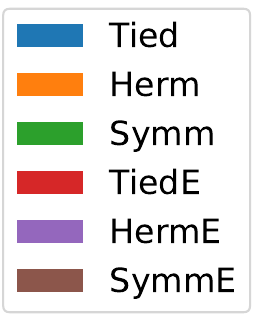}}}
\caption{For all seven molecular systems, we plot $\infty$-norm propagation errors (\ref{eqn:inftyerr}) for all six LSMR-trained models in both field-free/training (left) and field-on/test (right) settings.   Overall, the $8$-fold symmetry-preserving model SymmE, trained on an ensemble of trajectories, is either the best (e.g., for \paranitrocomma the largest system) or incurs error that only slightly exceeds that of the best model.}
\label{fig:LinftyPropErrors}
\end{figure*}

\subsection{Metrics}
\label{sect:metrics}
For detailed tables/plots of final values of the training loss for all models and all molecular systems, see SM Section \ref{sect:supptrainingloss}.  After training, we assess model errors using three metrics: propagation error, Hamiltonian error, and commutator error.

\subsubsection{Propagation Error}
\label{sect:properr}
Propagation error measures the ability of the learned Hamiltonian $\widetilde{H}(P)$ to generate predicted trajectories $\widetilde{P}(t)$ that match true trajectories $P(t)$.   Let $P(t)$ denote the true trajectory obtained by solving (\ref{eqn:TDHF}) using the true Hamiltonian $H(P)$, and let $\widetilde{P}(t)$ denote the solution of the TDHF equation 
\begin{equation}
\label{eqn:TDHFtilde}
\imath \frac{d \widetilde{P}(t)}{d t} =  \big[   \widetilde{H}(\widetilde{P}(t),t), \widetilde{P}(t) \big],
\end{equation}
with a Hamiltonian $\widetilde{H}(P,t)$ that equals the learned field-free Hamiltonian $\widetilde{H}(P)$ plus an optional time-dependent term. As explained in Section \ref{sect:meth}, we use the CI4 scheme to solve both (\ref{eqn:TDHF}) and (\ref{eqn:TDHFtilde}) for $J=20000$ time steps at $\Delta t = 8.268 \times 10^{-4}$, starting both trajectories at the same initial condition $\widetilde{P}(0) = P(0)$.  We then compute the $\infty$-norm error
\begin{equation}
\label{eqn:inftyerr}
\| P - \widetilde{P} \|_{\infty} = \max_{1 \leq j \leq J} \max_{1 \leq a,b \leq N} |P(t_j)_{ab} - \widetilde{P}(t_j)_{ab} |,
\end{equation}
the worst-case error across all matrix entries and all time.

We compute propagation error in both field-free and field-on settings.  In the field-free setting, we solve (\ref{eqn:TDHF}) and (\ref{eqn:TDHFtilde}) using the true and learned field-free Hamiltonians, respectively.  We view the resulting field-free propagation errors  as \emph{training set} errors.  In the field-on setting, we add a time-dependent electric field term to both the true and learned field-free Hamiltonians---see Section \ref{sect:traintest}.  As the true field-on trajectory is not used for training, we view the resulting field-on propagation errors as \emph{test set} errors.

As mentioned in Section \ref{sect:eightfoldsymmdef}, the reduced number of parameters in the $8$-fold symmetry-preserving method enables use of the exact Hessian-based formula (\ref{eqn:thetanormal2}) to train models for the four smaller molecular systems.  Thus we can use these four smaller systems to study how much error (if any) we introduce by training the $8$-fold model via the iterative LSMR method versus the exact Hessian-based approach.

In Figure \ref{fig:HessWash}, we show that in terms of propagation error (\ref{eqn:inftyerr}) on both the field-free/training set (left) and the field-on/test set (right), the LSMR-trained models do no worse---and in some cases outperform---the Hessian-trained models.  These results justify our use of the LSMR method to train the larger molecular systems, for which a comparison against the Hessian-based approach is impossible.  For the remainder of this paper, we focus our attention on LSMR-trained models.  Further results for Hessian-trained models can be found in SM Section \ref{sect:suppproperrors}.

In Figure \ref{fig:LinftyPropErrors}, we plot the propagation error (\ref{eqn:inftyerr}) for all LSMR-trained models and all seven molecular systems, on both the field-free/training set (left) and the field-on/test set (right).  All reported errors are orders of magnitude lower than what we achieved in prior results \cite{bhat2020machine,Gupta2022}. In nearly all cases, and especially for the three largest systems, ensemble-trained models (TiedE, HermE, and SymmE) outperform single trajectory-trained models (Tied, Herm, and Symm), often by several orders of magnitude.  For the two small systems \heh and \lih in 6-31G, the only model that outperforms ensemble-trained models is Symm, the $8$-fold symmetry-preserving model trained on a single field-free trajectory.

Among all models,  the $8$-fold symmetry-preserving model SymmE, trained on an ensemble of trajectories, is nearly the best (if not the best) for all molecular systems.  For \paranitrocomma the largest system we considered, the difference between SymmE and other models is especially striking.

\begin{figure*}[t]
\centering
\includegraphics[height=5.3cm]{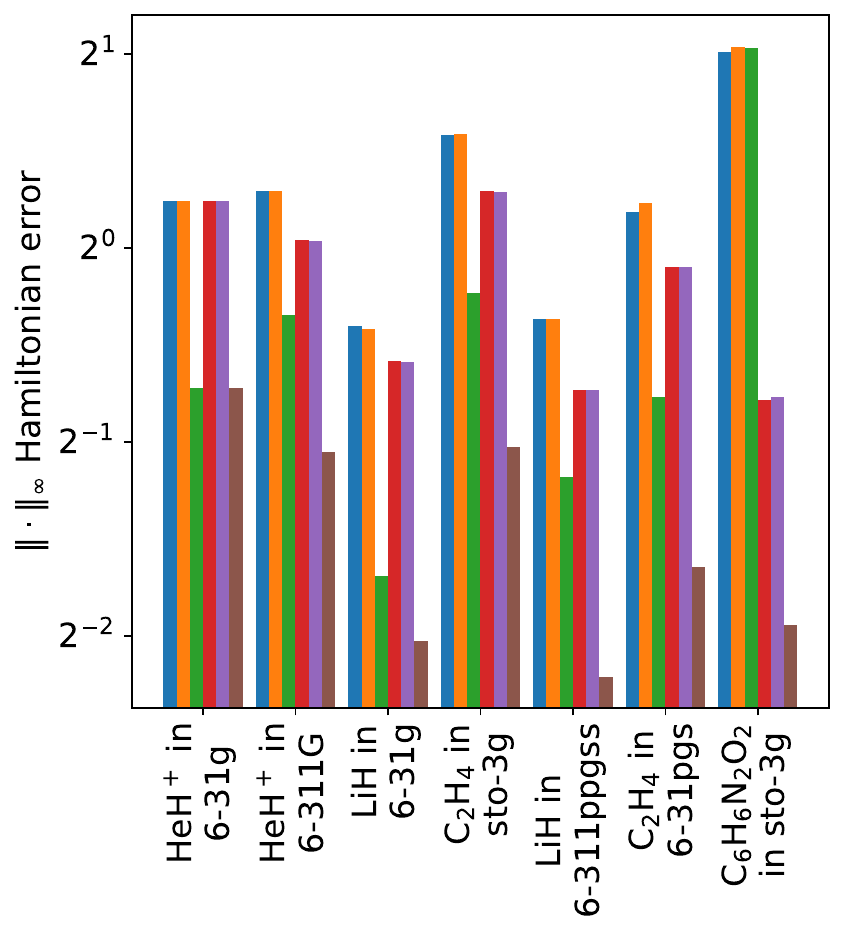}
\includegraphics[height=5.3cm]{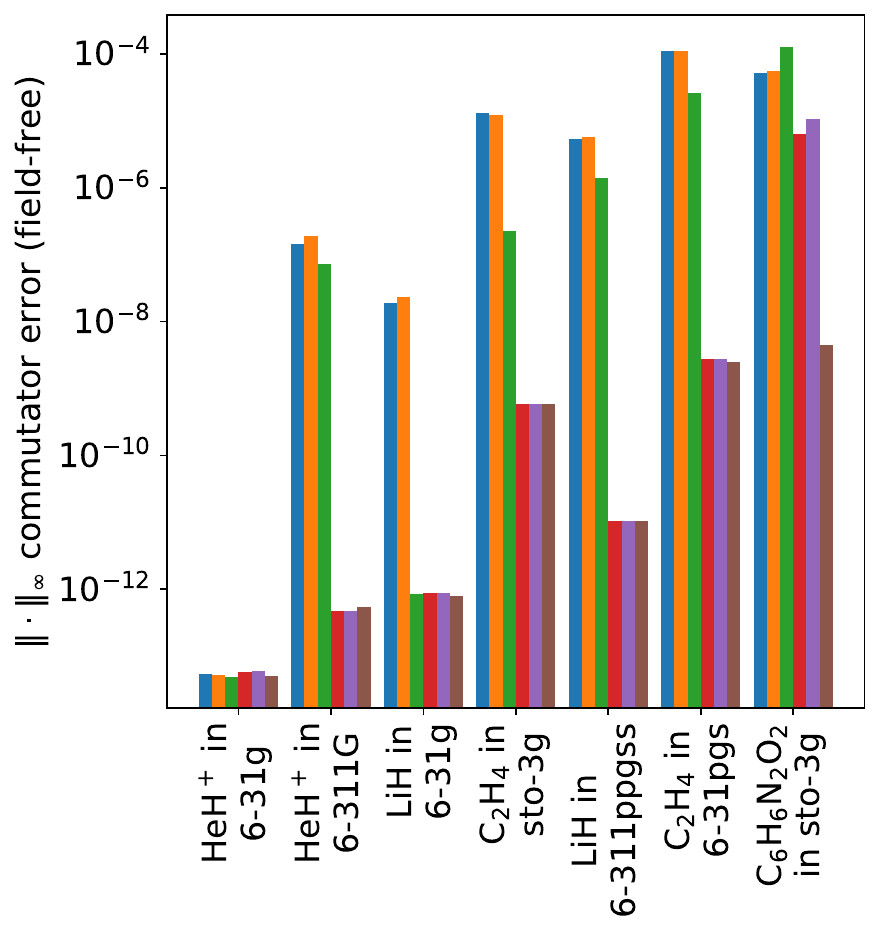}
\includegraphics[height=5.3cm]{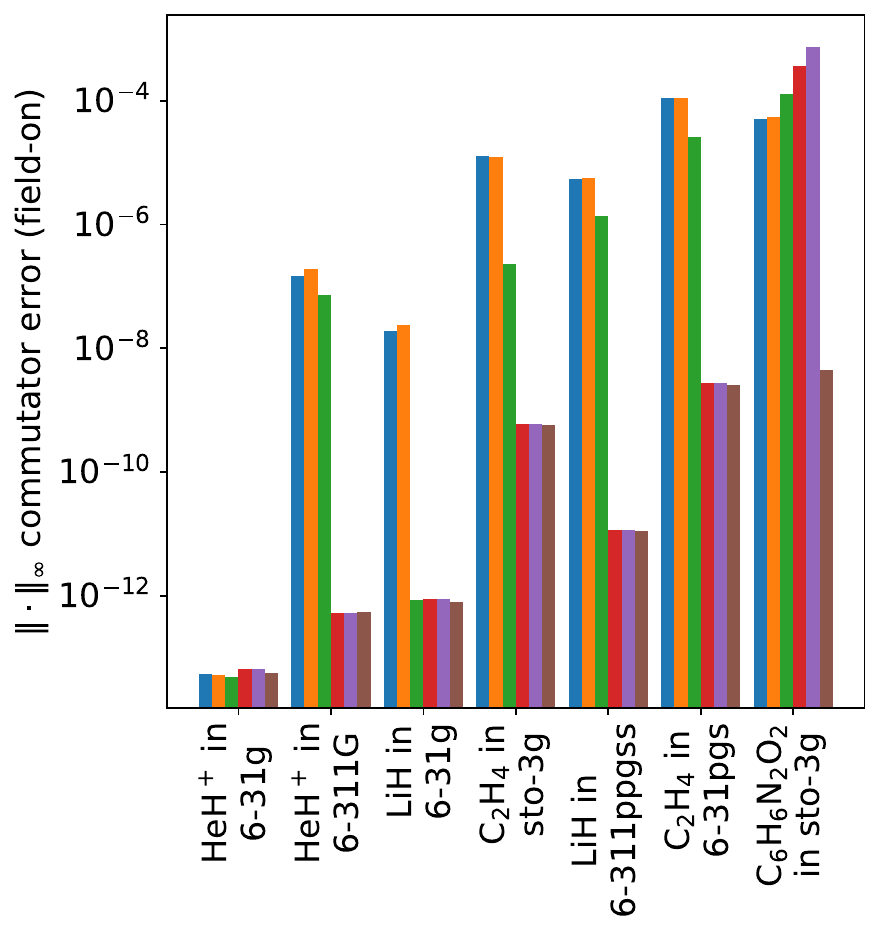}
\raisebox{2cm}{\includegraphics[width=0.1\textwidth]{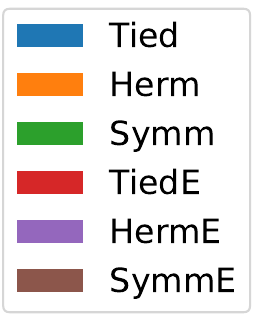}}
\caption{We plot Hamiltonian and commutator errors for all molecular systems and all LSMR-trained models.  The left plot shows that SymmE, the $8$-fold symmetry-preserving model (\ref{eqn:eightfoldmodel}) trained on the ensemble of field-free trajectories, has the lowest Hamiltonian error (\ref{eqn:hamerrors}) for each molecular system.  The middle and right plots show $\infty$-norm commutator error (\ref{eqn:commerror}) in the field-free/training and field-on/test cases, respectively.  The commutator errors reveal that all ensemble-trained models perform similarly well, except for \paranitrocomma for which SymmE shows a large advantage over all other models.}
\label{fig:HamCommErrors}
\end{figure*}

As long as $\Delta t > 0$, the training loss (\ref{eqn:loss}) measures something different than propagation error on the training set.  When we compute the loss, we are measuring our ability to predict a discrete-time version of the derivative $\dot{P}(t_j)$ using \emph{the true value of} $P(t_j)$ at each time.  This is equivalent to predicting a small number of time steps (at most, the width of the finite-difference stencil used to compute $\dot{P}$) ahead in time.  Thus the loss measures aggregate \emph{short-term} propagation error.  When we compute the propagation error metrics (\ref{eqn:inftyerr}) and (\ref{eqn:tdmae}), we use the predicted trajectory $\widetilde{P}(t_j)$, which can potentially accumulate errors at each time step.  There is no $t_j > 0$ at which we reset $\widetilde{P}(t_j)$ to equal $P(t_j)$.  Thus the metrics (\ref{eqn:inftyerr}) and (\ref{eqn:tdmae}) measure aggregate \emph{long-term} propagation error.

In SM Section \ref{sect:suppproperrors}, we plot---for each model and each molecular system---the mean absolute errors (MAE)
\begin{equation}
\label{eqn:tdmae}
\text{MAE}(t_j) = \frac{1}{N^2} \sum_{a,b}  |P(t_j)_{ab} - \widetilde{P}(t_j)_{ab} |,
\end{equation}
a scalar time series showing how the error evolves in time.  We use MAE as it conveys the average number of digits of accuracy in each entry of a predicted electron density matrix.  The MAE results agree with the $\infty$-norm results in Figure \ref{fig:LinftyPropErrors}.

\subsubsection{Hamiltonian and Commutator Errors}
\label{sect:hamcommerr}
For each of the three models (\ref{eqn:tiedmodel}), (\ref{eqn:hermrepmodel}), and (\ref{eqn:eightfoldmodel}), there exist parameters $\btheta^\circ$ such that the model perfectly matches the ground-truth Hamiltonian (\ref{eqn:gthamCO})---for detailed derivations, see SM Section \ref{sect:gtvals}.  Hence the $\infty$-norm error between trained ($\btheta^\star$) and true ($\btheta^\circ$) parameters is one way to measure the error between trained and true \emph{Hamiltonians}:
\begin{equation}
\label{eqn:hamerrors}
\| \btheta^\star - \btheta^\circ \|_{\infty} = \max_{1 \leq j \leq |\btheta|} | \theta^\star_j - \theta^\circ_j |.
\end{equation}

In the left-most plot in Figure \ref{fig:HamCommErrors}, we find that SymmE, the $8$-fold symmetry-preserving model (\ref{eqn:eightfoldmodel}) trained on the ensemble of field-free trajectories, has the smallest Hamiltonian error (\ref{eqn:hamerrors}) across all systems and all LSMR-trained models.

Examining the equations of motion (\ref{eqn:TDHF}), we see the possibility that some part of the Hamiltonian may not affect the dynamics at all.  Specifically, if $\widetilde{H} = \widetilde{H}_{\text{comm}} + \widetilde{H}_{\text{rem}}$ where $\widetilde{H}_{\text{comm}}$ commutes with $P$, then $[\widetilde{H},P] = [\widetilde{H}_{\text{rem}},P]$ where $\widetilde{H}_{\text{rem}}$ stands for the remaining part of $\widetilde{H}$.  If the errors in the learned Hamiltonian $\widetilde{H}$ concentrate in $\widetilde{H}_{\text{comm}}$, then those errors will cancel out of the commutator.  To quantify the extent to which this happens, we compute commutator errors in the $\infty$-norm:
\begin{multline}
\label{eqn:commerror}
\| [H(P(t),t), P(t)] - [\widetilde{H}(P(t),t), P(t)] \|_{\infty} \\
= \| [H(P(t),t) - \widetilde{H}(P(t),t), P(t)] \|_{\infty}
\end{multline}
The Hamiltonian only appears in the equation of motion (\ref{eqn:TDHF}) through the commutator. If the commutator error is small, the learned Hamiltonian $\widetilde{H}$ is accurate for practical purposes.

In the middle (field-free/training) and right-most (field-on/test) plots of Figure \ref{fig:HamCommErrors},  we plot the commutator error for all molecular systems and all LSMR-trained models.  We see that the TiedE and HermE models have nearly the same commutator error as the SymmE model, for all molecular systems except for \paranitro---here SymmE maintains its superior performance.  These results explain why the TiedE, HermE, and SymmE models all have low and nearly equal propagation errors (see Figure \ref{fig:LinftyPropErrors}), even though the raw Hamiltonian error of the TiedE and HermE models can be large (left-most plot of Figure \ref{fig:HamCommErrors}).

The field-free and field-on commutator errors show extremely similar trends.  This is because, in the field-on case, the electric field terms in $H$ and $\widetilde{H}$ cancel perfectly, so that the commutator error reduces to $\|[ H(P(t)) -\widetilde{H}(P(t)), P(t) ]\|$ where $H$ and $\widetilde{H}$ are the field-free Hamiltonians and $P(t)$ is the field-on trajectory.

\begin{figure*}[p]
\centering
\subfloat[\heh in 6-31G]{%
\includegraphics[width=0.31\textwidth]{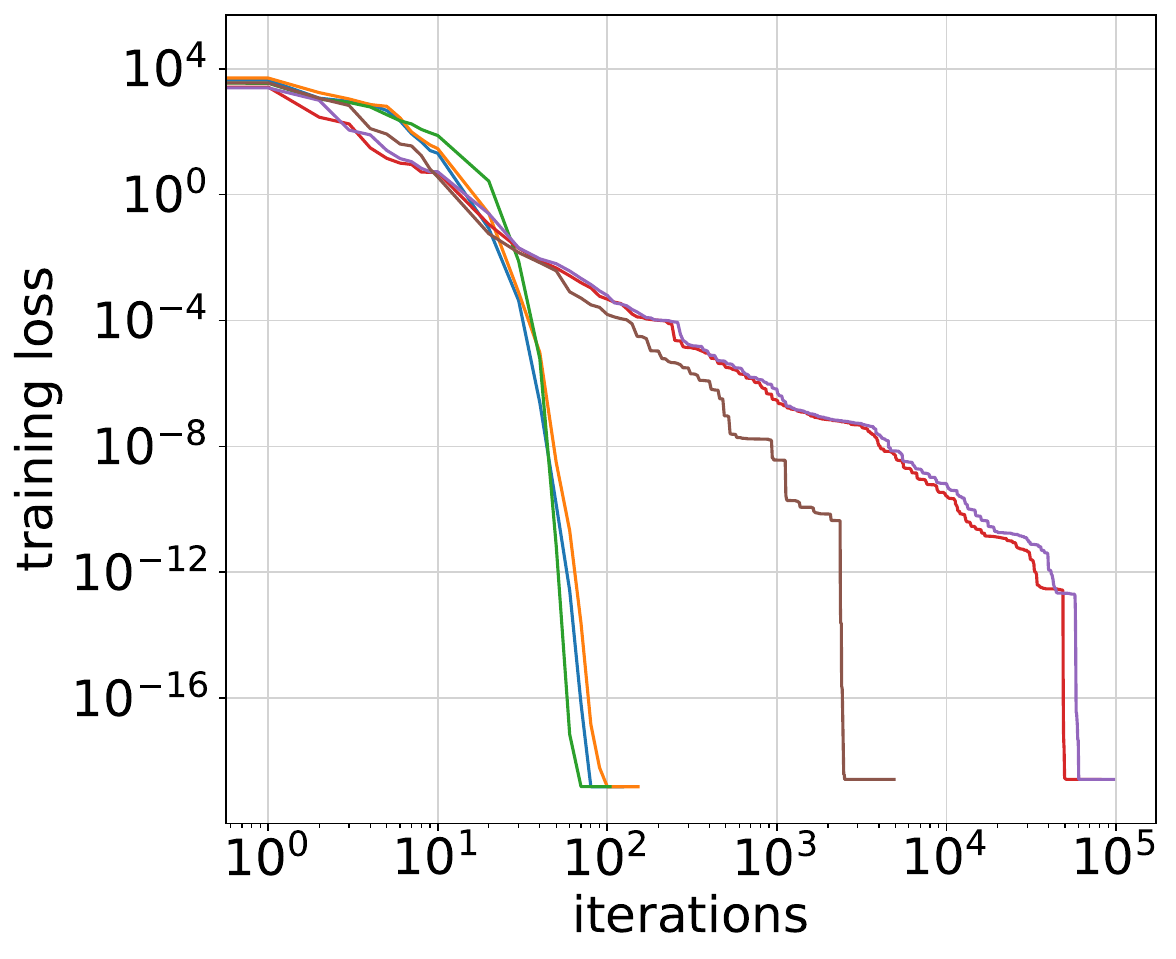}%
}
\hfill
\subfloat[\heh in 6-311G]{%
\includegraphics[width=0.31\textwidth]{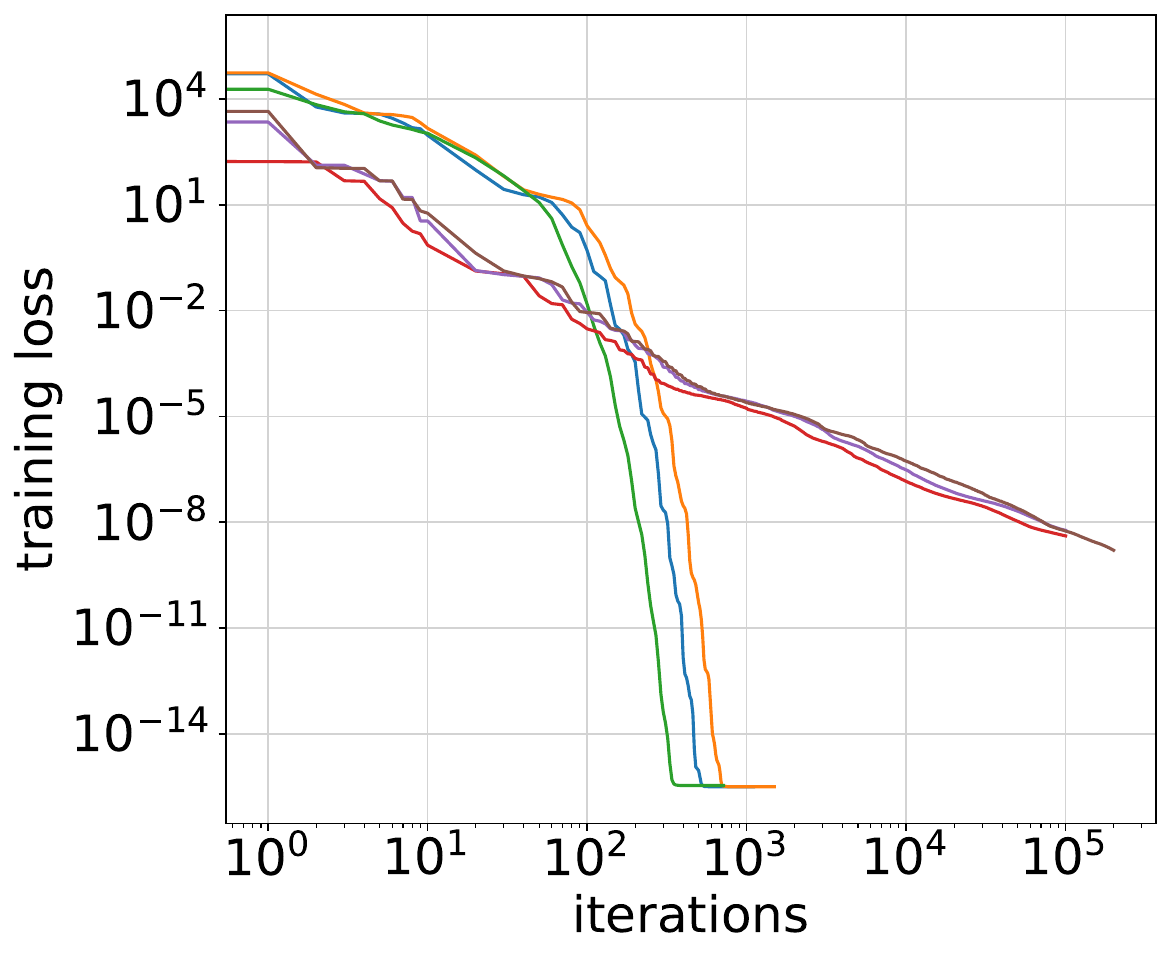}%
}
\hfill
\subfloat[\lih in 6-31G]{%
\includegraphics[width=0.31\textwidth]{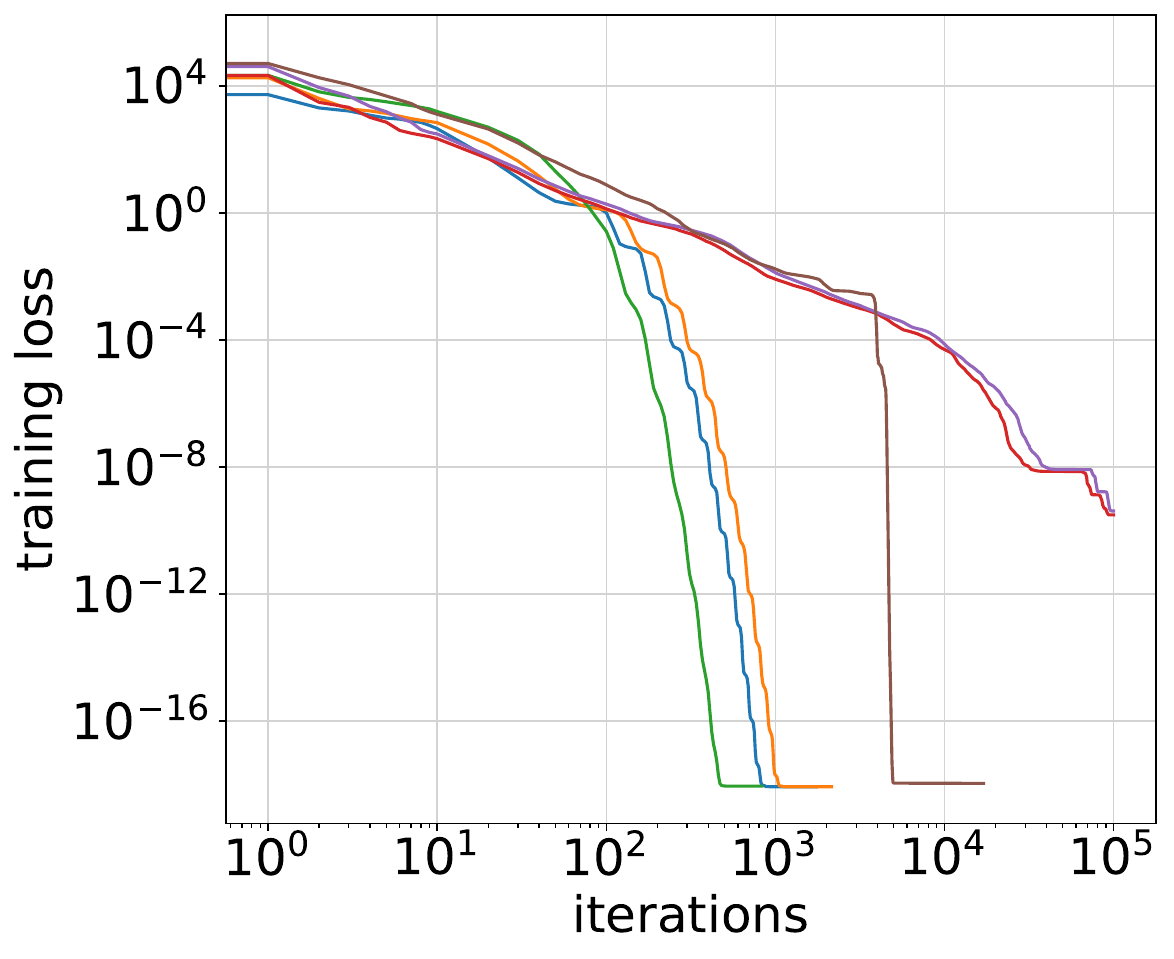}%
}
\\
\subfloat[\ethylene in 6-31G]{%
\includegraphics[width=0.31\textwidth]{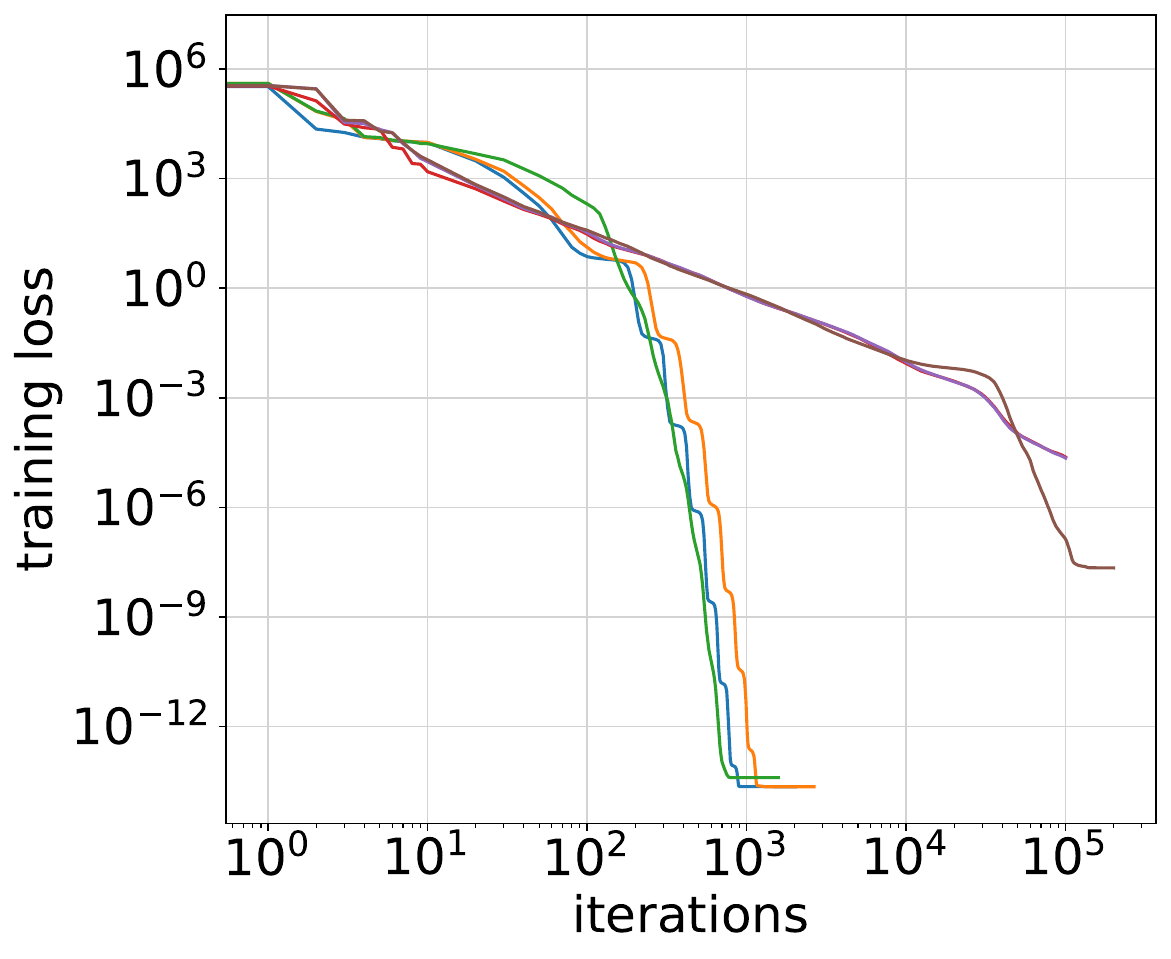}%
}
\hfill
\subfloat[\lih in 6-311++$\text{G}^{**}$]{%
\includegraphics[width=0.31\textwidth]{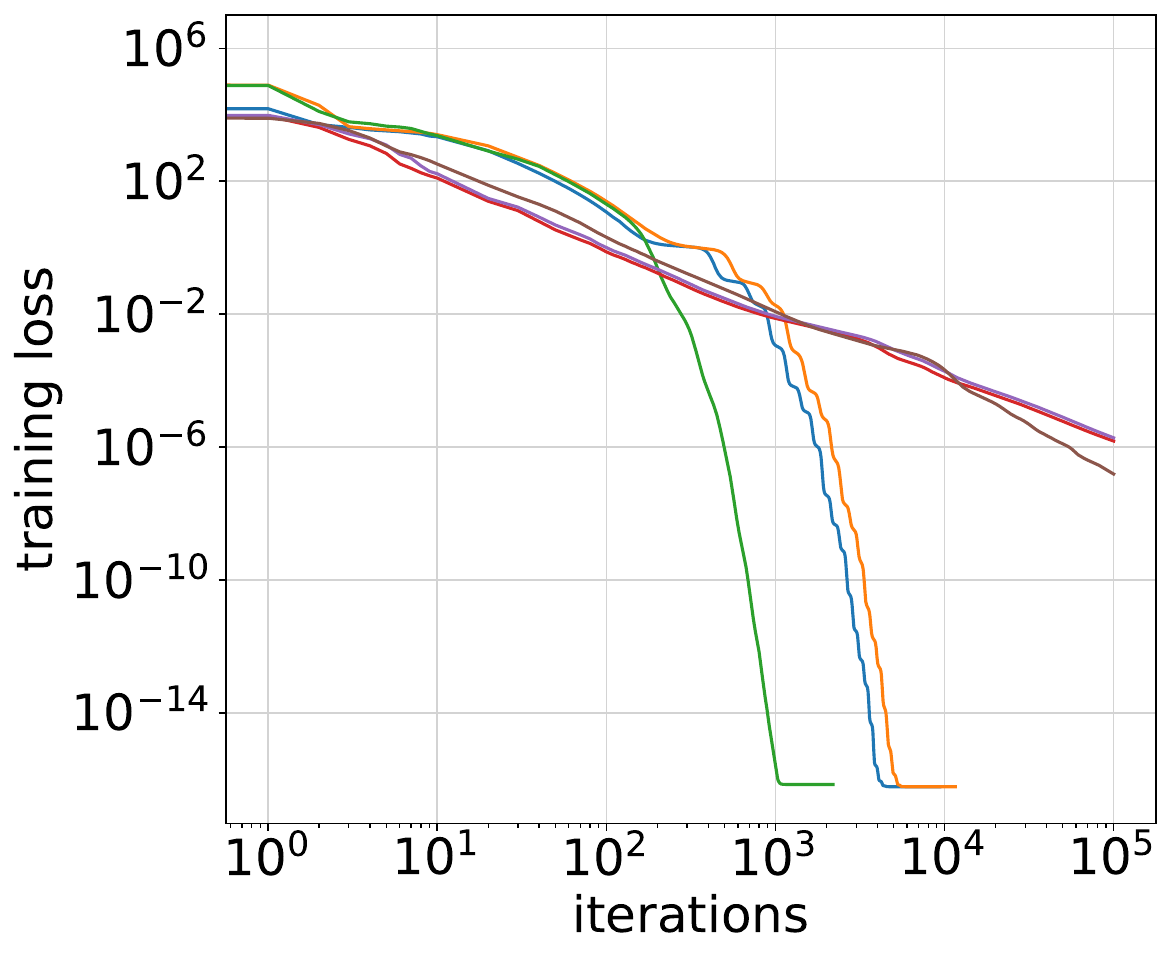}%
}
\hfill
\subfloat[\ethylene with 6-31+$\text{G}^*$]{%
\includegraphics[width=0.31\textwidth]{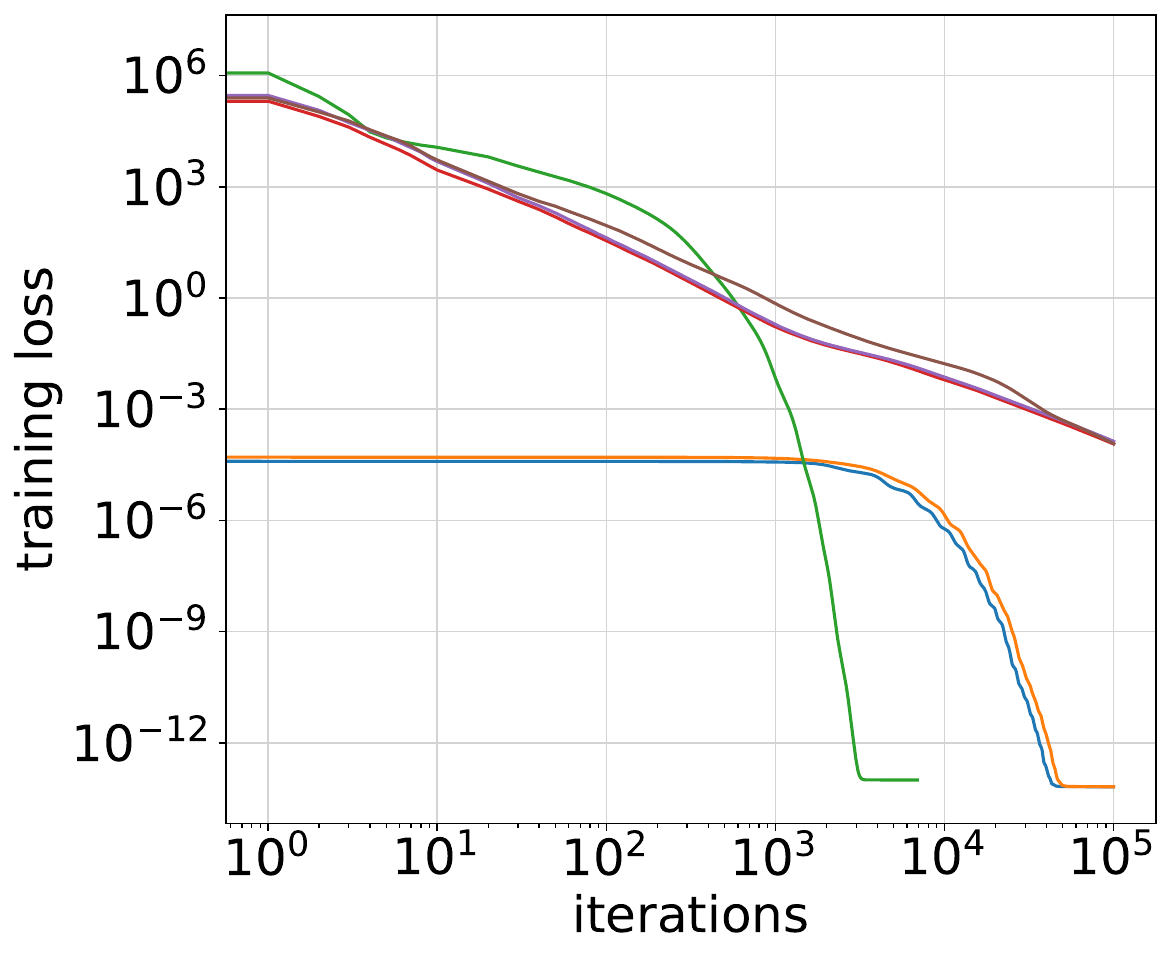}%
}
\\
\subfloat[\paranitro in STO-3G]{%
\includegraphics[width=0.31\textwidth]{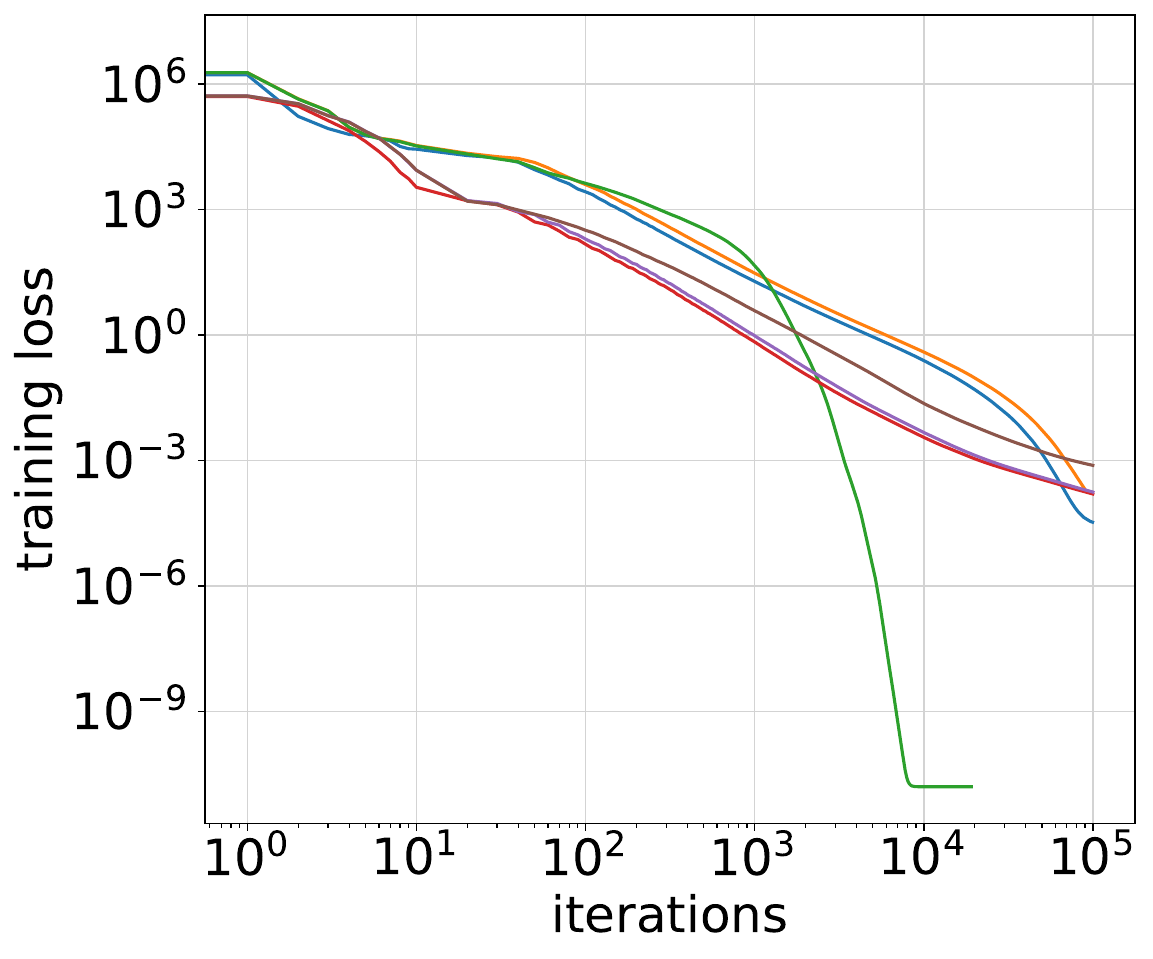}%
}
\hspace{1cm}
\subfloat{%
\raisebox{0.875cm}{\includegraphics[width=0.15\textwidth,clip,trim=10 10 10 10]{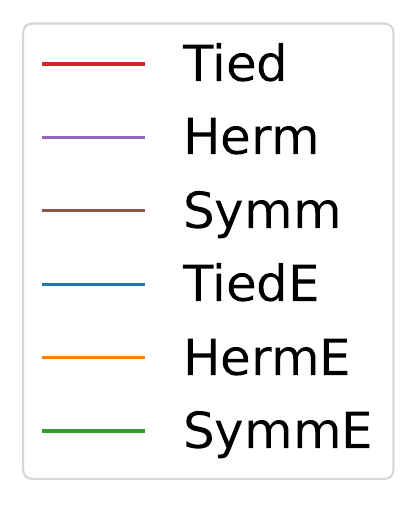}}%
}
\caption{We plot the training loss versus iteration number for six models and seven molecular systems.  Here Tied, Herm, and Symm refer to the models (\ref{eqn:tiedmodel}), (\ref{eqn:hermrepmodel}) and (\ref{eqn:eightfoldmodel}), respectively.  The suffix ``E'' indicates that the model has been trained on an ensemble of field-free trajectories; otherwise, the model has been trained on a single field-free trajectory.  \emph{Overall, we see that the $8$-fold symmetry-preserving model with the ensemble training set (SymmE) achieves the best results, and is significantly better for the larger molecular systems (e-g).} For the four smaller molecular systems (a-d), ensemble training reduces the number of iterations by orders of magnitude for all models.}
\label{fig:trainiter}
\end{figure*}

Further tables of Hamiltonian and commutator error results can be found in SM Section \ref{sect:supphamcommerrors}.

\subsubsection{Iteration Count}
In Figure \ref{fig:trainiter}, for the models trained with the iterative LSMR method, we plot training loss as a function of the number of iterations.  Across all molecular systems, the SymmE model achieves the best results.  For larger molecular systems, the SymmE model requires orders of magnitude fewer number of iterations to achieve small values of the training loss.  For smaller molecular systems, any of the models trained on the ensemble of field-free trajectories (TiedE, HermE, and SymmE) yield the best performance in terms of rapid minimization of training loss, i.e., within $10^3$ iterations.

\section{Discussion and Conclusion}
\label{sect:discussion}
As emphasized in Section \ref{sect:intro}, we undertook this study to develop strategies to learn quantum potentials from time-dependent data.  We focused on time-dependent Hartree-Fock (TDHF) systems because we know the full form of the field-free Hamiltonian, we have no problem using TDHF to generate data for large molecular systems, and because the lessons we learn here can be cross-applied to more complex approaches such as time-dependent density functional theory (TDDFT), time-dependent reduced density matrix functional theory (TDRDMFT), or Kadanoff-Baym theory.

One clear conclusion is the importance of both the quality and quantity of the training data.  For the largest molecular systems under study, we were able to achieve acceptably low values of field-on (test set) propagation error only by training on data generated via the fourth-order CI4 scheme with a sufficiently small time step ($\Delta t$).  With the coarser second-order MMUT scheme and/or a larger time step, we would not have been able to achieve these results.

In our prior work, we always trained our models on \emph{single} field-free trajectories \cite{bhat2020machine,Gupta2022}.  The results presented in Section \ref{sect:results} show the importance of training with an ensemble of field-free trajectories.  In Section \ref{sect:traintest}, we described the construction of this ensemble in detail.  In particular, when we constructed random initial conditions for the cloud of trajectories in the ensemble, we ensured that these initial conditions were Hermitian and idempotent.  Indeed, if we train on ensembles generated from trajectories that ignore these constraints, the resulting models perform poorly.  Especially for larger molecular systems, training on physically meaningful ensembles  constrains our models to stay closer to the true Hamiltonian than is possible with a single trajectory training set.

We explored three models in this work, two of which preserve the Hermitian symmetry of the Hamiltonian, and the third of which preserves a deeper $8$-fold symmetry.  Taking a holistic view of the training loss versus number of iterations, the propagation error, the Hamiltonian error, and the commutator error, we can build a strong case that the best model is the $8$-fold symmetry-preserving, ensemble-trained model (SymmE).  In short, building \emph{more} symmetries (i.e., more physics) into the machine learning model yields better results, especially as molecular system size increases.

We have focused on a method (TDHF) in which the underlying symmetries are known exactly.  In cases where we seek to learn a term (e.g., an exchange-correlation potential or self-energy) whose full symmetries are unknown, the present paper offers two approaches that complement prior methods to learn symmetries from data\cite{liu2022machine,PhysRevD.105.096031}.  First, if there is a symmetry that one suspects should be present, one can easily generalize our methods to create an appropriately symmetry-constrained machine learning model.  The only substantive changes to Algorithm \ref{algo:pop} will concern the dimensions of the tensors and the elements of the symmetry group $\{\sigma^m\}$.  For nonlinear neural networks, our techniques can be incorporated in a linear output layer.  One can then test empirically whether the resulting model yields results that are significantly better than non-symmetrized models.

Second, our results show that preserving only the Hermitian symmetry leads to predictive accuracy in terms of long-term propagation error: the TiedE and HermE models perform well in propagation error and commutator error tests.  Though they cannot compare with SymmE in terms of computational cost (training loss versus iteration number) or $\infty$-norm Hamiltonian error, the TiedE and HermE models required less code development time than the SymmE model.  Generalizing to more complex situations such as TDDFT or TDRDMFT, we hypothesize that a machine learning model that preserves only an outer layer of symmetries may be sufficient for a first round of results.  In subsequent rounds, these models can be refined by incorporating more symmetries, especially if the goal is to learn the true Hamiltonian.  

Both of these approaches, along with careful attention to generating an accurate  ensemble of trajectories, will ensure that the model can extrapolate beyond the training data, and also increase the chances that one can distill new physical insights from, e.g., a machine-learned closure model for theories such as TDDFT and TDRDMFT.  The scalability of our approach could be improved by reducing memory requirements with tensor train techniques or by density matrix dimensionality reduction, allowing application to larger molecular systems. However, as our goal is to use this method to derive new insights into how to learn an unknown molecular potential, such as the time-dependence of the exchange-correlation potential in TDDFT, molecules of the size explored here provide a rich playing field for exploration and study for future work. 

\begin{acknowledgments}
This research was sponsored by the Office of Naval Research and was accomplished under Grant Number W911NF-23-1-0153. The views and conclusions contained in this document are those of the authors 
and should not be interpreted as representing the official policies, either expressed or implied, of the Army Research 
Office or the U.S. Government. The U.S. Government is authorized to reproduce and distribute reprints for 
Government purposes notwithstanding any copyright notation herein.  This research was also supported by the 
    U.S. Department of Energy, Office of Science, Basic Energy Sciences under Award Number DE-SC0020203. This research used resources of the National Energy Research Scientific Computing Center (NERSC), a U.S. Department of Energy Office of Science User Facility located at Lawrence Berkeley National Laboratory, operated under Contract No. DE-AC02-05CH11231 using NERSC awards BES-m2530 and ASCR-m4577.  We acknowledge computational time on the Pinnacles cluster at UC Merced (supported by NSF OAC-2019144).  We also acknowledge use of resources available through the National Research Platform (NRP) at the University of California, San Diego. NRP has been developed, and is supported in part, by funding from the National Science Foundation, from awards 1730158, 1540112, 1541349, 1826967, 2112167, 2100237, and 2120019, as well as additional funding from community partners.
\end{acknowledgments}

\section*{Data Availability Statement}
All data and code that support the findings of this study are openly available in the GitHub repository \emph{learningTDHF} at\\
\url{https://github.com/hbhat4000/learningTDHF/}.

\section{References}
\nocite{*}
\bibliography{BhatGuptaIsbornAML}

\section{Supplementary Information}

\subsection{Propagation Schemes}
\label{sect:propschemes}
We summarize two methods to numerically integrate the TDHF equation (\ref{eqn:TDHF}) on the equispaced temporal grid $t_j = j \Delta t$ with fixed time step $\Delta t > 0$.  Here $P_{n}$ denotes our numerical approximation to $P(t_n)$.  Throughout this section, $\exp(M)$ means the matrix exponential of $M$.

\subsubsection{Modified Midpoint Unitary Transformation (MMUT)}
This scheme was introduced by \cite{Li2005} and is widely used to solve the TDHF system (\ref{eqn:TDHF}).  In our implementation, we begin with one step of size $\Delta t$:
\begin{equation}
\label{eqn:MMUTfirststep}
\begin{split}
K_0 &= (-i \Delta t) H(0, P_0) \\
P_1 &= \exp(K_0) P_0 \exp(-K_0).
\end{split}
\end{equation}
For subsequent steps, we employ a leap-frog type iteration with steps of size $2 \Delta t$:
\begin{equation}
\label{eqn:MMUTsubsequentsteps}
\begin{split}
K_{n} &= (-2 i \Delta t) H( n \Delta t, P_n ) \\
P_{n+1} &= \exp(K_n) P_{n-1} \exp(-K_n).
\end{split}
\end{equation}

\subsubsection{Casas and Iserles (2006)}
The Appendix of \citep{Casas2006} explains how to form the following fourth-order scheme to integrate (\ref{eqn:TDHF}), which we call the CI4 scheme:
\begin{equation}
\label{eqn:CasasIserles2006}
\begin{split}
k_1 &= (-i \Delta t) H(t_n, P_n)\\
Q_1 &= k_1\\
u_2 &= \frac{1}{2} Q_1 \\
k_2 &= (-i \Delta t) H\left( t_n + \frac{\Delta t}{2}, \exp(u_2) P_n \exp(-u_2) \right)\\
Q_2 &= k_2 - k_1\\
u_3 &= \frac{1}{2} Q_1 + \frac{1}{4} Q_2 \\
k_3 &= (-i \Delta t) H\left( t_n + \frac{\Delta t}{2}, \exp(u_3) P_n \exp(-u_3) \right)\\
Q_3 &= k_3 - k_2\\
u_4 &= Q_1 + Q_2\\
k_4 &= (-i \Delta t) H(t_n + \Delta t, \exp(u_4) P_n \exp(-u_4))\\
Q_4 &= k_4 - 2 k_2 + k_1 \\
u_5 &= \frac{1}{2} Q_1 + \frac{1}{4} Q_2 + \frac{1}{3} Q_3 - \frac{1}{24} Q_4 - \frac{1}{48} [Q_1, Q_2] \\
k_5 &= (-i \Delta t) H\left( t_n + \frac{\Delta t}{2}, \exp(u_5) P_n \exp(-u_5) \right)\\
Q_5 &= k_5 - k_2 \\
u_6 &= Q_1 + Q_2 + \frac{2}{3} Q_3 + \frac{1}{6} Q_4 - \frac{1}{6} [Q_1, Q_2] \\
k_6 &= (-i \Delta t) H(t_n + \Delta t, \exp(u_6) P_n \exp(-u_6))\\
Q_6 &= k_6 - 2 k_2 + k_1 \\
v &= Q_1 + Q_2 + \frac{2}{3} Q_5 + \frac{1}{6} Q_6 \\
 &\qquad \qquad \qquad \quad - \frac{1}{6} \left[ Q_1, Q_2 - Q_3 + Q_5 + \frac{1}{2} Q_6 \right] \\
P_{n+1} &= \exp(v) P_n \exp(-v)
\end{split}
\end{equation}

\subsection{Ground Truth Parameter Values}
\label{sect:gtvals}
For each of the models we presented in Section \ref{sect:model} of the manuscript, we seek a set of parameters $\btheta^\circ$ such that the model coincides with
\begin{equation}
\label{eqn:gthamCO2}
H(P)_{ab} = H^\text{core}_{ab} + \sum_{cd} \mathcal{E}_{abcd} P_{cd},
\end{equation}
the ground truth Hamiltonian in the CO basis.  We will be able to find $\btheta^\circ$ for the Hermitian representation model (\ref{eqn:hermrepmodel}) and the $8$-fold symmetry-preserving model (\ref{eqn:eightfoldmodel}), but not for the tied regression model (\ref{eqn:tiedmodel}).  In (\ref{eqn:gthamCO2}), $\mathcal{E}$ is the $2$-electron tensor in the CO basis, defined in the main paper in (\ref{eqn:mathcale}) and reproduced here for convenience:
\begin{equation}
\label{eqn:Econvenience}
\mathcal{E}_{abcd} = \sum_{ijkl} \overline{\mathcal{X}_{ia}} \mathcal{X}_{jb} \left[ (i j| l k) - \frac{1}{2} (i k | l j) \right] \mathcal{X}_{kc} \overline{\mathcal{X}_{l d}}.
\end{equation}
Then note that
\[
\overline{\mathcal{E}_{bacd}} = \sum_{ijkl} \overline{\mathcal{X}_{ja}}  \mathcal{X}_{ib} \left[ (i j| l k) - \frac{1}{2} (i k | l j) \right] \mathcal{X}_{l d} \overline{\mathcal{X}_{kc}} .
\]
Let us assume the use of a real set of atomic orbitals, so that the overlap matrix $S$ can be diagonalized by a real orthogonal matrix.  This implies that $\mathcal{X}$ and $\mathcal{E}$ are both real, and that the $2$-electron tensor $(i j | k l)$ satisfies the permutation symmetry (\ref{eqn:eightfold}).  Then, using this symmetry, we have
\begin{multline}
\label{eqn:Esymm}
\mathcal{E}_{bacd} = \overline{\mathcal{E}_{bacd}} = \sum_{ijkl} \overline{\mathcal{X}_{ja}}  \mathcal{X}_{ib} \left[ (j i| k l) - \frac{1}{2} (j l | k i) \right] \mathcal{X}_{l d} \overline{\mathcal{X}_{kc}} \\
= \mathcal{E}_{a b d c}.
\end{multline}
To see the last equality, relabel the dummy indices of summation according to: $i \to j'$, $j \to i'$, $k \to l'$, and $l \to k'$.  Then the right-hand side will be precisely of the form (\ref{eqn:Econvenience}) except with $d$ and $c$ interchanged.  \emph{The symmetry (\ref{eqn:Esymm}) is what guarantees that (\ref{eqn:gthamCO2}) is Hermitian:}
\begin{align*}
\overline{H(P)}_{ba} &= H^\text{core}_{ba} + \sum_{cd} \overline{\mathcal{E}_{bacd}} \overline{P_{cd}} \\
 &= H^\text{core}_{ab} + \sum_{cd} \mathcal{E}_{abdc} P_{dc} = H(P)_{ab},
\end{align*}
where in the second line we have used two facts: (i) $H^\text{core}$ is real and symmetric and (ii) $P$ is Hermitian.  We make this point to illustrate how $H(P)$ can be symmetric without requiring an explicit symmetrization step.  Starting from (\ref{eqn:gthamCO2}), if we separate out the real and imaginary parts of $P$ via $P = P^R + \imath P^I$, we arrive at
\begin{equation}
\label{eqn:reconcile}
H(P)_{ab} = H^\text{core}_{ab} + \sum_{cd} \mathcal{E}_{abcd} P^R_{cd} + \imath \sum_{cd} \mathcal{E}_{abcd} P^I_{cd}.
\end{equation}
This serves as a useful form against which to reconcile our models.

\subsubsection{Tied Model}
\label{sect:tiedgt}
Before proceeding, for real $N \times N$ matrices $A$ and $B$, let us introduce the Frobenius inner product
\begin{equation}
\label{eqn:froip}
\langle A, B \rangle = \sum_{c d} A_{cd} B_{cd}.
\end{equation}
With respect to this inner product, one can check that the subspace of real symmetric $N \times N$ matrices is orthogonal to the subspace of real antisymmetric $N \times N$ matrices.  If we express the tied regression model (\ref{eqn:tiedmodel}) as one equation, we obtain
\begin{multline*}
\widetilde{H}(P)_{ab} = H^\text{core}_{ab} + \frac{1}{2} \sum_{cd} P^R_{cd} \left[ \beta_{cdab} + \beta_{cdba} \right] \\
+ \frac{\imath}{2} \sum_{cd} P^I_{cd} \left[ \beta_{cdab} - \beta_{cdba} \right] 
\end{multline*}
Comparing this with (\ref{eqn:reconcile}), we see that for $\bbeta$ to reproduce the ground truth, we must have
\begin{subequations}
\label{eqn:tiedrecon}
\begin{align}
\label{eqn:tiedrecon1}
\sum_{cd} P^R_{cd} \left\{ \mathcal{E}_{abcd} - \frac{1}{2} \left[ \beta_{cdab} + \beta_{cdba} \right] \right\} &= 0 \\
\label{eqn:tiedrecon2}
\sum_{cd} P^I_{cd} \left\{ \mathcal{E}_{abcd} - \frac{1}{2} \left[ \beta_{cdab} - \beta_{cdba} \right] \right\} &= 0.
\end{align}
\end{subequations}
All quantities in these equations---$P^R$, $P^I$, $\mathcal{E}$, and $\bbeta$---are real.  We know that $P$ is Hermitian; therefore, $P^R$ is symmetric, while $P^I$ is antisymmetric.  Putting these facts together, we see that a \emph{sufficient condition} for $\bbeta$  is that, for each fixed $(a,b)$,
\begin{subequations}
\label{eqn:tiedsuff}
\begin{gather}
\label{eqn:tiedsuff1}
\left\{ \mathcal{E}_{abcd} - \frac{1}{2} \left[ \beta_{cdab} + \beta_{cdba} \right] \right\} \text{ is antisymmetric in $(c,d)$ } \\
\label{eqn:tiedsuff2}
\left\{ \mathcal{E}_{abcd} - \frac{1}{2} \left[ \beta_{cdab} - \beta_{cdba} \right] \right\} \text{ is symmetric in $(c,d)$ }.
\end{gather}
\end{subequations}
If (\ref{eqn:tiedsuff1}) holds, then the left-hand side of (\ref{eqn:tiedrecon1}) will be the Frobenius inner product between a symmetric matrix $P^R$ and an antisymmetric matrix (for each fixed $(a,b)$).  This inner product will be zero.

Similarly, if (\ref{eqn:tiedsuff2}) holds, then the left-hand side of (\ref{eqn:tiedrecon2}) will be the Frobenius inner product between an antisymmetric matrix $P^I$ and a symmetric matrix (for each fixed $(a,b)$).  This inner product will also be zero.

Now consider the choice
\begin{equation}
\label{eqn:tiedgt}
\bbeta_{cdab}^{\circ} = \mathcal{E}_{abcd}.
\end{equation}
Thanks to (\ref{eqn:Esymm}), this choice of $\bbeta$ implies that
\begin{subequations}
\label{eqn:betachoice}
\begin{align}
\frac{1}{2} \left[ \beta_{cdab}^{\circ} + \beta_{cdba}^{\circ} \right] &= \frac{1}{2} \left[ \mathcal{E}_{abcd} + \mathcal{E}_{bacd} \right] = \frac{1}{2} \left[ \mathcal{E}_{abcd} + \mathcal{E}_{abdc} \right] \\
\frac{1}{2} \left[ \beta_{cdab}^{\circ} - \beta_{cdba}^{\circ} \right] &= \frac{1}{2} \left[ \mathcal{E}_{abcd} - \mathcal{E}_{bacd} \right] = \frac{1}{2} \left[ \mathcal{E}_{abcd} - \mathcal{E}_{abdc} \right] 
\end{align}
\end{subequations}
For each fixed $(a,b)$, the right-hand sides are, respectively, symmetric and antisymmetric with respect to $(c,d)$.  Now note that
\begin{equation}
\label{eqn:symmantisymm}
\mathcal{E}_{abcd} = \underbrace{ \frac{1}{2} \left[ \mathcal{E}_{abcd} + \mathcal{E}_{abdc} \right] }_{\text{symmetric part of $\mathcal{E}$}} + \underbrace{ \frac{1}{2} \left[ \mathcal{E}_{abcd} - \mathcal{E}_{abdc} \right] }_{\text{antisymmetric part of $\mathcal{E}$}}
\end{equation}
Putting (\ref{eqn:betachoice}) and (\ref{eqn:symmantisymm}) together, we see that 
\begin{subequations}
\label{eqn:tiedexpr}
\begin{align}
\label{eqn:tiedexpr1}
\mathcal{E}_{abcd} - \frac{1}{2} \left[ \beta_{cdab}^{\circ} + \beta_{cdba}^{\circ} \right] = \frac{1}{2} \left[ \mathcal{E}_{abcd} - \mathcal{E}_{abdc} \right] \\
\label{eqn:tiedexpr2}
\mathcal{E}_{abcd} - \frac{1}{2} \left[ \beta_{cdab}^{\circ} - \beta_{cdba}^{\circ} \right] = \frac{1}{2} \left[ \mathcal{E}_{abcd} + \mathcal{E}_{abdc} \right]
\end{align}
\end{subequations}
This shows that the choice (\ref{eqn:tiedgt}) satisfies the sufficient condition (\ref{eqn:tiedsuff}).  Therefore, with (\ref{eqn:tiedgt}), the tied regression model (\ref{eqn:tiedmodel}) matches the ground truth Hamiltonian perfectly.

\subsubsection{Hermitian Representation Model}
\label{sect:hermgt}
We reproduce the Hermitian representation model here for convenience:
\begin{subequations}
\label{eqn:hermrepmodel2}
\begin{align}
\beta_{cdab} &= \sum_{k=1}^{N(N+1)/2} B^R_{ab k} v_{cd k} \\
\gamma_{cdab} &= \sum_{k=1}^{N(N-1)/2} B^I_{ab k} w_{cd k} \\
\widetilde{H}(P)_{ab} &= H^\text{core}_{ab} + \sum_{cd} P^R_{cd} \beta_{cdab} + \imath \sum_{cd} P^I_{cd} \gamma_{cdab} 
\end{align}
\end{subequations}
Here $B^R$ and $B^I$ are tensors that give, respectively, bases for the spaces of real symmetric and real antisymmetric matrices.  Comparing (\ref{eqn:hermrepmodel2}) against (\ref{eqn:reconcile}), we see that we must solve for tensors $\mathbf{v}$ and $\mathbf{w}$ such that
\begin{subequations}
\label{eqn:hermgtcond}
\begin{align}
\sum_{cd} P^R_{cd} \left\{ \mathcal{E}_{abcd} - \sum_k B^R_{abk} v_{cdk} \right\} &= 0 \\
\sum_{cd} P^I_{cd} \left\{ \mathcal{E}_{abcd} - \sum_k B^I_{abk} w_{cdk} \right\} &= 0
\end{align}
\end{subequations}
Based on the discussion in Section \ref{sect:tiedgt} and especially (\ref{eqn:betachoice}), we claim that it is sufficient to solve for $\mathbf{v}^\circ$ and $\mathbf{w}^\circ$ such that
\begin{subequations}
\label{eqn:hermgt}
\begin{align}
\label{eqn:hermgt1}
\frac{1}{2} \left[ \mathcal{E}_{abcd} + \mathcal{E}_{bacd} \right] = \frac{1}{2} \left[ \mathcal{E}_{abcd} + \mathcal{E}_{abdc} \right] &= \sum_k B^R_{abk} v_{cdk}^{\circ} \\
\label{eqn:hermgt2}
\frac{1}{2} \left[ \mathcal{E}_{abcd} - \mathcal{E}_{bacd} \right] = \frac{1}{2} \left[ \mathcal{E}_{abcd} - \mathcal{E}_{abdc} \right] &= \sum_k B^I_{abk} w_{cdk}^{\circ}
\end{align}
\end{subequations}
The left-hand sides of (\ref{eqn:hermgt}) are, respectively, symmetric and antisymmetric with respect to interchange of $a$ and $b$.  This implies that we can represent them using $B^R$ and $B^I$, respectively.  In each equality, the middle expressions are, respectively, symmetric and antisymmetric with respect to interchange of $c$ and $d$.  By arguments made in Section \ref{sect:tiedgt}, this will cause (\ref{eqn:hermgtcond}) to be satisfied.   Therefore, with (\ref{eqn:hermgt}), the Hermitian representation model (\ref{eqn:hermrepmodel}) matches the ground truth Hamiltonian perfectly.

Actually solving (\ref{eqn:hermgt}) is trivial.  Let us say that $(a,b)$ is an upper-triangular index if $a \leq b$.  As $k$ goes from $1$ to $N(N+1)/2$, each $k$ will correspond to a unique upper-triangular index $(a',b')$; we then set $v^\circ_{:,:,k}$ equal to $(\mathcal{E}_{a',b',:,:} + \mathcal{E}_{b',a',:,:})/2$---the left-hand side of (\ref{eqn:hermgt1}) with $a=a'$, $b=b'$, and $(c,d)$ allowed to be free.

Similarly, let us say that $(a,b)$ is a strictly upper-triangular index if $a < b$.  As $k$ goes from $1$ to $N(N-1)/2$, each $k$ will correspond to a unique strictly upper-triangular index $(a',b')$; we set $w^\circ_{:,:,k}$ equal to $(\mathcal{E}_{a',b',:,:} - \mathcal{E}_{b',a',:,:})/2$---the left-hand side of (\ref{eqn:hermgt2}) with $a=a'$, $b=b'$, and $(c,d)$ allowed to be free.

\subsubsection{Eight-Fold Symmetry-Preserving Model}
\label{sect:eightfoldgt}
When we implement (\ref{eqn:eightfoldmodel}) numerically, we use a slight modification of Algorithm \ref{algo:pop} that enables us to directly populate the sparse tensor
\begin{equation}
\label{eqn:Sprime}
\mathscr{S}'_{ijklm} = \mathscr{S}_{ijlkm} - (1/2) \mathscr{S}_{ikljm}.
\end{equation}
This allows us to express (\ref{eqn:eightfoldmodel}) as
\begin{equation}
\label{eqn:efficienteightfold}
\widetilde{H}(P)_{ij} = H^\text{core}_{ij} + \sum_{k l} \left[ \sum_m \mathscr{S}'_{ijklm} \beta_m \right] P_{kl},
\end{equation}
which is more efficient to compute as it avoids an extra tensor contraction/transpose. Form (\ref{eqn:efficienteightfold}) is close to the CO basis ground truth Hamiltonian (\ref{eqn:gthamCO}), which we reproduce here as well:
\[
H(P)_{ij} = H^{\text{core}}_{ij} + \sum_{kl} \mathcal{E}_{ijkl} P_{kl}.
\]
To explain why there exists $\bbeta$ such that the model matches the ground truth, we first take the $2$-electron tensor and express it in the CO basis via
\begin{equation}
\label{eqn:Tdef}
\mathcal{T}_{abcd} =\sum_{ijkl} \mathcal{X}_{ia} \mathcal{X}_{jb}  (i j| k l) \mathcal{X}_{kc} \mathcal{X}_{l d}.
\end{equation}
As we did above, we assume the use of a real set of atomic orbitals, so that (i) $\mathcal{X}$ is real and (ii) the $2$-electron tensor $(i j | k l)$ satisfies the permutation symmetry (\ref{eqn:eightfold}).  Now we claim that $\mathcal{T}$ also satisfies the permutation symmetry (\ref{eqn:eightfold}).

Recall from Section \ref{sect:eightfoldsymmdef} the definition of $\sigma^m$, the $m$-th permutation present in the symmetry (\ref{eqn:eightfold}).  Suppose we apply $\sigma^m$ to the indices $(a,b,c,d)$ in $\mathcal{T}$.  With the shorthand $(a',b',c',d') = \sigma^m(a,b,c,d)$, we can write the result as
\[
\mathcal{T}_{a' b' c' d'} =\sum_{ijkl} \mathcal{X}_{i a'} \mathcal{X}_{j b'}  (i j| k l) \mathcal{X}_{k c'} \mathcal{X}_{l d'}.
\]
Let us also permute the indices $(i,j,k,l)$ using $\sigma^m(i,j,k,l) = (i',j',k',l')$.  As the $2$-electron tensor is invariant under $\sigma^m$, we  obtain
\[
\mathcal{T}_{a' b' c' d'} =\sum_{ijkl} \mathcal{X}_{i' a'} \mathcal{X}_{j' b'}  (i' j'| k' l') \mathcal{X}_{k' c'} \mathcal{X}_{l' d'}.
\]
As summing over $(i,j,k,l)$ is equivalent to summing over $(i',j',k',l')$, the right-hand side of the previous expression equals $\mathcal{T}_{a b c d}$.  In short, changing from the AO to the CO basis does not alter the symmetries enjoyed by the $2$-electron tensor.  This means that $\mathcal{T}$ can be represented using the basis $\mathscr{S}$ constructed in Section \ref{sect:eightfoldsymmdef}, i.e., there exists $\bbeta^\circ$ such that
\[
\sum_m \mathscr{S}_{ijklm} \beta^\circ_m = \mathcal{T}_{ijkl}.
\]
With real $\mathcal{X}$, the expressions (\ref{eqn:Econvenience}), (\ref{eqn:Tdef}), and (\ref{eqn:Sprime}) together imply that
\begin{multline*}
\mathcal{E}_{abcd} = \mathcal{T}_{abdc} - \frac{1}{2} \mathcal{T}_{acdb} = \sum_m \left[ \mathscr{S}_{abdcm} - \frac{1}{2} \mathscr{S}_{acdbm} \right] \beta_m^\circ \\
= \sum_m \mathscr{S}'_{ijklm} \beta_m^\circ,
\end{multline*}
as desired.  Computing $\bbeta^\circ$ is simple.  For each $m$, let $(i,j,k,l)$ be any $4$-tuple in the unique orbit that corresponds to $m$.  Then set $\beta^\circ_{ijklm} = \mathcal{T}_{ijkl}$.

\subsection{Derivatives of Models}
\label{sect:modelderivatives}
Picking up from where Section \ref{sect:training} of the manuscript ends, let us first give mathematical details regarding derivatives of the residual (\ref{eqn:res}).  In this section, we use the notation
\[
\widetilde{H}^{(j)} = \widetilde{H}(P^{(j)})
\]
and we omit writing out the $\btheta$-dependence of $S$ and $\widetilde{H}$ unless absolutely necessary.  Then we can express the residual (\ref{eqn:res}) as
\begin{equation}
\label{eqn:resnew}
S^{(j)}_{k \ell} = \imath \dot{P}^{(j)}_{k \ell} - \left[ \sum_r \widetilde{H}^{(j)}_{kr} P^{(j)}_{r \ell} - P^{(j)}_{k r} \widetilde{H}^{(j)}_{r \ell} \right].
\end{equation}
Taking a derivative, we see that the $\dot{P}$ term drops out and we are left with
\begin{equation}
\label{eqn:resderiv}
\frac{ \partial S^{(j)}_{k \ell}}{ \partial \theta_m } = - \left[ \sum_r \frac{ \partial \widetilde{H}^{(j)}_{kr}}{ \partial \theta_m } P^{(j)}_{r \ell} - P^{(j)}_{k r} \frac{ \partial \widetilde{H}^{(j)}_{r \ell}}{ \partial \theta_m } \right].
\end{equation}
As this is true for all of our models, we use it as a starting point for each of the subsequent derivations.

\subsubsection{Tied Model}
For the tied regression model (\ref{eqn:tiedmodel}), recall that $\btheta = \bbeta$, where $\bbeta$ can be viewed as a tensor of shape $N \times N \times N \times N$. Using (\ref{eqn:tiedmodel}) and the notation $P^{(j)} = P^{R,j} + \imath P^{I,j}$, it is straightforward to derive
\begin{equation}
\label{eqn:tiedhamderiv}
\frac{ \partial \widetilde{H}^{(j)}_{kr}}{ \partial \beta_{abcd}} = \frac{1}{2} \left[ P^{R,j}_{ab}(\delta_{ck} \delta_{dr} + \delta_{cr} \delta_{dk}) + \imath P^{I,j}_{ab}(\delta_{ck} \delta_{dr} - \delta_{cr} \delta_{dk}) \right].
\end{equation}
Using this and (\ref{eqn:resderiv}), we can derive a first expression for the right Jacobian-vector product
\begin{multline*}
\sum_{abcd} \frac{ \partial S^{(j)}_{k \ell}}{ \partial \beta_{abcd}} \xi_{abcd} = \\
-\sum_r \left[ \sum_{ab} \frac{1}{2} P^{R,j}_{ab} ( \xi_{abkr} + \xi_{abrk} ) + \frac{\imath}{2} P^{I,j}_{ab} ( \xi_{abkr} - \xi_{abrk} ) \right] P^{(j)}_{r \ell} \\
+\sum_r \left[ \sum_{ab} \frac{1}{2} P^{R,j}_{ab} ( \xi_{abr\ell} + \xi_{a b \ell r} ) + \frac{\imath}{2} P^{I,j}_{ab} ( \xi_{abr \ell} - \xi_{a b \ell r} ) \right] P^{(j)}_{k r}.
\end{multline*}
When we actually compute this product, we first compute the intermediate quantities
\begin{equation}
\label{eqn:zetadef}
\zeta^{R,j}_{kr} = \sum_{ab} P^{R,j}_{ab} \xi_{abkr}, \quad \text{ and } \quad \zeta^{I,j}_{kr} = \sum_{ab} P^{I,j}_{ab} \xi_{abkr}.
\end{equation}
These expressions are tensor contractions that can be computed efficiently via \texttt{einsum} in both NumPy \citep{harris2020array} and CuPy \citep{cupy_learningsys2017}, a drop-in GPU-enabled replacement of NumPy.  With this, we can express the right Jacobian-vector product as
\begin{multline}
\label{eqn:tiedrightjvp}
\sum_{abcd} \frac{ \partial S^{(j)}_{k \ell}}{ \partial \beta_{abcd}} \xi_{abcd} = \\
-\frac{1}{2} \sum_r (\zeta^{R,j}_{kr} + \zeta^{R,j}_{rk}) P^{(j)}_{r \ell} - P^{(j)}_{kr} (\zeta^{R,j}_{r \ell} + \zeta^{R,j}_{\ell r}) \\
-\frac{\imath}{2} \sum_r (\zeta^{I,j}_{kr} - \zeta^{I,j}_{rk}) P^{(j)}_{r \ell} - P^{(j)}_{kr} (\zeta^{I,j}_{r \ell} - \zeta^{I,j}_{\ell r}).
\end{multline}
For fixed $j$, the expressions on the right-hand side are matrix-matrix products, which can be computed efficiently in CuPy.  Using (\ref{eqn:resderiv}) and (\ref{eqn:tiedhamderiv}), the left product needed by LSMR is
\begin{multline*}
\sum_{jk\ell} \xi_{jk\ell} \overline{\frac{ \partial S^{(j)}_{k \ell}}{ \partial \beta_{abcd}}} = 
\sum_{jk \ell r} \xi_{jkl} \biggl[ \biggl( -\frac{1}{2} P^{R,j}_{ab} \left( \delta_{ck} \delta_{dr} + \delta_{cr} \delta_{dk} \right) \\
+ \frac{\imath}{2} P^{I,j}_{ab} \left( \delta_{ck} \delta_{dr} - \delta_{cr} \delta_{dk} \right) \biggr) \overline{P^{(j)}_{r\ell}} \\
+ \overline{P^{(j)}_{kr}} \biggl( \frac{1}{2} P^{R,j}_{ab} (\delta_{cr} \delta_{d \ell} + \delta_{c\ell} \delta_{dr} ) \\
- \frac{\imath}{2} P^{I,j}_{ab} (\delta_{cr} \delta_{d \ell} - \delta_{c\ell} \delta_{dr} ) \biggr) \biggr].
\end{multline*}
From this, we obtain our final expression for the left Jacobian-vector product, which we write in a form suitable for \texttt{einsum} implementation:
\begin{multline}
\label{eqn:tiedleftjvp}
\sum_{jk\ell} \xi_{jk\ell} \overline{\frac{ \partial S^{(j)}_{k \ell}}{ \partial \beta_{abcd}}} 
= \frac{1}{2} \sum_j P^{R,j}_{ab} \biggl(  -\sum_{\ell} \xi_{j c \ell} \overline{P^{(j)}_{d\ell}} +\xi_{j d \ell} \overline{P^{(j)}_{c\ell}}  \\
+ \sum_{k} \xi_{jkd} \overline{P^{(j)}_{kc}} + \xi_{jkc} \overline{P^{(j)}_{kd}} \biggr) \\
+ \frac{\imath}{2} \sum_j P^{I,j}_{ab} \biggl( \sum_{\ell} \xi_{j c \ell} \overline{P^{(j)}_{d\ell}} - \xi_{j d \ell} \overline{P^{(j)}_{c\ell}} \\
- \sum_{k} \xi_{jkd} \overline{P^{(j)}_{kc}} - \xi_{jkc} \overline{P^{(j)}_{kd}} \biggr).
\end{multline}

\subsubsection{Hermitian Representation Model}
For the Hermitian representation model (\ref{eqn:hermrepmodel}), the parameters are $\btheta = (\mathbf{v}, \mathbf{w})$.  The form of the model leads to extremely compact expressions for the derivatives of the Hamiltonian:
\begin{equation}
\label{eqn:Hermhamderivs}
\frac{ \partial \widetilde{H}^{(s)}_{k\ell}}{\partial v_{abc}} = P^{R,s}_{ab} B^R_{k\ell c} \quad \text{ and } \quad \frac{ \partial \widetilde{H}^{(s)}_{k\ell}}{\partial w_{abc}} = \imath P^{I,s}_{ab} B^I_{k\ell c}.
\end{equation}
With this and (\ref{eqn:resderiv}), we can compute the first right Jacobian-vector product
\begin{align}
\sum_{abc} \frac{ \partial S^{(s)}_{k \ell}}{\partial v_{abc}} \xi_{abc} &= \sum_{abc}\sum_{r} -\frac{ \partial \widetilde{H}^{(s)}_{kr}}{\partial v_{abc}} \xi_{abc} P_{r \ell}^{(s)} + P_{kr}^{(s)} \frac{ \partial \widetilde{H}^{(s)}_{r\ell}}{\partial v_{abc}} \xi_{abc} \nonumber \\
 &= \sum_{abcr} -P^{R,s}_{ab} B^R_{krc} \xi_{abc} P_{r \ell}^{(s)} + P_{kr}^{(s)} P^{R,s} B^R_{r\ell c} \xi_{abc} \nonumber \\
\label{eqn:hermvrightjvp}
 &= \sum_{r} -\eta^{R,s}_{kr}  P_{r \ell}^{(s)} + P_{kr}^{(s)} \eta^{R,s}_{r\ell}
\end{align}
where $\eta^{R,s}_{kr} = \sum_{abc} P^{R,s}_{ab} B^R_{krc} \xi_{abc}$. By (\ref{eqn:Hermhamderivs}), we see that the second right Jacobian-vector product is the imaginary counterpart of the above: starting from (\ref{eqn:hermvrightjvp}), change all instances of $R$ to $I$ and multiply by an overall factor of $\imath$ to obtain
\begin{equation}
\label{eqn:hermwrightjvp}
\sum_{abc} \frac{ \partial S^{(s)}_{k \ell}}{\partial w_{abc}} \xi_{abc} = \imath \sum_{r} -\eta^{I,s}_{kr}  P_{r \ell}^{(s)} + P_{kr}^{(s)} \eta^{I,s}_{r\ell}.
\end{equation}
where $\eta^{I,s}_{kr} = \sum_{abc} P^{I,s}_{ab} B^I_{krc} \xi_{abc}$.  For the first left Jacobian-vector product, we compute
\begin{align}
\sum_{sk \ell} \xi_{sk \ell} \overline{\frac{ \partial S^{(s)}_{k \ell}}{\partial v_{abc}} } &= \sum_{sk \ell} \sum_r - \overline{\frac{ \partial \widetilde{H}^{(s)}_{kr}}{\partial v_{abc}}} \xi_{sk \ell} \overline{P_{r \ell}^{(s)}} + \overline{P_{kr}^{(s)}}  \overline{\frac{ \partial \widetilde{H}^{(s)}_{r\ell}}{\partial v_{abc}}} \xi_{s k \ell} \nonumber  \\
\label{eqn:hermvleftjvp}
&= \sum_{sk\ell r} -P^{R,s}_{ab} B^R_{krc} \xi_{sk\ell} \overline{P^{(s)}_{r\ell}} + \overline{P^{(s)}_{kr}} P^{R,s}_{ab} B^R_{r \ell c} \xi_{s k \ell}.
\end{align}
To obtain the second left Jacobian-vector product from this, we again substitute $I$ for $R$ but now multiply by $-\imath$ to account for the complex conjugate.  This yields
\begin{equation}
\label{eqn:hermwleftjvp}
\sum_{sk \ell} \xi_{sk \ell} \overline{\frac{ \partial S^{(s)}_{k \ell}}{\partial w_{abc}} } = \imath \sum_{sk\ell r} P^{I,s}_{ab} B^I_{krc} \xi_{sk\ell} \overline{P^{(s)}_{r\ell}} -\overline{P^{(s)}_{kr}} P^{I,s}_{ab} B^I_{r \ell c} \xi_{s k \ell}.
\end{equation}

\subsubsection{Eightfold Symmetry-Preserving Model}
\label{sect:eightfoldderivs}
For the $8$-fold symmetry-preserving model (\ref{eqn:eightfoldmodel}), we have $\btheta = \bbeta$ of dimension $n_T$ given by (\ref{eqn:BHNT}); differentiating yields
\begin{equation}
\label{eqn:eightfoldderiv}
\frac{ \partial \widetilde{H}^{(s)}_{ij}}{\partial \beta_m} = \sum_{k \ell} \left[ \mathscr{S}_{ij \ell km} - \frac{1}{2} \mathscr{S}_{ikj \ell m} \right] P^{(s)}_{k \ell}.
\end{equation}
By (\ref{eqn:resderiv}), the right Jacobian-vector product is then
\begin{multline*}
\sum_m \frac{ \partial S^{(s)}_{ij}}{\partial \beta_m} \xi_m = \sum_{r k \ell m} \biggl( -\left[ \mathscr{S}_{ir\ell k m} - \frac{1}{2} \mathscr{S}_{ikr \ell m} \right] P^{(s)}_{k \ell} P^{(s)}_{rj} \xi_m \\
+ P^{(s)}_{ir} \left[ \mathscr{S}_{rj\ell km} - \frac{1}{2} \mathscr{S}_{rkj \ell m} \right] P^{(s)}_{k \ell} \xi_m \biggr).
\end{multline*}
We note the repeated object
\[
\nu_{i r k \ell} = \sum_m \left[ \mathscr{S}_{ir\ell km} - \frac{1}{2} \mathscr{S}_{ikr \ell m} \right] \xi_m = \sum_m \mathscr{S}_{ir k \ell m}' \xi_m.
\]
This is computable in CuPy via contraction of the sparse tensor $\mathscr{S}'$---also used to compute (\ref{eqn:efficienteightfold})---against the vector $\xi$.  The output has the same dimension as the ground-truth $4$-index, two-electron tensor $(i j \mid k \ell)$.  Hence $\nu$ can be stored as a dense tensor.  With this, we obtain an efficiently computable form for the right Jacobian-vector product:
\begin{equation}
\label{eqn:efrightjvp}
\sum_m \frac{ \partial S^{(s)}_{ij}}{\partial \beta_m} \xi_m = \sum_{r k \ell} -\nu_{i r k \ell} P^{(s)}_{k \ell} P^{(s)}_{rj} + P^{(s)}_{ir} P^{(s)}_{k \ell} \nu_{r j k \ell}.
\end{equation}
The left Jacobian-vector product benefits from the extra sum over $s$.  We can define
\begin{equation*}
\Xi^1_{irk\ell} = \sum_{sj} \overline{ P^{(s)}_{k \ell} P^{(s)}_{rj}} \xi_{sij} \quad \text{ and } \quad
\Xi^2_{rjk\ell} = \sum_{sj} \overline{ P^{(s)}_{k \ell} P^{(s)}_{ir}} \xi_{sij},
\end{equation*}
after which we have
\begin{equation}
\label{eqn:efleftjvp}
\sum_{sij} \xi_{sij} \overline{ \frac{ \partial S^{(s)}_{ij}}{\partial \beta_m} } = -\sum_{irk\ell} \Xi^1_{irk\ell} \mathscr{S}'_{i r k \ell m} + \sum_{rjk\ell} \Xi^2_{r j k \ell} \mathscr{S}'_{r j k \ell m}.
\end{equation}
For this particular model, we also give expressions for the gradient and Hessian of the loss.   We do this because we use the exact Hessian-based formula (\ref{eqn:thetanormal2}) to train the $8$-fold symmetry-preserving model for four of the molecular systems under study.  We begin by expressing the loss (\ref{eqn:loss}) as
\[
\mathcal{L}(\bbeta) = \sum_{s=1}^{n_{\text{train}}} \sum_{ij} S^{(s)}(\bbeta)_{ij} \overline{S^{(s)}(\bbeta)_{ij}}.
\]
Then we see that the gradient of the loss is
\begin{equation}
\label{eqn:gradloss8fold}
\frac{ \partial \mathcal{L} }{ \partial \beta_m } = 2 \Re \sum_{s=1}^{n_{\text{train}}} \sum_{i j} S^{(s)}_{ij} \overline{\frac{ \partial S^{(s)}_{ij} }{ \partial \beta_m }} .
\end{equation}
We see that the summation here has precisely the same form as the left Jacobian-vector product (\ref{eqn:efleftjvp}), with $\xi_{sij} = S^{(s)}_{ij}$.  We use this in our code to efficiently compute the gradient of the loss.

For the Hessian, note that as the model is linear in the parameters $\bbeta$, the gradient of the residual does not itself depend on $\bbeta$.  Therefore, the Hessian of the loss is
\begin{equation}
\label{eqn:hessloss8fold}
\frac{ \partial^2 \mathcal{L} }{ \partial \beta_n \partial \beta_m } = 2 \Re \sum_{s=1}^{n_{\text{train}}} \sum_{i j} \frac{ \partial S^{(s)}_{ij}}{ \partial \beta_n } \overline{\frac{ \partial S^{(s)}_{ij} }{ \partial \beta_m }} .
\end{equation}
In our code, we compute this Hessian matrix one row at a time: for each fixed $n$, the expression on the right-hand side is again a left Jacobian-vector product.

\subsection{Supporting Results}
\label{sect:supporting}

\subsubsection{Training Loss}
\label{sect:supptrainingloss}
Here we give details, tables and figures that supplement the results in the main text. First note that for all molecular systems, for all models, when we use LSMR, we set tolerances equal to $10^{-16}$.  We use $10^5$ maximum iterations except when training $8$-fold symmetry-preserving models on the small molecular systems, in which case we use $2 \times 10^5$ maximum iterations.  This has no effect on the results, as can be seen from Figure \ref{fig:trainiter}.

\begin{figure*}
\centering
\includegraphics[width=3.5in]{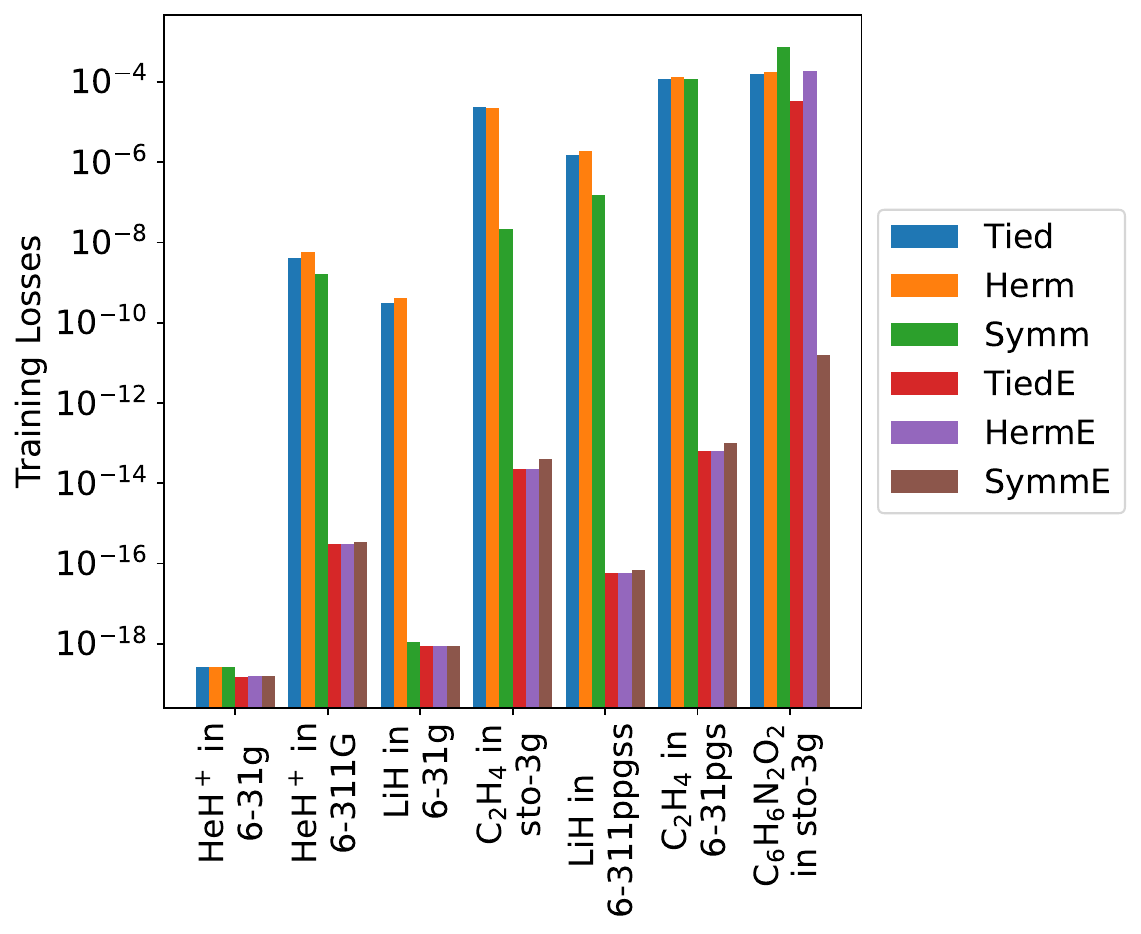}
\caption{Final converged values of the training loss for all LSMR-trained models, for all seven molecular systems.}
\label{fig:TrainingLosses}
\end{figure*}

We begin by reporting in Table \ref{tab:TrainingLosses} the final values of the training loss for all LSMR- and Hessian-trained models.  In Figure \ref{fig:TrainingLosses}, we plot the final values of the training loss for all LSMR-trained models.  The results complement the plots of training loss versus iteration number in Figure \ref{fig:trainiter} in the main text.

\begin{table*}
    \centering\footnotesize
    \caption{Training losses for all models, including both LSMR- and Hessian-trained models, across all molecular systems.  For each molecular system, the smallest value of training loss is boldfaced.}
    \label{tab:TrainingLosses}
    \begin{tabular}{l @{\hspace{1.5\tabcolsep}} c @{\hspace{1.5\tabcolsep}} c @{\hspace{1.5\tabcolsep}} c @{\hspace{1.5\tabcolsep}} c @{\hspace{1.5\tabcolsep}} c @{\hspace{1.5\tabcolsep}} c @{\hspace{1.5\tabcolsep}} c}
        \toprule
        \thead{Model} & \thead{$\text{HeH}^+$ in \\6-31G} & \thead{$\text{HeH}^+$ in \\ 6-311G} & \thead{$\text{LiH}$ in \\6-31G} & \thead{$\text{C}_2 \text{H}_4$ in \\STO-3G} & \thead{$\text{LiH}$ in \\6-311ppgss} & \thead{$\text{C}_2 \text{H}_4$ in\\ 6-31pgs} & \thead{$\text{C}_6 \text{H}_6 \text{N}_2 \text{O}_2$\\ in STO-3G} \\
        \midrule
        Tied & $2.63 \! \times \! 10^{-19}$ & $4.08 \! \times \! 10^{-9}$ & $3.14 \! \times \! 10^{-10}$ & $2.36 \! \times \! 10^{-5}$ & $1.52 \! \times \! 10^{-6}$ & $1.15 \! \times \! 10^{-4}$ & $1.57 \! \times \! 10^{-4}$ \\
        Herm & $2.63 \! \times \! 10^{-19}$ & $5.87 \! \times \! 10^{-9}$ & $4.16 \! \times \! 10^{-10}$ & $2.23 \! \times \! 10^{-5}$ & $1.90 \! \times \! 10^{-6}$ & $1.36 \! \times \! 10^{-4}$ & $1.75 \! \times \! 10^{-4}$ \\
        Symm & $2.63 \! \times \! 10^{-19}$ & $1.60 \! \times \! 10^{-9}$ & $1.11 \! \times \! 10^{-18}$ & $2.20 \! \times \! 10^{-8}$ & $1.54 \! \times \! 10^{-7}$ & $1.18 \! \times \! 10^{-4}$ & $7.60 \! \times \! 10^{-4}$ \\
        TiedE & $\mathbf{1.54 \! \times \! 10^{-19}}$ & $3.15 \! \times \! 10^{-16}$ & $8.64 \! \times \! 10^{-19}$ & $2.25 \! \times \! 10^{-14}$ & $\mathbf{5.97 \! \times \! 10^{-17}}$ & $6.51 \! \times \! 10^{-14}$ & $3.34 \! \times \! 10^{-5}$ \\
        HermE & $1.54 \! \times \! 10^{-19}$ & $\mathbf{3.15 \! \times \! 10^{-16}}$ & $\mathbf{8.64 \! \times \! 10^{-19}}$ & $\mathbf{2.25 \! \times \! 10^{-14}}$ & $5.97 \! \times \! 10^{-17}$ & $\mathbf{6.51 \! \times \! 10^{-14}}$ & $1.88 \! \times \! 10^{-4}$ \\
        SymmE & $1.55 \! \times \! 10^{-19}$ & $3.42 \! \times \! 10^{-16}$ & $9.01 \! \times \! 10^{-19}$ & $4.01 \! \times \! 10^{-14}$ & $6.96 \! \times \! 10^{-17}$ & $9.92 \! \times \! 10^{-14}$ & $\mathbf{1.58 \! \times \! 10^{-11}}$ \\
        SymmH & $1.75 \! \times \! 10^{-8}$ & $5.64 \! \times \! 10^{-6}$ & $9.17 \! \times \! 10^{-15}$ & $2.20 \! \times \! 10^{-8}$ & na & na & na \\
        SymmHE & $1.55 \! \times \! 10^{-19}$ & $3.42 \! \times \! 10^{-16}$ & $9.01 \! \times \! 10^{-19}$ & $4.01 \! \times \! 10^{-14}$ & na & na & na \\
        \bottomrule
    \end{tabular}
\end{table*}

\subsubsection{Propagation Errors}
\label{sect:suppproperrors}
Next, we report both field-free and field-on $\| \cdot \|_{\infty}$ propagation errors in Table \ref{tab:properrFF} and Table \ref{tab:properrFO}, respectively.   For each molecular system, the smallest propagation error value is boldfaced.  These results correspond to those plotted in Figure \ref{fig:HessWash} and Figure \ref{fig:LinftyPropErrors} in the main text.

We see that the Hessian-trained models (SymmH, SymmHE) never achieve the best (lowest) error, across all molecular systems, and across both field-free and field-on settings.  This supports the finding in the main text that we lose nothing by training our models via the iterative LSMR algorithm instead of the exact Hessian-based approach.  Among LSMR models, for the larger molecular systems, we see a dramatic difference in the performance of models trained on a single trajectory (Tied, Herm, and Symm) versus those trained on an ensemble of field-free trajectories (TiedE, HermE, and SymmE).  The SymmE model is either the best model or within a reasonably small tolerance of the best model.  For \paranitrocomma the largest molecular system, SymmE is clearly the best among all models.

\begin{figure*}[p]
\centering
\includegraphics[width=0.45\textwidth]{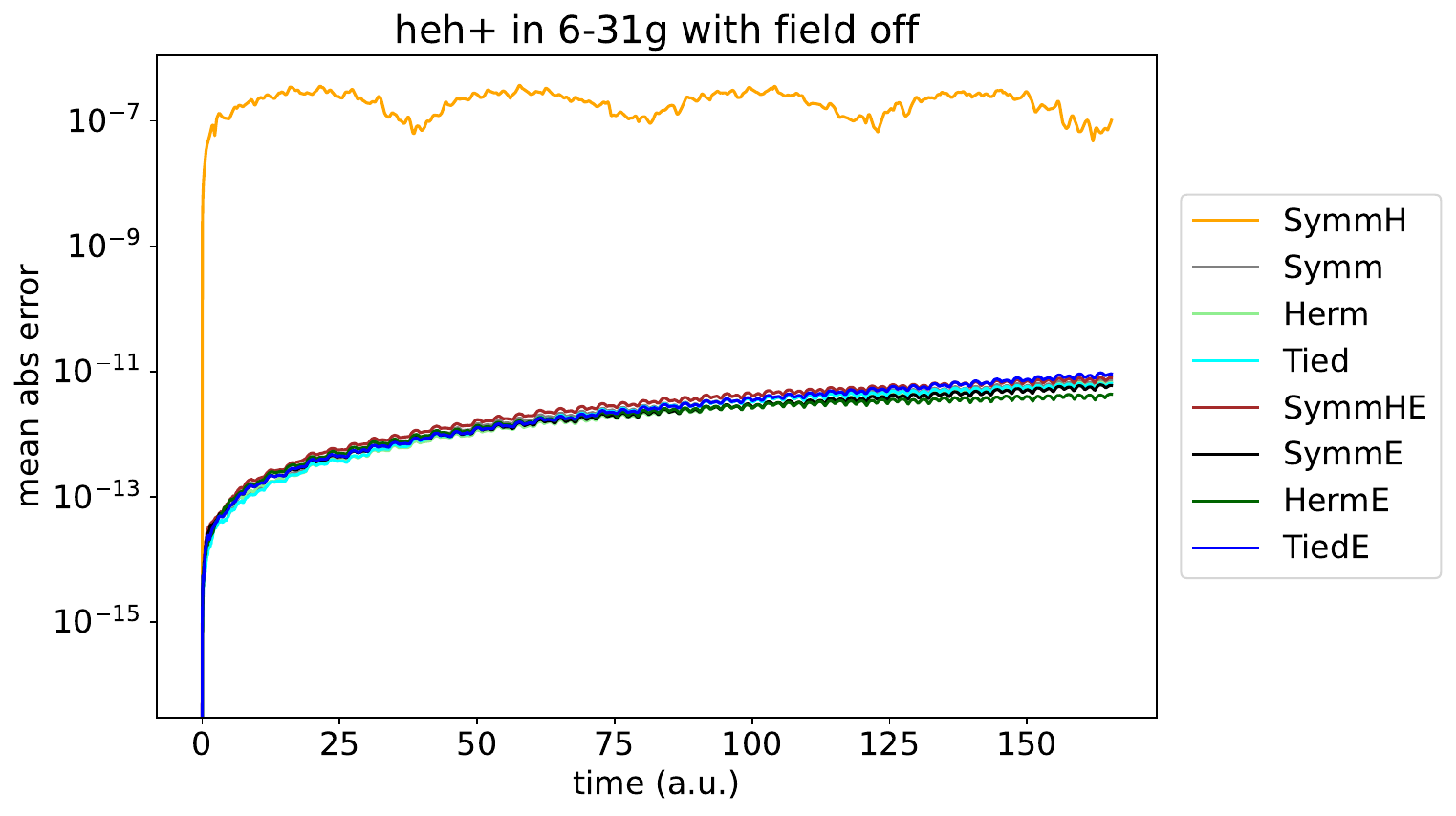}
\includegraphics[width=0.45\textwidth]{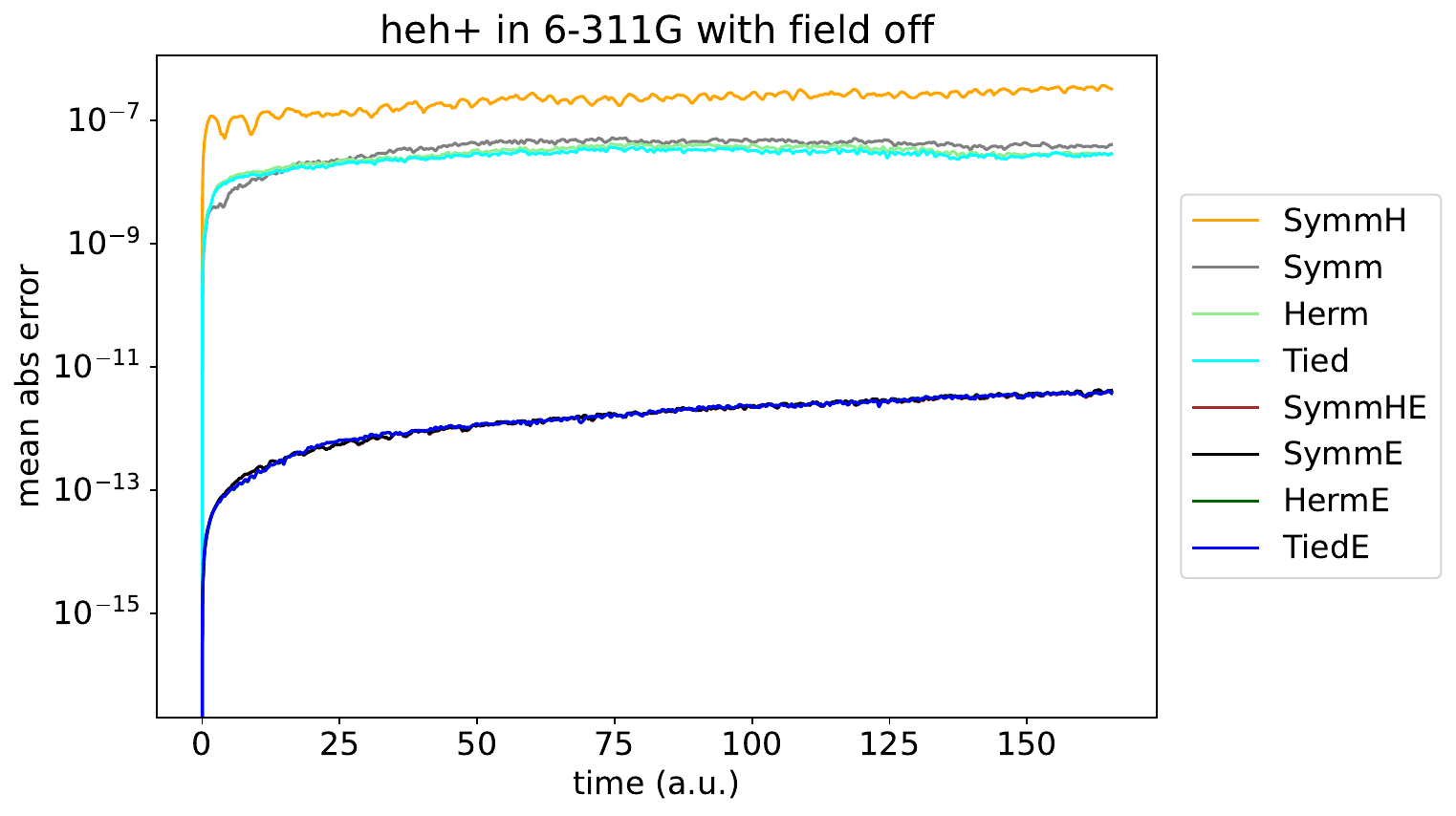} \\
\includegraphics[width=0.45\textwidth]{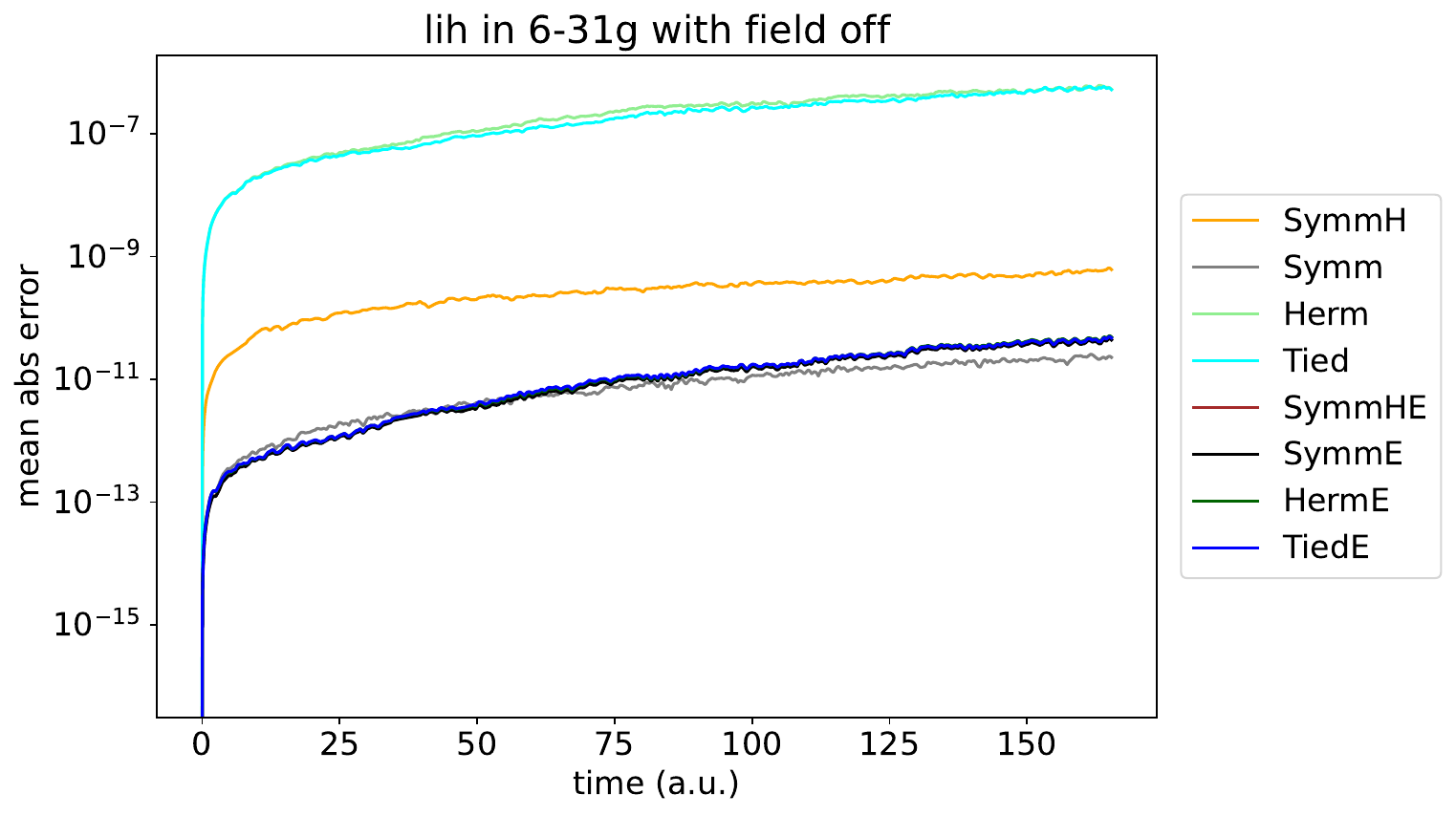}
\includegraphics[width=0.45\textwidth]{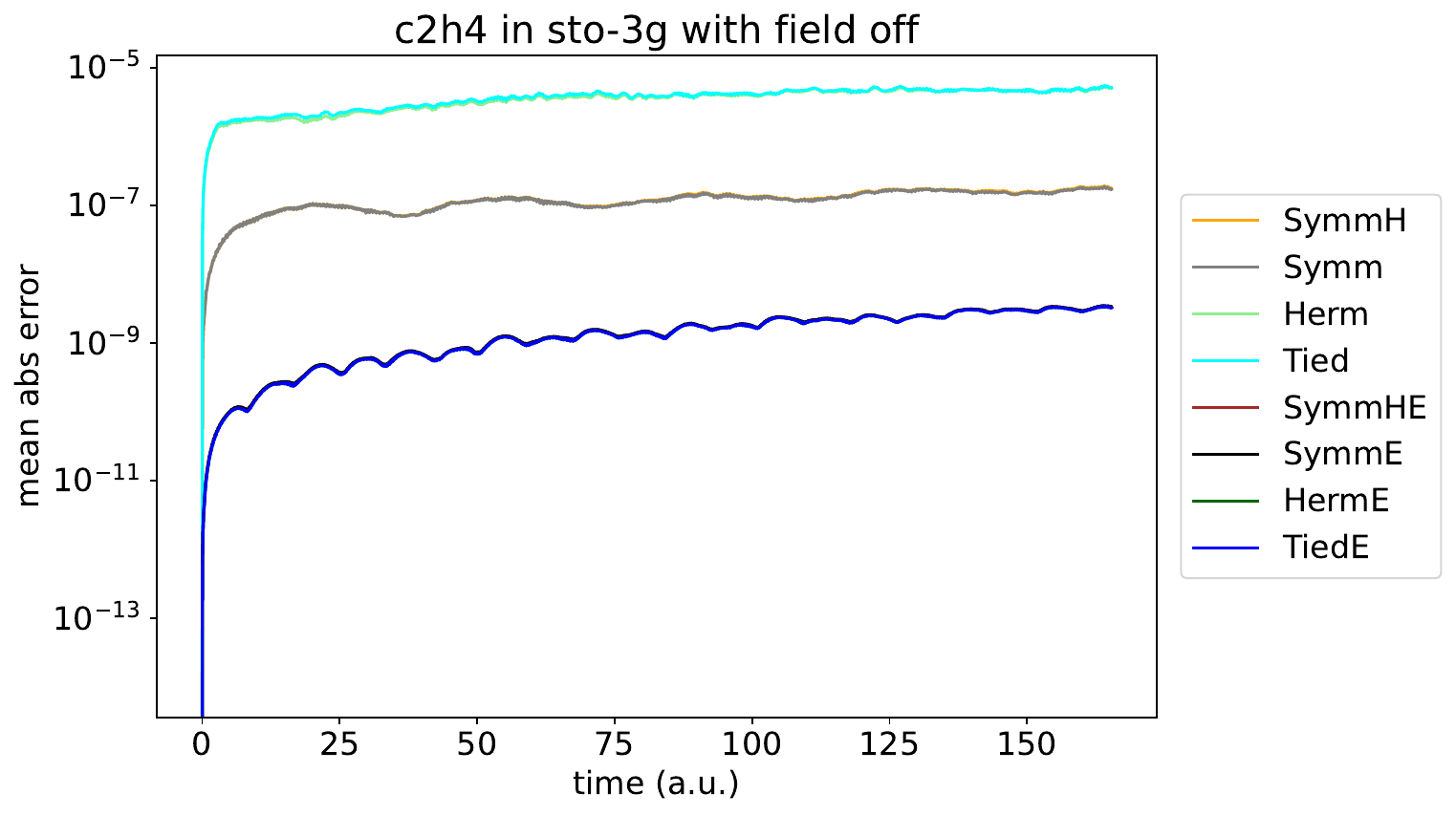} \\
\includegraphics[width=0.45\textwidth]{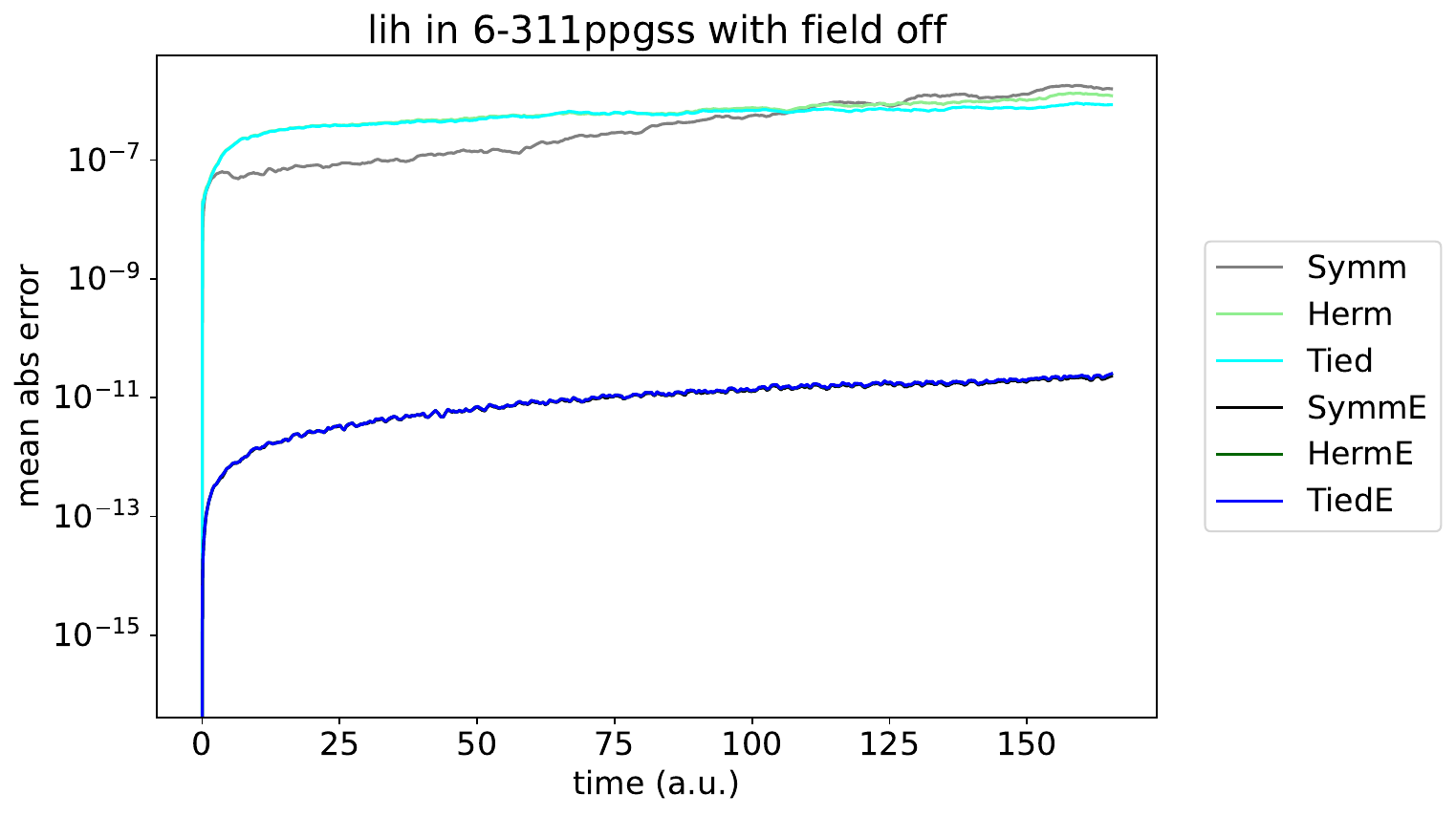}
\includegraphics[width=0.45\textwidth]{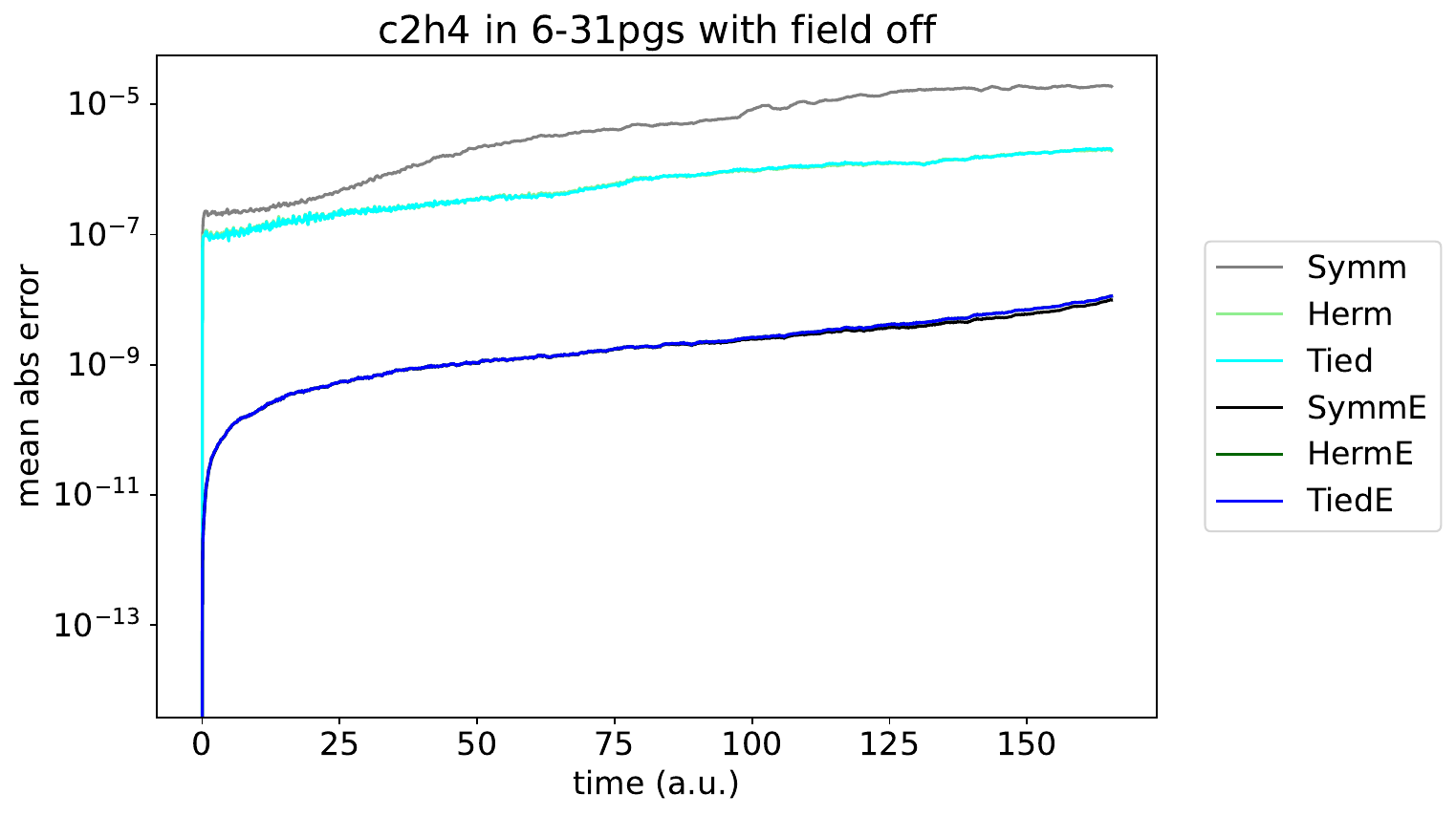} \\
\includegraphics[width=0.45\textwidth]{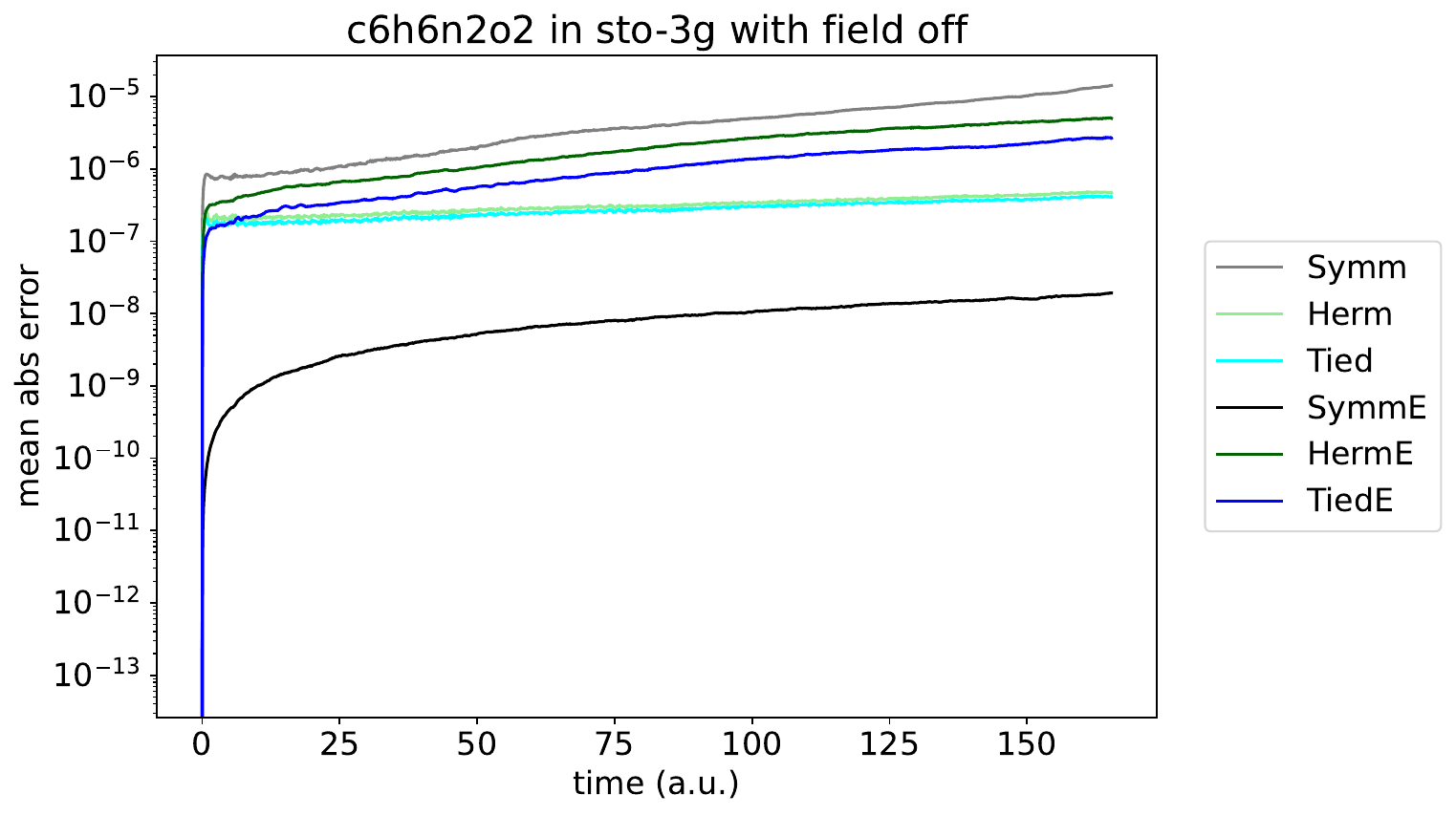}
\caption{We plot the field-free mean absolute propagation error (\ref{eqn:tdmae}) for all models and all molecular systems.}
\label{fig:MAEproperrFF}
\end{figure*}

\begin{figure*}[p]
\centering
\includegraphics[width=0.45\textwidth]{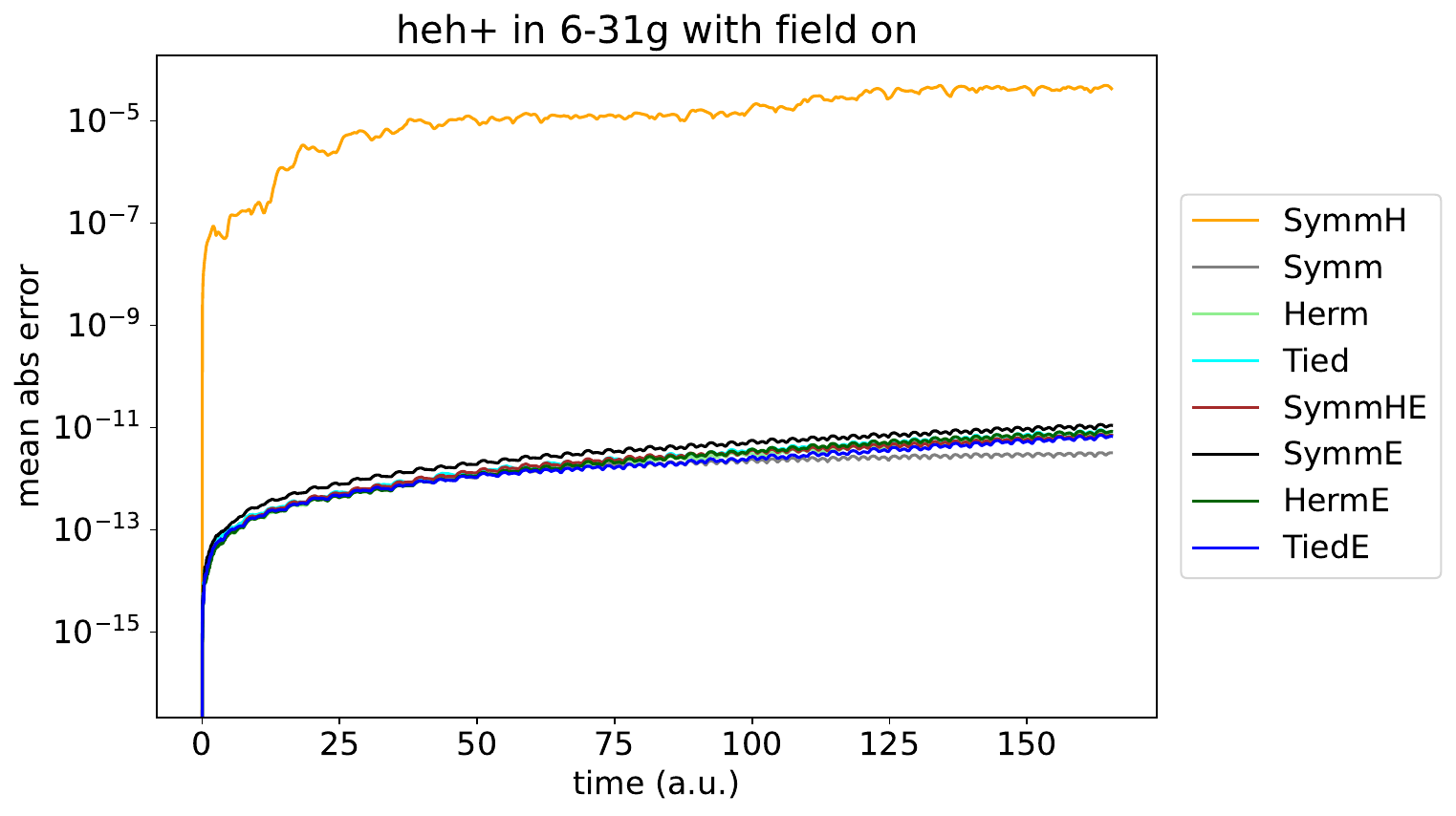}
\includegraphics[width=0.45\textwidth]{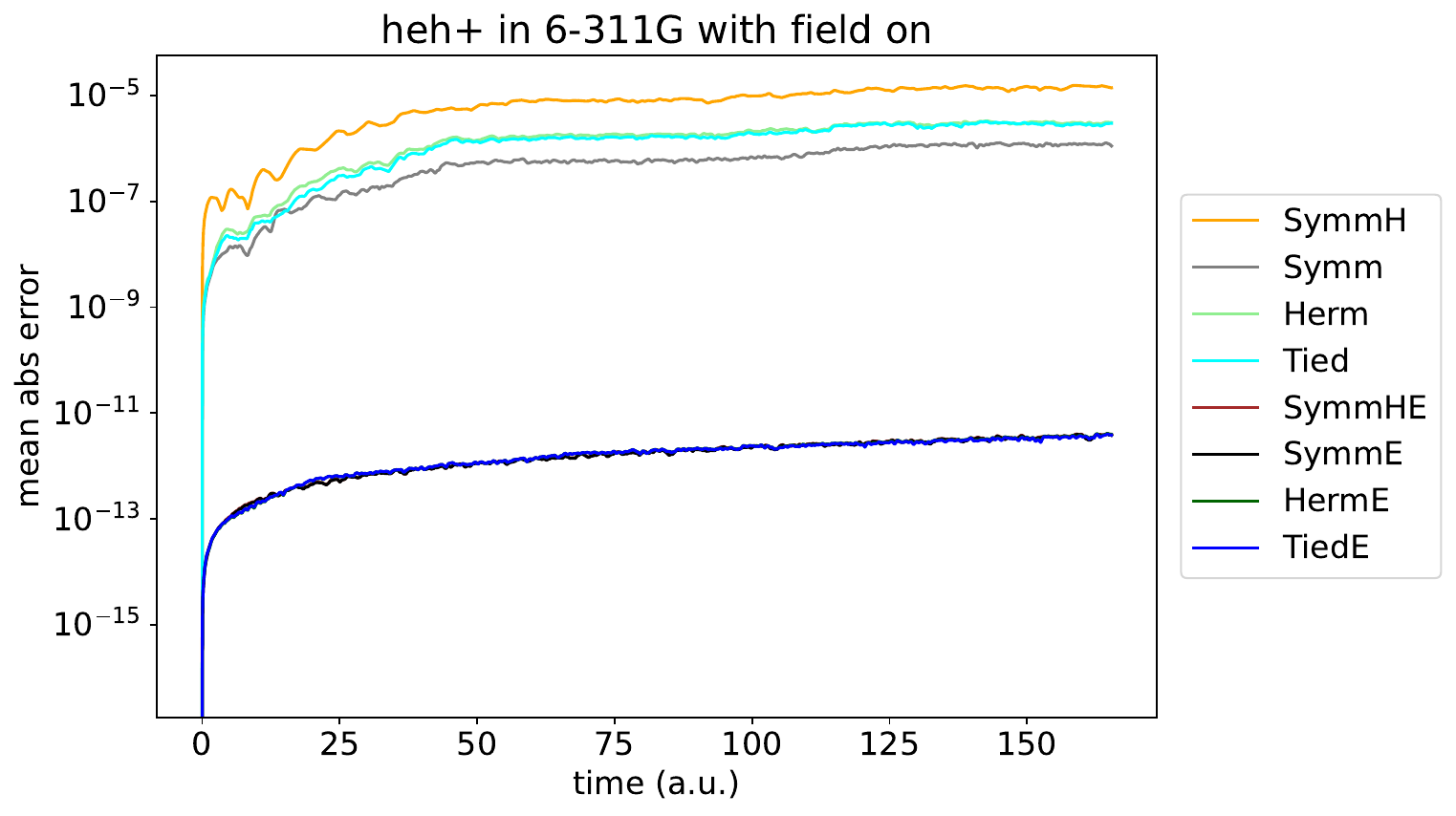} \\
\includegraphics[width=0.45\textwidth]{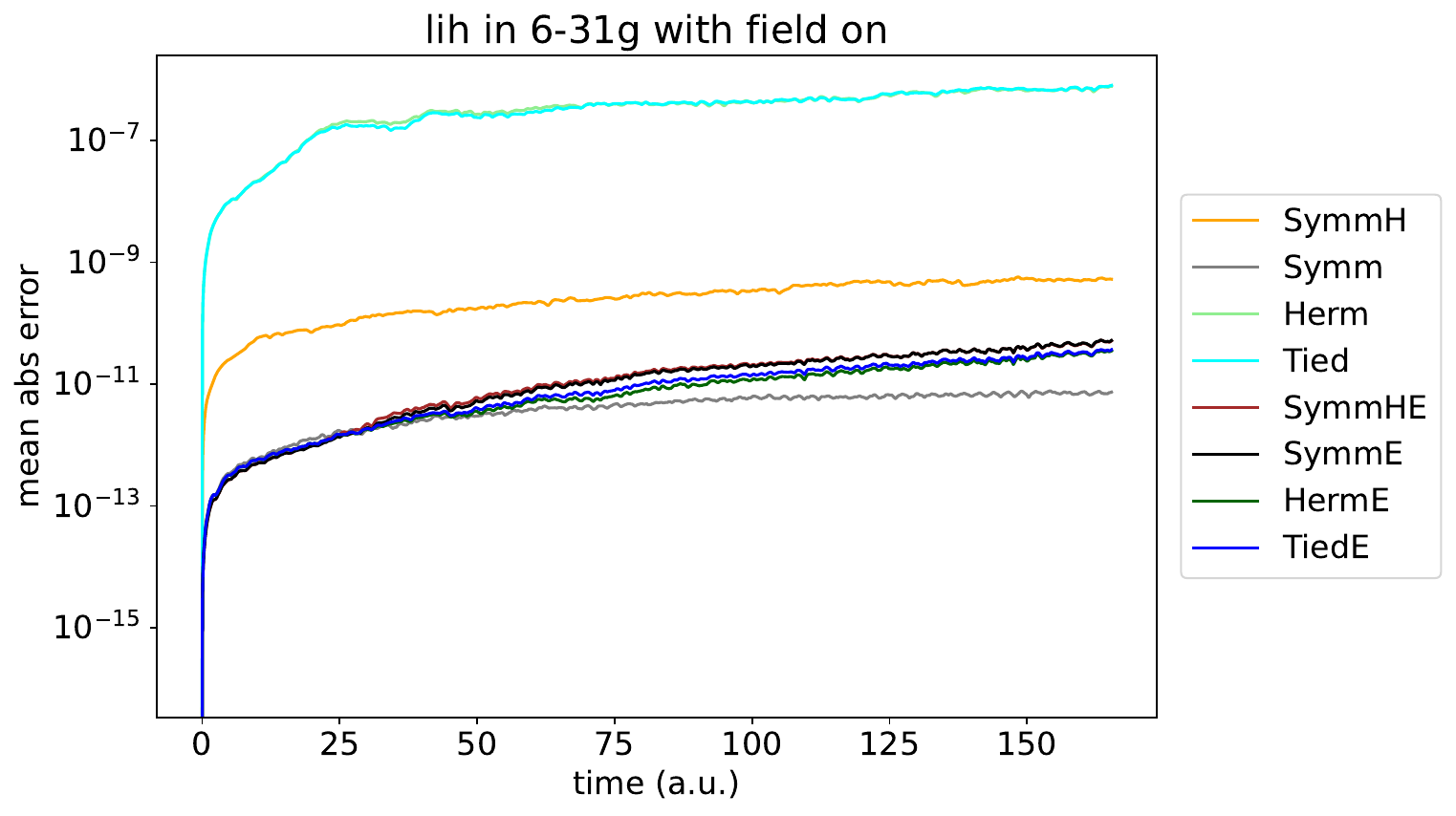}
\includegraphics[width=0.45\textwidth]{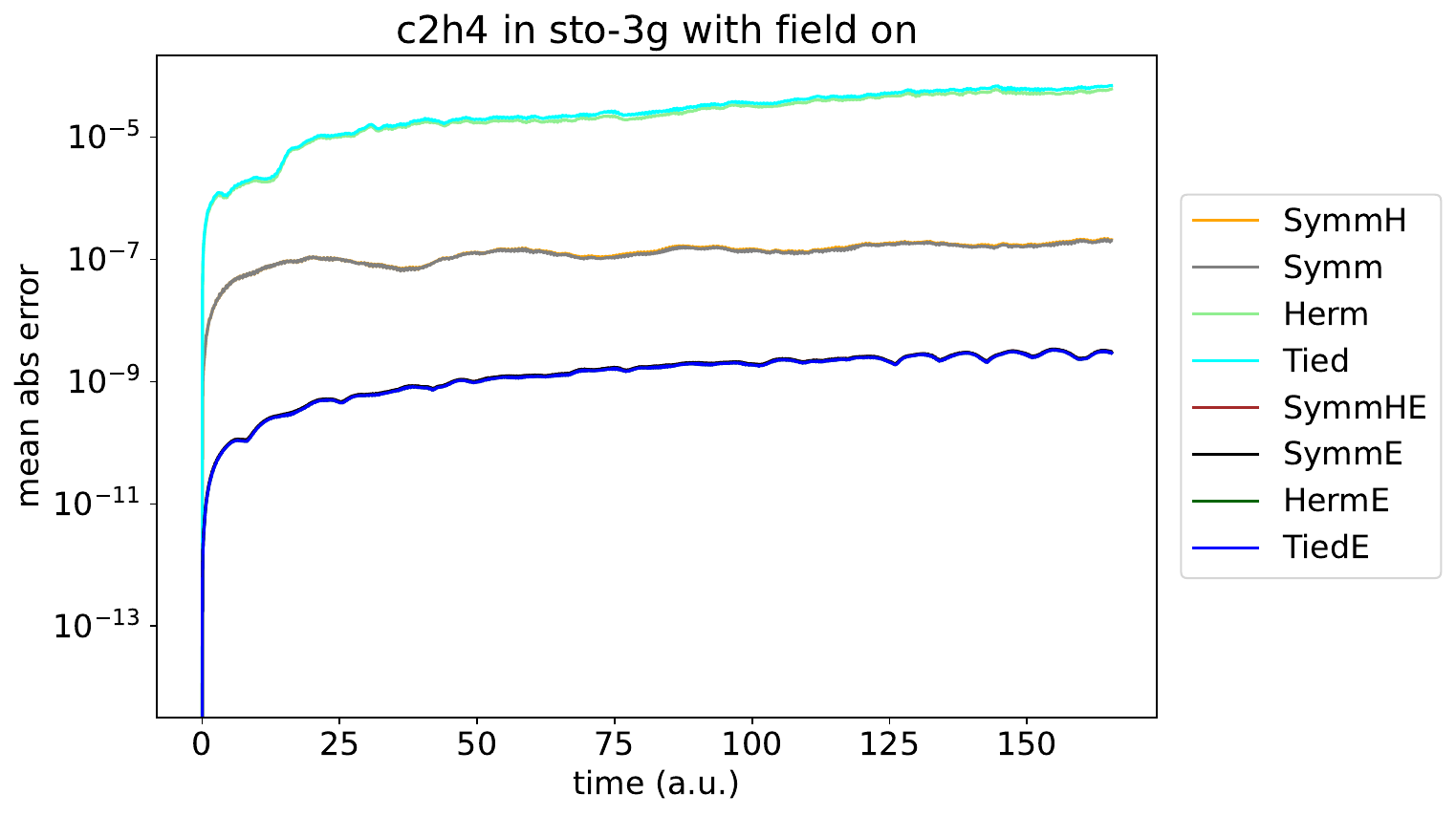} \\
\includegraphics[width=0.45\textwidth]{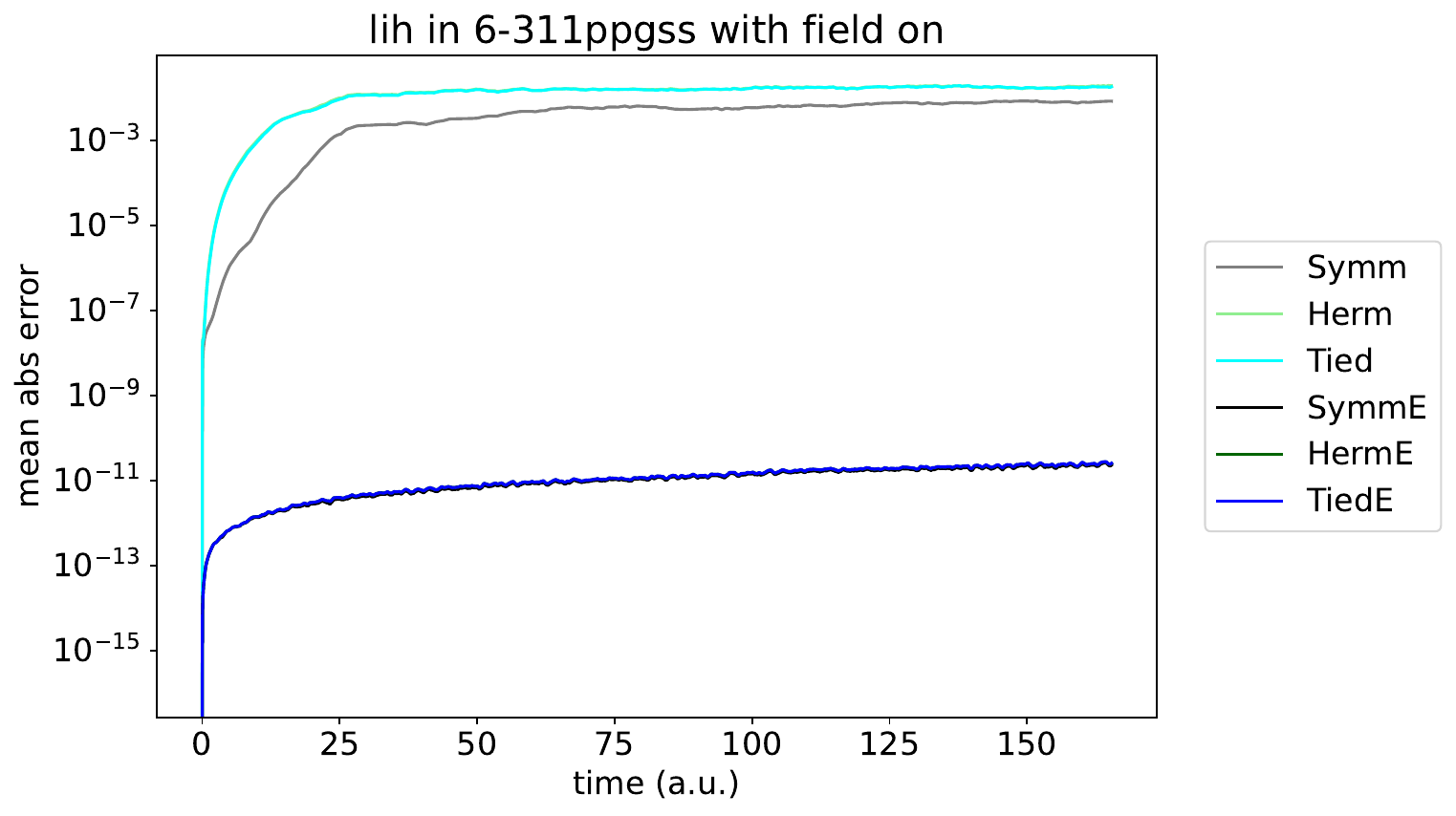}
\includegraphics[width=0.45\textwidth]{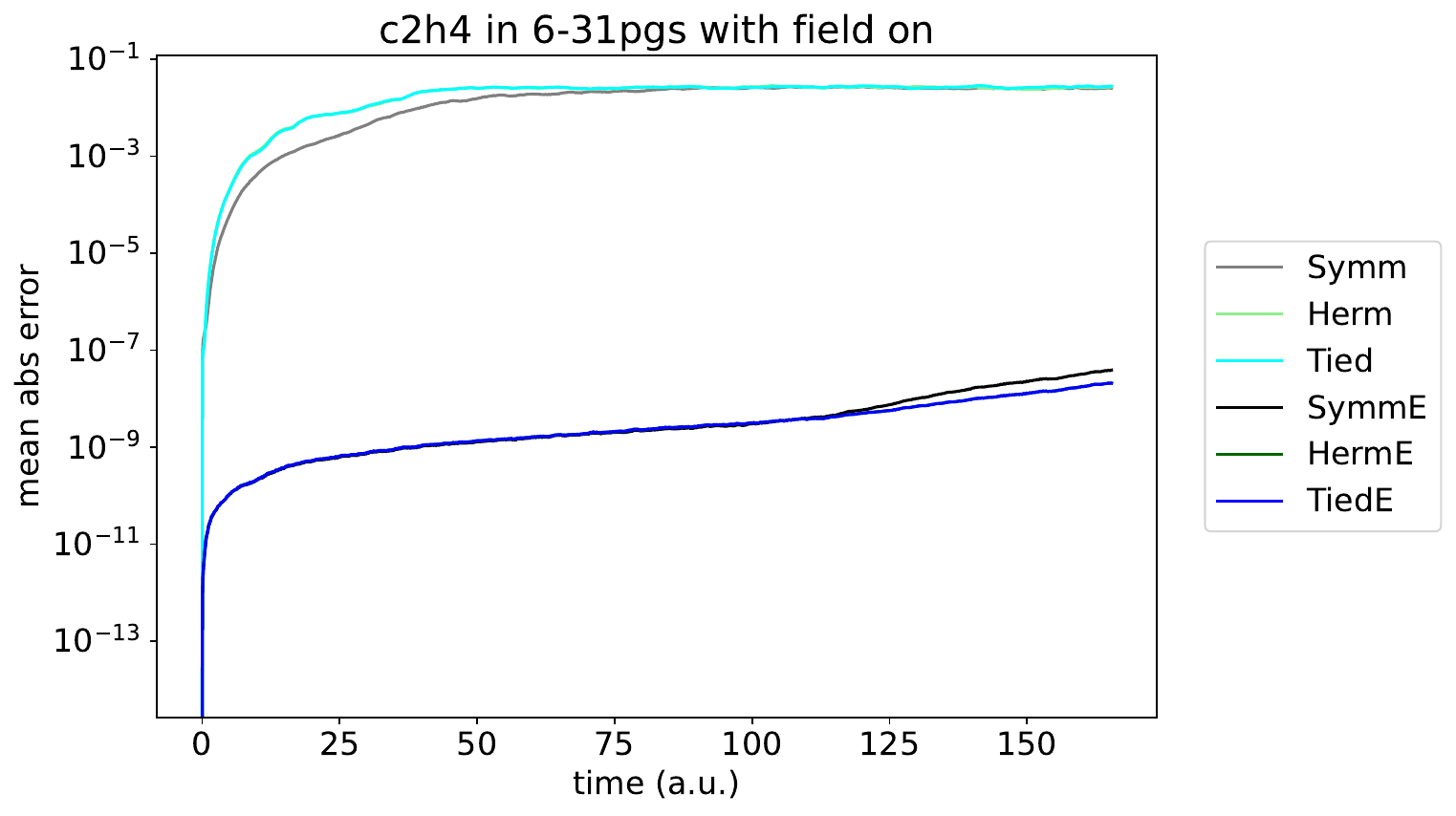} \\
\includegraphics[width=0.45\textwidth]{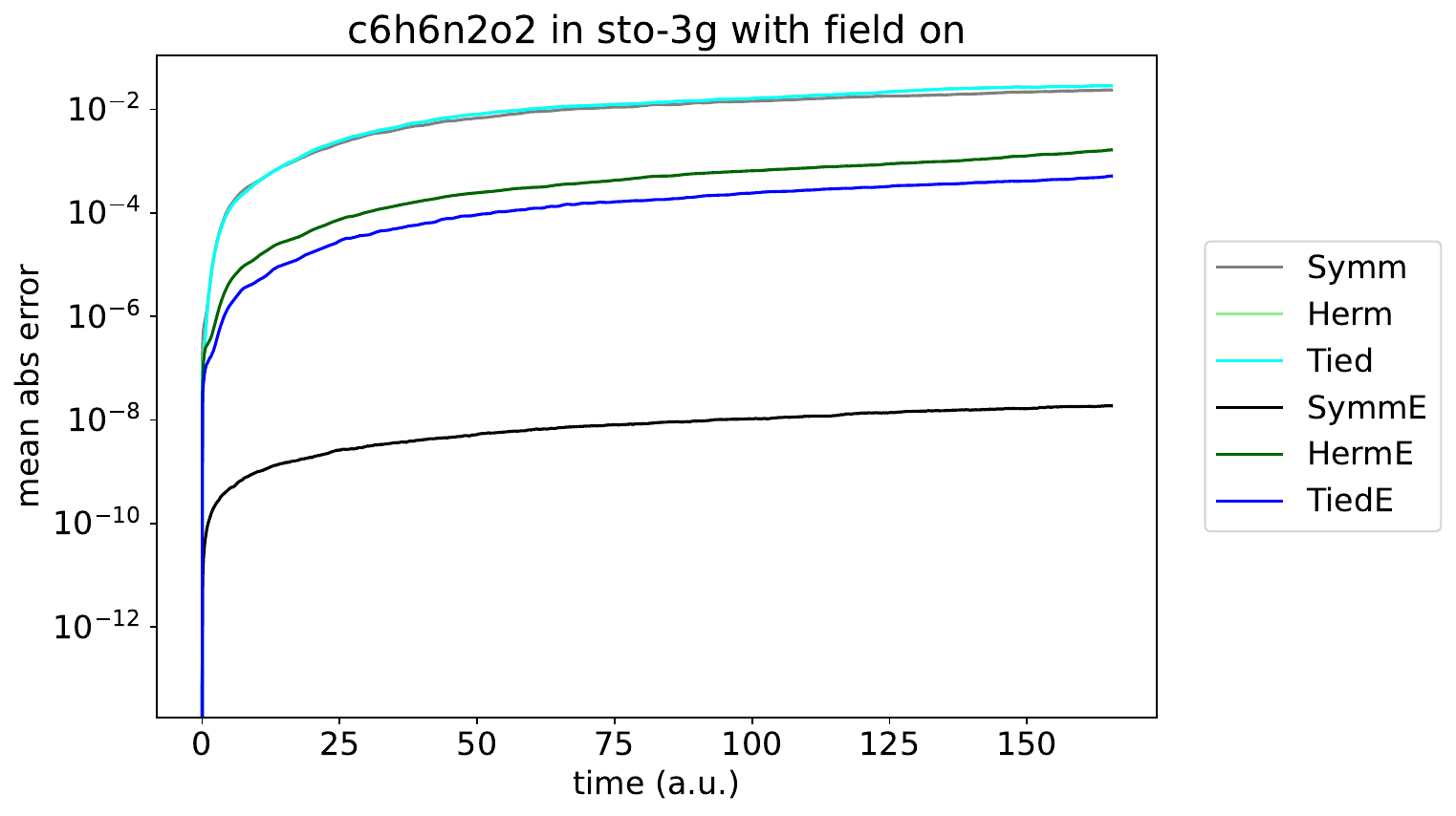}
\caption{We plot the field-on mean absolute propagation error (\ref{eqn:tdmae}) for all models and all molecular systems.}
\label{fig:MAEproperrFO}
\end{figure*}

Next, in Figure \ref{fig:MAEproperrFF} and Figure \ref{fig:MAEproperrFO}, respectively, we plot the field-off and field-on propagation errors using the MAE (mean absolute error) metric (\ref{eqn:tdmae}) described in the main text.  The horizontal axis is given in atomic units (a.u.) rather than the raw number of time steps.  Each plot includes $2 \times 10^5$ time steps worth of data for each MAE time series.  The results further support the conclusions given above and in the main text.

\begin{table*}
    \centering\footnotesize
    \caption{$ \| \cdot \|_{\infty} $ Propagation Errors (Field-Free). For each molecular system, the smallest propagation error value is boldfaced.}
    \label{tab:properrFF}
    \begin{tabular}{lccccccc}
        \toprule
        \thead{Model} & \thead{$\text{HeH}^+$ in \\6-31G} & \thead{$\text{HeH}^+$ in \\ 6-311G} & \thead{$\text{LiH}$ in \\6-31G} & \thead{$\text{C}_2 \text{H}_4$ in \\STO-3G} & \thead{$\text{LiH}$ in \\6-311ppgss} & \thead{$\text{C}_2 \text{H}_4$ in\\ 6-31pgs} & \thead{$\text{C}_6 \text{H}_6 \text{N}_2 \text{O}_2$\\ in STO-3G} \\
        \midrule
        Tied & $2.34 \! \times \! 10^{-11}$ & $4.92 \! \times \! 10^{-7}$ & $5.36 \! \times \! 10^{-6}$ & $6.06 \! \times \! 10^{-5}$ & $1.69 \! \times \! 10^{-5}$ & $7.63 \! \times \! 10^{-5}$ & $8.99 \! \times \! 10^{-6}$ \\
        Herm & $2.44 \! \times \! 10^{-11}$ & $5.74 \! \times \! 10^{-7}$ & $5.21 \! \times \! 10^{-6}$ & $6.17 \! \times \! 10^{-5}$ & $2.45 \! \times \! 10^{-5}$ & $6.65 \! \times \! 10^{-5}$ & $8.59 \! \times \! 10^{-6}$ \\
        Symm & $2.59 \! \times \! 10^{-11}$ & $8.27 \! \times \! 10^{-7}$ & $\mathbf{3.17 \! \times \! 10^{-10}}$ & $1.84 \! \times \! 10^{-6}$ & $3.79 \! \times \! 10^{-5}$ & $3.96 \! \times \! 10^{-4}$ & $2.23 \! \times \! 10^{-4}$ \\
        TiedE & $3.36 \! \times \! 10^{-11}$ & $\mathbf{4.96 \! \times \! 10^{-11}}$ & $4.55 \! \times \! 10^{-10}$ & $\mathbf{6.96 \! \times \! 10^{-8}}$ & $1.36 \! \times \! 10^{-9}$ & $1.91 \! \times \! 10^{-7}$ & $7.04 \! \times \! 10^{-5}$ \\
        HermE & $\mathbf{1.55 \! \times \! 10^{-11}}$ & $5.00 \! \times \! 10^{-11}$ & $4.64 \! \times \! 10^{-10}$ & $6.96 \! \times \! 10^{-8}$ & $1.36 \! \times \! 10^{-9}$ & $1.91 \! \times \! 10^{-7}$ & $1.32 \! \times \! 10^{-4}$ \\
        SymmE & $2.19 \! \times \! 10^{-11}$ & $6.05 \! \times \! 10^{-11}$ & $4.31 \! \times \! 10^{-10}$ & $7.06 \! \times \! 10^{-8}$ & $\mathbf{1.24 \! \times \! 10^{-9}}$ & $\mathbf{1.71 \! \times \! 10^{-7}}$ & $\mathbf{6.02 \! \times \! 10^{-7}}$ \\
        SymmH & $9.60 \! \times \! 10^{-7}$ & $4.66 \! \times \! 10^{-6}$ & $5.34 \! \times \! 10^{-9}$ & $1.85 \! \times \! 10^{-6}$ & na & na & na \\
        SymmHE & $2.89 \! \times \! 10^{-11}$ & $5.89 \! \times \! 10^{-11}$ & $4.44 \! \times \! 10^{-10}$ & $7.06 \! \times \! 10^{-8}$ & na & na & na \\
        \bottomrule
    \end{tabular}
\end{table*}

\begin{table*}
    \centering\footnotesize
    \caption{$ \| \cdot \|_{\infty} $ Propagation Errors (Field-On)}
    \label{tab:properrFO}
    \begin{tabular}{lccccccc}
        \toprule
        \thead{Model} & \thead{$\text{HeH}^+$ in \\6-31G} & \thead{$\text{HeH}^+$ in \\ 6-311G} & \thead{$\text{LiH}$ in \\6-31G} & \thead{$\text{C}_2 \text{H}_4$ in \\STO-3G} & \thead{$\text{LiH}$ in \\6-311ppgss} & \thead{$\text{C}_2 \text{H}_4$ in\\ 6-31pgs} & \thead{$\text{C}_6 \text{H}_6 \text{N}_2 \text{O}_2$\\ in STO-3G} \\
        \midrule
        Tied & $3.11 \! \times \! 10^{-11}$ & $5.40 \! \times \! 10^{-5}$ & $5.40 \! \times \! 10^{-6}$ & $9.93 \! \times \! 10^{-4}$ & $3.42 \! \times \! 10^{-1}$ & $7.72 \! \times \! 10^{-1}$ & $3.23 \! \times \! 10^{-1}$ \\
        Herm & $2.68 \! \times \! 10^{-11}$ & $5.01 \! \times \! 10^{-5}$ & $6.08 \! \times \! 10^{-6}$ & $9.36 \! \times \! 10^{-4}$ & $3.49 \! \times \! 10^{-1}$ & $6.31 \! \times \! 10^{-1}$ & $3.34 \! \times \! 10^{-1}$ \\
        Symm & $\mathbf{1.16 \! \times \! 10^{-11}}$ & $1.53 \! \times \! 10^{-5}$ & $\mathbf{6.10 \! \times \! 10^{-11}}$ & $1.84 \! \times \! 10^{-6}$ & $1.29 \! \times \! 10^{-1}$ & $5.62 \! \times \! 10^{-1}$ & $3.22 \! \times \! 10^{-1}$ \\
        TiedE & $2.48 \! \times \! 10^{-11}$ & $\mathbf{4.77 \! \times \! 10^{-11}}$ & $4.20 \! \times \! 10^{-10}$ & $\mathbf{8.63 \! \times \! 10^{-8}}$ & $1.41 \! \times \! 10^{-9}$ & $\mathbf{5.88 \! \times \! 10^{-7}}$ & $3.71 \! \times \! 10^{-3}$ \\
        HermE & $3.08 \! \times \! 10^{-11}$ & $4.88 \! \times \! 10^{-11}$ & $4.01 \! \times \! 10^{-10}$ & $\mathbf{8.63 \! \times \! 10^{-8}}$ & $1.41 \! \times \! 10^{-9}$ & $\mathbf{5.88 \! \times \! 10^{-7}}$ & $1.74 \! \times \! 10^{-2}$ \\
        SymmE & $4.05 \! \times \! 10^{-11}$ & $5.83 \! \times \! 10^{-11}$ & $5.85 \! \times \! 10^{-10}$ & $8.86 \! \times \! 10^{-8}$ & $\mathbf{1.31 \! \times \! 10^{-9}}$ & $1.45 \! \times \! 10^{-6}$ & $\mathbf{4.91 \! \times \! 10^{-7}}$ \\
        SymmH & $1.42 \! \times \! 10^{-4}$ & $2.43 \! \times \! 10^{-4}$ & $5.74 \! \times \! 10^{-9}$ & $1.88 \! \times \! 10^{-6}$ & na & na & na \\
        SymmHE & $2.59 \! \times \! 10^{-11}$ & $5.96 \! \times \! 10^{-11}$ & $5.76 \! \times \! 10^{-10}$ & $8.86 \! \times \! 10^{-8}$ & na & na & na \\
        \bottomrule
    \end{tabular}
\end{table*}

%

\subsubsection{Hamiltonian and Commutator Errors}
\label{sect:supphamcommerrors}
We report in Table \ref{tab:LinftyHamError} all Hamiltonian errors (\ref{eqn:hamerrors}) that we plotted in the left-most plot of Figure \ref{fig:HamCommErrors} in the main text.

In Table \ref{tab:LinftyCommErrorFF} and Table \ref{tab:LinftyCommErrorFO}, we report, respectively, the field-free and field-on commutator errors (\ref{eqn:commerror}) that we plotted in the middle and right-most plots of Figure \ref{fig:HamCommErrors} in the main text.

\begin{table*}
    \centering\footnotesize
    \caption{$\| \cdot \|_\infty$ Hamiltonian error}
    \label{tab:LinftyHamError}
    \begin{tabular}{lccccccc}
        \toprule
        \thead{Model} & \thead{$\text{HeH}^+$ in \\6-31G} & \thead{$\text{HeH}^+$ in \\ 6-311G} & \thead{$\text{LiH}$ in \\6-31G} & \thead{$\text{C}_2 \text{H}_4$ in \\STO-3G} & \thead{$\text{LiH}$ in \\6-311ppgss} & \thead{$\text{C}_2 \text{H}_4$ in\\ 6-31pgs} & \thead{$\text{C}_6 \text{H}_6 \text{N}_2 \text{O}_2$\\ in STO-3G} \\
        \midrule
        Tied & $1.18 \! \times \! 10^{0}$ & $1.23 \! \times \! 10^{0}$ & $7.57 \! \times \! 10^{-1}$ & $1.50 \! \times \! 10^{0}$ & $7.78 \! \times \! 10^{-1}$ & $1.14 \! \times \! 10^{0}$ & $2.02 \! \times \! 10^{0}$ \\
        Herm & $1.18 \! \times \! 10^{0}$ & $1.23 \! \times \! 10^{0}$ & $7.49 \! \times \! 10^{-1}$ & $1.50 \! \times \! 10^{0}$ & $7.78 \! \times \! 10^{-1}$ & $1.18 \! \times \! 10^{0}$ & $2.06 \! \times \! 10^{0}$ \\
        Symm & $\mathbf{6.07 \! \times \! 10^{-1}}$ & $7.89 \! \times \! 10^{-1}$ & $3.10 \! \times \! 10^{-1}$ & $8.53 \! \times \! 10^{-1}$ & $4.42 \! \times \! 10^{-1}$ & $5.87 \! \times \! 10^{-1}$ & $2.05 \! \times \! 10^{0}$ \\
        TiedE & $1.18 \! \times \! 10^{0}$ & $1.03 \! \times \! 10^{0}$ & $6.67 \! \times \! 10^{-1}$ & $1.22 \! \times \! 10^{0}$ & $6.02 \! \times \! 10^{-1}$ & $9.36 \! \times \! 10^{-1}$ & $5.81 \! \times \! 10^{-1}$ \\
        HermE & $1.18 \! \times \! 10^{0}$ & $1.03 \! \times \! 10^{0}$ & $6.66 \! \times \! 10^{-1}$ & $1.22 \! \times \! 10^{0}$ & $6.02 \! \times \! 10^{-1}$ & $9.36 \! \times \! 10^{-1}$ & $5.88 \! \times \! 10^{-1}$ \\
        SymmE & $\mathbf{6.07 \! \times \! 10^{-1}}$ & $\mathbf{4.83 \! \times \! 10^{-1}}$ & $\mathbf{2.46 \! \times \! 10^{-1}}$ & $\mathbf{4.92 \! \times \! 10^{-1}}$ & $\mathbf{2.16 \! \times \! 10^{-1}}$ & $\mathbf{3.20 \! \times \! 10^{-1}}$ & $\mathbf{2.60 \! \times \! 10^{-1}}$ \\
        SymmH & $8.43 \! \times \! 10^{-1}$ & $8.24 \! \times \! 10^{-1}$ & $3.10 \! \times \! 10^{-1}$ & $8.53 \! \times \! 10^{-1}$ & na & na & na \\
        SymmHE & $\mathbf{6.07 \! \times \! 10^{-1}}$ & $\mathbf{4.83 \! \times \! 10^{-1}}$ & $\mathbf{2.46 \! \times \! 10^{-1}}$ & $\mathbf{4.92 \! \times \! 10^{-1}}$ & na & na & na \\
        \bottomrule
    \end{tabular}
\end{table*}

\begin{table*}
    \centering\footnotesize
    \caption{$\| \cdot \|_\infty$ commutator error (field-free)}
    \label{tab:LinftyCommErrorFF}
    \begin{tabular}{lccccccc}
        \toprule
        \thead{Model} & \thead{$\text{HeH}^+$ in \\6-31G} & \thead{$\text{HeH}^+$ in \\ 6-311G} & \thead{$\text{LiH}$ in \\6-31G} & \thead{$\text{C}_2 \text{H}_4$ in \\STO-3G} & \thead{$\text{LiH}$ in \\6-311ppgss} & \thead{$\text{C}_2 \text{H}_4$ in\\ 6-31pgs} & \thead{$\text{C}_6 \text{H}_6 \text{N}_2 \text{O}_2$\\ in STO-3G} \\
        \midrule
        Tied & $5.42 \! \times \! 10^{-14}$ & $1.45 \! \times \! 10^{-7}$ & $1.92 \! \times \! 10^{-8}$ & $1.30 \! \times \! 10^{-5}$ & $5.38 \! \times \! 10^{-6}$ & $1.11 \! \times \! 10^{-4}$ & $5.14 \! \times \! 10^{-5}$ \\
        Herm & $5.21 \! \times \! 10^{-14}$ & $1.91 \! \times \! 10^{-7}$ & $2.37 \! \times \! 10^{-8}$ & $1.23 \! \times \! 10^{-5}$ & $5.73 \! \times \! 10^{-6}$ & $1.11 \! \times \! 10^{-4}$ & $5.52 \! \times \! 10^{-5}$ \\
        Symm & $\mathbf{4.94 \! \times \! 10^{-14}}$ & $7.39 \! \times \! 10^{-8}$ & $8.59 \! \times \! 10^{-13}$ & $2.27 \! \times \! 10^{-7}$ & $1.40 \! \times \! 10^{-6}$ & $2.58 \! \times \! 10^{-5}$ & $1.30 \! \times \! 10^{-4}$ \\
        TiedE & $5.84 \! \times \! 10^{-14}$ & $\mathbf{4.68 \! \times \! 10^{-13}}$ & $8.78 \! \times \! 10^{-13}$ & $5.93 \! \times \! 10^{-10}$ & $\mathbf{1.04 \! \times \! 10^{-11}}$ & $2.79 \! \times \! 10^{-9}$ & $6.38 \! \times \! 10^{-6}$ \\
        HermE & $5.86 \! \times \! 10^{-14}$ & $\mathbf{4.68 \! \times \! 10^{-13}}$ & $8.78 \! \times \! 10^{-13}$ & $5.93 \! \times \! 10^{-10}$ & $\mathbf{1.04 \! \times \! 10^{-11}}$ & $2.79 \! \times \! 10^{-9}$ & $1.08 \! \times \! 10^{-5}$ \\
        SymmE & $5.11 \! \times \! 10^{-14}$ & $5.43 \! \times \! 10^{-13}$ & $\mathbf{7.81 \! \times \! 10^{-13}}$ & $\mathbf{5.79 \! \times \! 10^{-10}}$ & $\mathbf{1.04 \! \times \! 10^{-11}}$ & $\mathbf{2.50 \! \times \! 10^{-9}}$ & $\mathbf{4.49 \! \times \! 10^{-9}}$ \\
        SymmH & $6.21 \! \times \! 10^{-7}$ & $3.23 \! \times \! 10^{-6}$ & $1.67 \! \times \! 10^{-10}$ & $2.28 \! \times \! 10^{-7}$ & na & na & na \\
        SymmHE & $5.11 \! \times \! 10^{-14}$ & $5.43 \! \times \! 10^{-13}$ & $\mathbf{7.81 \! \times \! 10^{-13}}$ & $\mathbf{5.79 \! \times \! 10^{-10}}$ & na & na & na \\
        \bottomrule
    \end{tabular}
\end{table*}

\begin{table*}
    \centering\footnotesize
    \caption{$\| \cdot \|_\infty$ commutator error (field-on)}
    \label{tab:LinftyCommErrorFO}
    \begin{tabular}{lccccccc}
        \toprule
        \thead{Model} & \thead{$\text{HeH}^+$ in \\6-31G} & \thead{$\text{HeH}^+$ in \\ 6-311G} & \thead{$\text{LiH}$ in \\6-31G} & \thead{$\text{C}_2 \text{H}_4$ in \\STO-3G} & \thead{$\text{LiH}$ in \\6-311ppgss} & \thead{$\text{C}_2 \text{H}_4$ in\\ 6-31pgs} & \thead{$\text{C}_6 \text{H}_6 \text{N}_2 \text{O}_2$\\ in STO-3G} \\
        \midrule
        Tied & $5.42 \! \times \! 10^{-14}$ & $1.45 \! \times \! 10^{-7}$ & $1.92 \! \times \! 10^{-8}$ & $1.30 \! \times \! 10^{-5}$ & $5.38 \! \times \! 10^{-6}$ & $1.11 \! \times \! 10^{-4}$ & $5.14 \! \times \! 10^{-5}$ \\
        Herm & $5.21 \! \times \! 10^{-14}$ & $1.91 \! \times \! 10^{-7}$ & $2.37 \! \times \! 10^{-8}$ & $1.23 \! \times \! 10^{-5}$ & $5.73 \! \times \! 10^{-6}$ & $1.11 \! \times \! 10^{-4}$ & $5.52 \! \times \! 10^{-5}$ \\
        Symm & $\mathbf{4.94 \! \times \! 10^{-14}}$ & $7.39 \! \times \! 10^{-8}$ & $8.59 \! \times \! 10^{-13}$ & $2.27 \! \times \! 10^{-7}$ & $1.40 \! \times \! 10^{-6}$ & $2.58 \! \times \! 10^{-5}$ & $1.30 \! \times \! 10^{-4}$ \\
        TiedE & $6.47 \! \times \! 10^{-14}$ & $\mathbf{5.28 \! \times \! 10^{-13}}$ & $8.78 \! \times \! 10^{-13}$ & $5.88 \! \times \! 10^{-10}$ & $1.14 \! \times \! 10^{-11}$ & $2.79 \! \times \! 10^{-9}$ & $3.64 \! \times \! 10^{-4}$ \\
        HermE & $6.43 \! \times \! 10^{-14}$ & $\mathbf{5.28 \! \times \! 10^{-13}}$ & $8.78 \! \times \! 10^{-13}$ & $5.88 \! \times \! 10^{-10}$ & $1.14 \! \times \! 10^{-11}$ & $2.79 \! \times \! 10^{-9}$ & $7.56 \! \times \! 10^{-4}$ \\
        SymmE & $5.60 \! \times \! 10^{-14}$ & $5.42 \! \times \! 10^{-13}$ & $\mathbf{7.81 \! \times \! 10^{-13}}$ & $\mathbf{5.79 \! \times \! 10^{-10}}$ & $\mathbf{1.11 \! \times \! 10^{-11}}$ & $\mathbf{2.50 \! \times \! 10^{-9}}$ & $\mathbf{4.48 \! \times \! 10^{-9}}$ \\
        SymmH & $2.70 \! \times \! 10^{-5}$ & $6.92 \! \times \! 10^{-6}$ & $1.43 \! \times \! 10^{-10}$ & $2.20 \! \times \! 10^{-7}$ & na & na & na \\
        SymmHE & $5.60 \! \times \! 10^{-14}$ & $5.42 \! \times \! 10^{-13}$ & $\mathbf{7.81 \! \times \! 10^{-13}}$ & $\mathbf{5.79 \! \times \! 10^{-10}}$ & na & na & na \\
        \bottomrule
    \end{tabular}
\end{table*}

\clearpage

\subsection{Coordinates}
\label{sect:coordinates}
For each molecule studied in the paper, we provide the Cartesian coordinates used:
\subsubsection{$\heh$}

{\setlength{\tabcolsep}{12pt}

\begin{tabular}{cccc}
  Atom & X & Y & Z \\ \hline
 H            &         $0.$   &     $0.$   &    $-0.386$ \\ 
 He           &         $0.$   &     $0.$   &     $0.386$ 
\end{tabular}

\subsubsection{$\lih$}
\begin{tabular}{cccc}
  Atom & X & Y & Z \\ \hline
 H                  &   $0.$    &    $0.$    &   $-0.765$ \\ 
 Li                 &   $0.$    &    $0.$    &    $0.765$ 
\end{tabular}

\subsubsection{$\ethylene$}
\begin{tabular}{cccc}
 Atom & X & Y & Z \\ \hline
 C        &             $0.$    &    $0.$      &  $0.6695$ \\
 C        &             $0.$    &    $0.$      & $-0.6695$ \\
 H        &             $0.$    &    $0.9289$  &  $1.2321$ \\
 H        &             $0.$    &   $-0.9289$  &  $1.2321$ \\
 H        &             $0.$    &    $0.9289$  & $-1.2321$ \\
 H        &             $0.$    &   $-0.9289$  & $-1.2321$ 
\end{tabular}

\subsubsection{$\paranitro$}
\setlength{\tabcolsep}{5pt}
\begin{tabular}{cccc}
  Atom & X & Y & Z \\ \hline
 C         &          $ -2.15314 $ & $  0.62401 $ & $  0.01284 $\\
 C         &          $ -1.48656 $ & $  1.87488 $ & $ -0.06819 $\\
 C         &          $ -0.04998 $ & $  1.92027 $ & $ -0.04738 $\\
 H         &          $ -1.90848 $ & $  2.54685 $ & $  0.72995 $\\
 H         &          $ -1.91724 $ & $  2.4975  $ & $ -0.89528 $\\
 C         &          $  0.67266 $ & $  0.6908  $ & $ -0.04161 $\\
 H         &          $  0.31093 $ & $  2.54359 $ & $  0.81952 $\\
 H         &          $  0.37127 $ & $  2.63697 $ & $ -0.7987 $\\
 C         &          $  0.009  $ & $  -0.5715  $ & $ -0.05165 $\\
 C         &          $ -1.42839 $ & $ -0.59477 $ & $ -0.05804 $\\
 H         &          $  0.41042 $ & $ -1.1899  $ & $  0.80036 $\\
 H         &          $  0.45446 $ & $ -1.25658 $ & $ -0.81887 $\\
 H         &          $ -1.80766 $ & $ -1.26992 $ & $  0.75914 $\\
 H         &          $ -1.8386 $ & $  -1.25338 $ & $ -0.86697 $\\
 N         &          $ -3.53937 $ & $  0.59033 $ & $ -0.21327 $\\
 H         &          $ -4.02567 $ & $  1.42109 $ & $  0.11027 $\\
 H         &          $ -3.98594 $ & $ -0.25957 $ & $  0.11778 $\\
 N         &          $  2.1583  $ & $  0.72591 $ & $ -0.0236 $\\
 O         &          $  2.70665 $ & $  1.83755 $ & $ -0.00548 $\\
 O         &          $  2.75876 $ & $ -0.35845 $ & $ -0.01779 $
\end{tabular}}

\end{document}